\documentclass[traditabstract]{aa}
\usepackage{verbatim} 
\usepackage{amsmath}
\usepackage{amssymb}
\usepackage{graphicx}
\usepackage{booktabs}
\usepackage[authoryear]{natbib}
\usepackage[colorlinks=true,linkcolor=blue,urlcolor=blue,citecolor=blue]{hyperref}
\usepackage{xspace}
\usepackage{siunitx}
\usepackage{mathrsfs}
\usepackage{placeins}
\usepackage{lscape}
\usepackage{relsize}

\makeatletter

\usepackage{txfonts}\usepackage{twoopt}
\bibpunct{(}{)}{;}{a}{}{,}
\usepackage{enumitem}

\titlerunning{Contact tracing of binary stars}
\authorrunning{J.~Henneco et al.}
\makeatother
\begin{document}
\title{Contact tracing of binary stars: Pathways to stellar mergers}
\author{
J.~Henneco\inst{\ref{HITS}}\thanks{jan.henneco@protonmail.com}, F.\,R.\,N.\,~Schneider\inst{\ref{HITS},\,\ref{ZAH}} \and E.~Laplace\inst{\ref{HITS}}
}
\institute{Heidelberger Institut f\"{u}r Theoretische Studien, Schloss-Wolfsbrunnenweg 35, 69118 Heidelberg, Germany\label{HITS} \and Zentrum f\"{u}r Astronomie der Universit\"{a}t Heidelberg, Astronomisches Rechen-Institut, M\"{o}nchhofstr. 12-14, 69120 Heidelberg, Germany\label{ZAH}
}
\date{Received xxxxx / Accepted xxxxx}

\abstract{%
    Stellar mergers are responsible for a wide variety of phenomena such as rejuvenated blue stragglers, highly magnetised stars, spectacular transients, iconic nebulae, and stars with peculiar surface chemical abundances and rotation rates. Before stars merge, they enter a contact phase. Here, we investigate which initial binary-star configurations lead to contact and classical common-envelope (CE) phases and assess the likelihood of a subsequent merger. To this end, we compute a grid of about 6000 detailed 1D binary evolution models with initial component masses of $0.5\text{--}20.0\,\mathrm{M}_{\odot}$ at solar metallicity. Both components are evolved, and rotation and tides are taken into account. We identify five mechanisms that lead to contact and mergers: runaway mass transfer, mass loss through the outer Lagrange point $\mathrm{L}_{2}$, expansion of the accretor, orbital decay because of tides, and non-conservative mass transfer. At least $40\%$ of mass-transferring binaries with initial primary-star masses of $5\text{--}20\,\mathrm{M}_{\odot}$ evolve into a contact phase; ${>}\,12\%$ and ${>}\,19\%$ likely merge and evolve into a CE phase, respectively. Because of non-conservative mass transfer in our models, classical CE evolution from late Case-B and Case-C binaries is only found for initial mass ratios $q_\mathrm{i}<0.15\text{--}0.35$. For larger mass ratios, we find stable mass transfer. In early Case-B binaries, contact occurs for initial mass ratios $q_\mathrm{i}<0.15\text{--}0.35$, while in Case-A mass transfer, this is the case for all $q_\mathrm{i}$ in binaries with the initially closest orbits and $q_\mathrm{i}<0.35$ for initially wider binaries. Our models predict that most Case-A binaries with mass ratios of $q<0.5$ upon contact mainly get into contact because of runaway mass transfer and accretor expansion on a thermal timescale,  with subsequent $\mathrm{L}_{2}$-overflow in more than half of the cases. Thus, these binaries likely merge quickly after establishing contact or remain in contact only for a thermal timescale. Contrarily, Case-A contact binaries with higher mass ratios form through accretor expansion on a nuclear timescale and can thus give rise to long-lived contact phases before a possible merger. Observationally, massive contact binaries are almost exclusively found with mass ratios $q>0.5$, confirming our model expectations. Because of non-conservative mass transfer with mass transfer efficiencies of $15\text{--}65\%$, $5\text{--}25\%$ and $25\text{--}50\%$ in Case-A, -B and -C mass transfer, respectively (for primary-star masses above $3\,\mathrm{M}_{\odot}$), our contact, merger and classical CE incidence rates are conservative lower limits. With more conservative mass transfer, these incidences would increase. Moreover, in most binaries, the non-accreted mass cannot be ejected, raising the question of the further evolution of such systems. The non-accreted mass may settle into circumstellar and circumbinary disks, but could also lead to further contact systems and mergers. Overall, contact binaries are a frequent and fascinating result of binary mass transfer of which the exact outcomes still remain to be understood and explored further.
}

\keywords{binaries: general -- stars: evolution -- stars: massive -- stars: low-mass -- methods: numerical}

\maketitle

\section{Introduction}\label{sec:introduction}
Mergers of non-compact stars frequently occur in the Universe \citep{Podsiadlowski1992a, Sana2012, deMink2014}. They can be caused by the evolution of the components in a binary system, during which the stars come into contact because of their radial expansion or orbital decay. Such orbital decay is not necessarily a result of the binary evolution itself, but can also be induced by von Zeipel-Kozai-Lidov oscillations \citep{vonZeipel1910, Lidov1962, Kozai1962, Naoz2016} caused by a third component or a circumbinary disk \citep[e.g.][]{Lubow2000, Perets2009, Toonen2020, Toonen2022}. A third option is dynamically driven mergers, which occur during close encounters between stars in dense stellar environments \citep[e.g.][]{Hills1976, PortegiesZwart1997, PortegiesZwart1999, PortegiesZwart2004}.

The products of stellar mergers can explain a multitude of objects. Examples include blue stragglers \citep[e.g.][]{Rasio1995a, Sills1997, Sills2001, Mapelli2006, Glebbeek2008, Ferraro2012, Schneider2015}, some of the most massive stars observed in the Universe \citep[e.g.][]{PortegiesZwart1999, Banerjee2012, Schneider2014, Boekholt2018}, B[e] supergiants \citep[e.g.][]{Podsiadlowski2006, Wu2020}, OBA stars with large-scale surface magnetic fields \citep{Schneider2019}, highly magnetic white dwarfs and magnetars \citep[e.g.][]{Tout2004, Ferrario2005, Wickramasinghe2005, Wickramasinghe2014,Shenar2023}, and $\alpha$-rich young stars \citep[e.g.][]{Chiappini2015,Martig2015,Izzard2018,Hekker2019}. Transients linked to stellar mergers include supernovae such as the core-collapse supernova SN~1987A \citep[e.g.][]{Podsiadlowski1990, Podsiadlowski1992b, Morris2007}, the Great Eruption of $\eta~$Car \citep{Frew2004, Gallagher1989, Iben1999, Podsiadlowski2006, Morris2006, Podsiadlowski2010, Fitzpatrick2012, PortegiesZwart2016, Smith2018c, Owocki2019, Hirai2021}, and luminous red novae such as V1309~Sco \citep{Tylenda2011,Stepien2011}, V838 Mon \citep{Soker2007} and V4332 Sgr \citep{Tylenda2005}.\\

3D simulations of stellar mergers with magnetohydrodynamics \citep[MHD,][]{Schneider2019} and smoothed particle hydrodynamics \citep[SPH,][]{Sills1997, Sills2001, Freitag2005, Dale2006, Suzuki2007, Gaburov2008, Antonini2011, Glebbeek2013, Ballone2023} codes provide useful insights into the merger events and merger products \citep{Schneider2020, Costa2022}. However, to obtain realistic initial conditions for these computationally expensive simulations, it is crucial to understand in which binary configurations stellar mergers are most commonly expected to occur. Moreover, it is important to characterise the interior structures of stars directly before they enter a merger phase, as these largely determine the merger outcome.

Before stars in a binary system merge, they go through a phase of contact \citep[e.g.][]{Langer2012}. Yet, not every contact phase necessarily leads to a merger. These contact phases can be (over-)contact binaries in which both stars (over-)fill their Roche lobe\footnote{For simplicity, we will use ``contact'' to refer to both ``contact'' and ``overcontact'' systems.}, or classical common-envelope (CE) phases, in which one star is engulfed by the envelope of the other star \citep[e.g.][]{Ivanova2013}. Hence, as a first step towards predicting the occurrence of mergers, it is important to find out in which binary configurations contact phases occur and what the outcomes of these phases are. Mapping the occurrence of contact binaries using detailed 1D binary evolution simulations has been done, e.g. by \citet{Pols1994}, \citet{Wellstein2001}, \citet{deMink2007}, \citet{Claeys2011} and \citet{Mennekens2017} for massive stars. \citet{Marchant2016} and \citet{Menon2021} additionally computed through contact binary phases, which yields information on their lifetime and stability (i.e. likelihood to merge). More recently, three extended grids of detailed binary evolution models, spanning masses of $0.5$-$300\,\mathrm{M}_{\odot}\,$, have been computed as part of the binary population synthesis code \texttt{POSYDON} \citep{Fragos2023}. These contain valuable information about the onset of contact phases. Using rapid binary population synthesis codes, \citet{deMink2014} and \citet{Schneider2014} evolved entire populations of binary systems and mapped the occurrence of contact phases.\\

In this work, we use a grid of low-mass and massive binary evolution models with component masses between 0.5 and 20.0 $\mathrm{M}_{\odot}$ at solar metallicity ($Z=0.0142$, \citealt{Asplund2009}) to trace the occurrence of contact phases over the complete range of initial mass ratios\footnote{In this work, the mass ratio $q$ is always defined as the mass of the less massive star over the mass of the more massive star.} and orbital separations. We evolve both components and include physical processes such as stellar winds, rotation and tidal interactions. It is known that binary systems can evolve towards tidal instabilities, which lead to rapid orbital decay and subsequent mergers \citep{Darwin1879}. These instabilities have been proposed to be responsible for the lack of observations of W~UMa type contact binaries at low mass ratios \citep{Rasio1995a} and the final spiral-in of the progenitor system of V1309 Sco \citep{Stepien2011}.

Including these physical processes allows us to arrive at a picture of the physical mechanisms leading to contact and their likelihood to lead to stellar mergers that is as complete as possible. Moreover, it illustrates the relative importance of, for example, tides, wind-mass loss, and mass-transfer efficiency on the evolution of binaries. We allow for non-conservative mass transfer and expect differences in the occurrence rate of contact binaries compared to works that assume conservative mass transfer \citep[e.g.][]{Pols1994, Wellstein2001, Menon2021} or that use fixed mass-transfer efficiencies \citep[e.g.][]{deMink2007, Claeys2011}. Lastly, by including low-mass and massive binaries in our grid, we can compare the onset of contact over a wide mass range.\\

This paper is structured as follows. In Sect.\,\ref{sec:methods}, we describe the computational setup of the grid of binary evolution models. Section \ref{sec:mechanisms} covers the physical mechanisms that have been identified to lead to contact phases, their likelihood to result in a merger, and the way in which they are traced throughout the evolution of the binary models. Our findings of the occurrence of contact phases and mergers over the whole mass range, as well as some notable cases, are given in Sect.\,\ref{sec:results1}. Our results are discussed in Sect.\,\ref{sec:discussion}, and summary and conclusions can be found in Sect.\,\ref{sec:conclusions}.

\section{Methods}\label{sec:methods}
We compute a grid of 5957 1D binary evolution models using the binary module of \texttt{MESA} \citep[release 12778; ][]{Paxton2011, Paxton2013, Paxton2015, Paxton2018, Paxton2019}. First, we describe the adopted single-star physics used in the stellar models in Sect.\,\ref{subsec:single_physics} before describing the binary star physics in Sect.\,\ref{subsec:binary_physics}. We briefly outline the setup of the grid in Sect.\,\ref{subsec:binary_grid} and list the stopping conditions of our binary evolution models in Sect.\,\ref{subsec:stopping_cond}. In Sect.\,\ref{sec:method_probs} we describe how we compute the birth probabilities for the binaries in our grid, which we use for population studies.

\subsection{Adopted stellar physics}\label{subsec:single_physics}
Each binary component is initialised from a precomputed zero-age main-sequence (ZAMS) model at solar metallicity, that is, $Z=0.0142$ and $Y=0.2703$ \citep{Asplund2009}. We use a blend of the OPAL \citep{Iglesias1993,Iglesias1996} and \citet{Ferguson2005} opacity tables appropriate for the chemical composition of \citet{Asplund2009}. We allow the stars to rotate using the shellular approximation as implemented in \texttt{MESA}, with a limit on the rotation rate at 97\% of the Roche critical rotation rate $\Omega_{\mathrm{c}} = \sqrt{GM/R_{\mathrm{eq}}^{3}} \simeq \sqrt{8GM/27R^3}$ \citep{Maeder2009}. In this expression, $M$ and $R$ are the mass and radius of the star, respectively, $G$ is the gravitational constant and $R_{\mathrm{eq}}$ is the equatorial radius of a rotationally deformed star. All models are hydrostatic, meaning that \texttt{MESA}'s implicit hydrodynamic solver is disabled. We use the \texttt{approx21} nuclear network.

\subsubsection{Mixing}
Convective mixing is handled via the mixing length theory \citep{BV1958, Cox1968} and a mixing length parameter of $\alpha_{\mathrm{mlt}} = 2.0$ \citep{Paxton2013}. The Ledoux criterion is used to identify regions in the star that are unstable to convection. The efficiency of semi-convective mixing is set to $\alpha_{\mathrm{sc}} = 10.0$ \citep{Schootemeijer2019}. Thermohaline mixing is also included with an efficiency of $\alpha_{\mathrm{th}} = 1.0$ \citep{Marchant2021}.

Convective boundary mixing (CBM) is included via the step-overshoot scheme, in which we allow the convective hydrogen-burning core to extend by $0.20\,H_{P}$ beyond the core boundary set by the Ledoux criterion, with $H_{P}$ the pressure scale height \citep{Martinet2021}. At the bottom of nonburning convective envelopes, we use step overshoot with a $0.05\,H_{P}$ extension of the convective region towards the centre. This corresponds to one-half of the upper limit typically inferred for the Sun \citep{Angelou2020}. For convectively burning cores beyond the main sequence (MS), overshooting is not yet understood properly and is known to have a large effect on the final fate of stars \citep{Herwig2000, Temaj2023}. Because of this, we follow \citet{Marchant2021} and use exponential overshoot extending only $0.005\,H_{P}$ beyond the edge of the convective region set by the Ledoux criterion.

To account for the rotational mixing of chemical elements and diffusion of angular momentum, the Goldreich-Schubert-Fricke instability, Eddington-Sweet circulation, and the secular and dynamic shear instabilities are included (see for example \citealt{Heger2000} for a detailed description). Additional diffusion of angular momentum is taken care of by the Spruit-Tayler dynamo. We scale the strength of the mixing of chemical elements by a factor $f_{\mathrm{c}} = 1/30$ as in \citet{Heger2000}. The sensitivity of the rotational instabilities to stabilising composition gradients, which is incorporated in the factor $f_{\mu}$, is set to $f_{\mu}=0.1$ \citep[see also][]{Pinsonneault1989}.

\subsubsection{Wind mass-loss prescription}\label{sec:wind_mass_loss}
Because our binary models contain low-, intermediate- and high-mass stars, and the masses can change significantly over the star's evolution, we employ a wind mass-loss prescription that covers this large mass range and all evolutionary stages.

We consider two distinct regimes, the hot-wind regime with surface temperature\footnote{In these wind mass-loss rate computations, the surface temperature $T_{\mathrm{surf}}$ is that of the outermost cell of the model.} $T_{\mathrm{surf}} \geqslant 11\,\mathrm{kK}$ and the cool-wind regime for stars with $T_{\mathrm{surf}} \leqslant 10\,\mathrm{kK}$. We use linear interpolation to determine the mass-loss rate in the temperature region between those two values.

Within the hot-wind regime, the mass-loss rate for stars with hydrogen envelopes (surface hydrogen mass fraction $X_{\mathrm{surf}} > 0.5$) is computed via the \citet{Vink2000} prescription. When $X_{\mathrm{surf}}$ drops below $0.4$, either the prescription for Wolf-Rayet (WR) stars from \citet{Sander2020} or the prescription for low-mass helium stars from \citet{Vink2017} is used, depending on whether the star's luminosity $L$ is higher or lower than a certain luminosity $L_{0}$ respectively. As described in \citet{Sander2020}, $L_{0}$ is the asymptotic limit below which no WR-like wind mass-loss is expected to occur. Its value is metallicity-dependent and obtained from stellar atmosphere models.  In the regime where $L > L_{0}$, we compute the mass-loss rate with both prescriptions and take the maximum value as the adopted wind mass-loss rate (J.~Vink, 2021, priv. comm.). All of the aforementioned prescriptions have a scaling factor of 1.0. For $0.4 \leqslant X_{\mathrm{surf}} \leqslant 0.5$, the mass-loss rate is determined via linear interpolation between the mass-loss rates from both regimes.

When a model reaches the cool-wind regime, the distinction is made between stars expected to become giants or supergiants. The cut is made at $\log_{10}\left(\mathscr{L}/\mathscr{L}_{\odot}\right) = 3.15$, where $\mathscr{L}$ is defined as in \citet{Langer2014},

\begin{align}
    \mathscr{L} = \frac{1}{4\pi\sigma G}\frac{L}{M}\quad .
\end{align}

Here, $\sigma$ is the Stefan-Boltzmann constant. This cut corresponds roughly to a mass of $10\,\mathrm{M}_{\odot}$ at the base of the (super-)giant branch. Models below the cut use the \citet{Reimers1975} wind prescription on the red giant branch (RGB) and the \citet{Bloecker1995} wind prescription on the asymptotic giant branch (AGB). Following \citet{Choi2016}, we use a scaling factor of $0.1$ for the former and $0.2$ for the latter. Models above the cut in $\mathscr{L}$ use the \citet{Nieuwenhuijzen1990} prescription with a scaling factor of 1.0.

We increase the scaling factor for the \citet{Bloecker1995} wind to 3.0 at the onset of thermal pulses (TP) during the AGB phase following \citet{Choi2016}. This increase aims to ease the computations through this phase by mimicking the enhanced mass loss during the TP-AGB phase while simultaneously avoiding the TPs themselves (by removing the envelope). Additionally, by removing part of the envelope, we aim to avoid the Hydrogen Recombination Instability (HRI) and the Fe-Peak Instability (FePI). These instabilities, which lead to envelope inflation over multiple orders of magnitude, can occur when the cold, expanded envelopes of AGB stars are modelled with a hydrostatic code (Rees et al., in prep.). The HRI is caused by the increased dynamical instability of the envelope because of hydrogen recombination \citep{Wagenhuber1994}, and the FePI occurs when luminosity at the base of the convective envelope exceeds the Eddington luminosity because of a local iron opacity bump \citep{Lau2012}. The physical mechanism behind these instabilities most likely leads to events of extreme mass loss. The timescales on which this envelope inflation occurs are too short to be captured correctly in \texttt{MESA}'s hydrostatic mode and can lead to numerical issues. The increase of the \citet{Bloecker1995} wind scaling factor is not successful in avoiding numerical issues in each model, especially in those models where the aforementioned instabilities occur. Because of this, we opt to disregard models in which binary mass transfer occurs after the TP-AGB phase. A more elaborate approach to compute through the TP-AGB phase and these instabilities is provided in, for example, Rees et al. (in prep.).

\subsection{Adopted binary physics}\label{subsec:binary_physics}
In our models, we evolve both binary components. This allows us to consider the behaviour of both the donor and the accretor star for tracing potential contact scenarios (see Sect.\,\ref{sec:mechanisms}). Only mass transfer from the initially more massive or \emph{primary} star onto the initially less massive or \emph{secondary} star is considered. Hence, in this work, references to primary (secondary) and donor (accretor) are equivalent. We will mostly employ the names primary (subscript ``1'') and secondary (subscript ``2'') star. We assume that all binaries are on circular orbits.

\subsubsection{Mass transfer and accretion}\label{methods:mass_transfer}
Whenever the primary star is on the MS, mass transfer is computed using the \texttt{contact} scheme \citep{Marchant2016}. In semi-detached binaries, this scheme uses \texttt{MESA}'s \texttt{roche\_lobe} scheme, which ensures that the donor stays within its Roche lobe. When both stars (over-)fill their Roche lobe it switches to a different solver for the mass-transfer rate suitable for contact binaries. In this scheme, only the computation of the mass-transfer rate is handled. Energy transport between the components of the contact binary and the tidal distortion are not taken into account (for the effect of including those, see \citealt{Fabry2022,Fabry2023}). For systems with post-MS primary masses smaller than $1.3\,\mathrm{M}_{\odot}$, the \texttt{Kolb} scheme \citep{Kolb1990} is used, because this scheme is better suited for envelopes with larger pressure scale heights \citep{Fragos2023}. Both schemes are solved implicitly.

The mass transfer efficiency $\beta$ is defined as the effective, overall change in mass of the accretor over the mass transferred from the donor to the accretor, $\beta \equiv -\dot{M}_{2}/\dot{M}_{\mathrm{trans}}$. In this definition, $\dot{M}_{\mathrm{trans}} < 0$ and $\dot{M}_{2} > 0$. $\dot{M}_{2}$ also includes the wind mass loss of the accretor. During conservative mass transfer in our models $\beta \approx 1$, because typically $|\dot{M}_{\mathrm{trans}}|$ is orders of magnitude larger than the absolute value of the wind mass-loss rate. When the accretor star spins up to its critical rotation rate $\Omega_{\mathrm{c},\,2}$, mass transfer is non-conservative and $\beta < 1$. When the accretor star's accretion timescale $\tau_{\mathrm{acc}} \equiv M_{2}/\dot{M}_{\mathrm{trans}}$ approaches the star's dynamical timescale $\tau_{\mathrm{dyn,\,2}}$, defined for a star with sound speed $c_{\mathrm{S}}$ as $\tau_{\mathrm{dyn,\,2}} \equiv R_{2}/c_{\mathrm{S}}$ \citep{Kippenhahn2012}, we limit the accretion rate to that of $0.1$ times the star's Kelvin-Helmholtz (or thermal) timescale $\tau_{\mathrm{KH}}$. This also results in non-conservative mass transfer, that is, $\beta < 1$. We do this because the accreting models tend not to converge numerically when $\tau_{\mathrm{acc}} \sim \tau_{\mathrm{dyn,\,2}}$. Models for which the accretion rate is limited are marked in the results. The accretion of angular momentum during mass transfer is computed following \citet{Lubow1975} and \citet{Ulrich1976}, which includes accretion through ballistic impact and a Keplerian disk.

\subsubsection{Tides and angular momentum loss}\label{sec:methods:tides}
Tidal synchronisation is computed uniformly over the components' structure using the convective synchronisation timescale from \citet{Hurley2002}. Upon initialisation of the binary models, the rotation periods of both components are equal to the orbital period.
Orbital angular momentum in our models evolves via mass loss from the system (through winds or isotropic re-emission) and spin-orbit coupling. In the former case, the lost mass is assumed to have the specific angular momentum of the star's orbit in which vicinity it is leaving the system.

\subsection{Binary-star grid}\label{subsec:binary_grid}

\begin{figure}
    \centering
    \resizebox{\hsize}{!}{\includegraphics{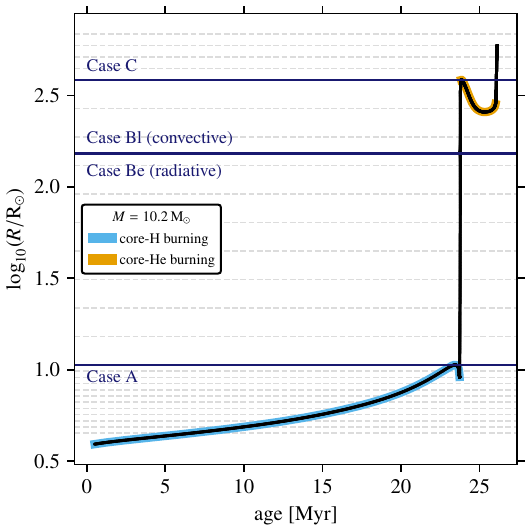}}
    \caption{Radial expansion of a $10.2\,\mathrm{M}_{\odot}$ star (solid black line). The radial evolution is divided into different cases based on the main phases of expansion (solid horizontal lines). Each dashed horizontal line indicates a sample point at which we put $R = R_{\mathrm{RL}}$.}
    \label{fig:radius_sampling}
\end{figure}

In our grid, we vary the initial mass of the primary star $M_{1,\,\mathrm{i}}$ (in units of $\mathrm{M}_{\odot}$), the initial mass ratio $q_{\mathrm{i}}=M_{2,\,\mathrm{i}}/M_{1,\,\mathrm{i}}$ with $M_{2,\,\mathrm{i}}$ the initial mass of the secondary star, and the initial separation $a_{\mathrm{i}}$ (in units of $\mathrm{R}_{\odot}$). Twenty primary masses are selected between $0.8$ and $20.0\,\mathrm{M}_{\odot}$ with logarithmic mass spacing $\Delta \log_{10}M_{1,\,\mathrm{i}} \approx 0.074$. At the high-mass end, we add three additional primary masses ($13.1$, $15.6$ and $18.4\,\mathrm{M}_{\odot}$) for increased resolution, bringing the total number of primary masses to 23. Values between $0.1$ and $0.9$ are considered for the mass ratios, with linear spacing and steps of 0.1. We impose a lower limit on the secondary mass of $0.5\,\mathrm{M}_{\odot}$\, to avoid fully convective companions, which are difficult to converge numerically when they accrete. We add an additional mass ratio at $q_{\mathrm{i}}=0.97$ to model twin systems. Lastly, we choose the initial binary separations $a_{\mathrm{i}}$ such that the first phase of mass transfer occurs during all stages of the primary star's evolution. These stages at which mass transfer occurs are called Case~A, B and C, for when the donor star is on the MS, before core-helium ignition, and after core-helium ignition, respectively\footnote{As demonstrated in \citet{Ge2015}, core-helium ignition can occur before the primary star reaches the base of the supergiant branch for $M_{\mathrm{i}}\gtrsim 15\,\mathrm{M}_{\odot}$. Although mass transfer at this point would be classified as Case C, it behaves in a very similar way to Case B. This does not, however, occur in our models.}. This distinction is related to the phases in which the (primary) star expands, as illustrated in Fig.\,\ref{fig:radius_sampling}. Case~B phases are further divided into early (Case~Be) and late (Case~Bl) phases. The distinction is based on the presence of a deep convective zone in the envelope of the star\footnote{For stars with convective envelopes during the main sequence, that is, initial masses ${<}\,1.3\,\mathrm{M}_{\odot}$, the distinction between early and late Case-B mass transfer is not made.}.

To ensure sufficient sampling of all these stages, we have used the radius evolution of single-star models in combination with the volume-equivalent Roche lobe radius $R_{\mathrm{RL}}$ for a given mass ratio $q=M_{2}/M_{1}$ and separation $a$, \citep{Eggleton1983},

\begin{align}\label{eq:eggleton_approx}
    \frac{R_{\mathrm{RL}}}{a} = \frac{0.49 q^{-2/3}}{0.6q^{-2/3} + \ln\left(1+q^{-1/3}\right)}\quad .
\end{align}

For a given primary mass $M_{1}$ and mass ratio $q$, we select a number of points along the radius evolution of the star, indicated by the dashed horizontal lines in Fig.\,\ref{fig:radius_sampling}. At each of these points, we assume the star fills its Roche lobe, $R = R_{\mathrm{RL}}$, and mass transfer ensues. From Eq.\,(\ref{eq:eggleton_approx}), we then get the separation $a$ at which this happens. Assuming that the orbit stays approximately constant until mass transfer starts, we can take this value for the separation $a$ as our initial value, $a_{\mathrm{i}} = a$. Although this method of sampling the initial separation space is adequate for closer binaries, it cannot accurately predict the division lines between later mass-transfer cases for a number of reasons (see, for example, the dividing line between Case~Bl and Case~C in Fig.\,\ref{fig:contact_Mp1016}). Firstly, the assumption that the orbital separation $a$ stays approximately constant does not hold in systems with strong wind mass loss and tidal interaction prior to mass transfer. Secondly, post-MS donors use \texttt{MESA}'s \texttt{Kolb} scheme for mass transfer \citep{Kolb1990}. This scheme can result in significant mass-transfer rates before the donor star formally overfills its Roche lobe, since it takes both the optically thick and thin Roche lobe overflow into account \citep{Paxton2015}.

\subsection{Stopping conditions}\label{subsec:stopping_cond}
In principle, we compute each binary model until one of the components, usually the initially more massive primary star, reaches the end of core carbon burning. This is defined as the moment when the central mass fractions of He and C fall below $10^{-6}$ and $10^{-2}$, respectively. 

Systems with MS primaries having predominantly radiative envelopes ($M_{1,\,\mathrm{i}} > 1.3\,\mathrm{M}_{\odot}$) use \texttt{MESA}'s \texttt{contact} mass-transfer scheme (see Sect.\,\ref{methods:mass_transfer}) and can therefore be modelled through such phases. During a contact phase, the computation terminates when the condition for mass overflow through the second Lagrange point ($\mathrm{L}_{2}$) or $\mathrm{L}_{2}$-overflow from \citet{Marchant2016} is met (see Sect.\,\ref{sec:l2over}). 

For primaries with mostly convective envelopes (post-MS or $M_{1,\,\mathrm{i}} \leq 1.3\,\mathrm{M}_{\odot}$), which use the \texttt{Kolb} mass-transfer scheme. The computation stops whenever the secondary star overfills its Roche lobe by more than $R_{\mathrm{RL},\,2}$. 

When the mass-transfer rate $\dot{M}_{\mathrm{trans}}$ reaches a value of $10\,\mathrm{M}_{\odot}\,\mathrm{yr}^{-1}$, the evolution is terminated. The timescale on which such fast mass transfer occurs nears typical dynamical timescales, which is not numerically feasible with our \texttt{MESA} setup. 

Lastly, reverse mass transfer in a semi-detached system, from the secondary to the primary star, is not considered in this study. The onset of reverse mass transfer is a stopping condition.

\subsection{Birth probabilities}\label{sec:method_probs}
To describe the models in our grid as a population (Sect.\,\ref{sec:population}--\ref{sec:comparison}), we compute their birth probability $p_{\mathrm{birth}}$ from the initial distribution function, which here is the product of the distribution functions of the initial primary mass $M_{1,\,\mathrm{i}}$, mass ratio $q_{\mathrm{i}}$ and period $P_{\mathrm{i}}$. For the $M_{1,\,\mathrm{i}}$ distribution, we use the \citet{Kroupa2001} initial mass function (IMF),
\begin{equation}
    \psi(M_{1,\,\mathrm{i}}) = C_{m}M^{\alpha}\quad,
\end{equation}
with $\alpha = -2.3$ as appropriate for primary star masses ${\geq}\,0.5\,\mathrm{M}_{\odot}$ and $C_{m}$ a normalisation constant. We assume the other two initial distribution functions to be uniform in $q_{\mathrm{i}}$ and $\log_{10}P_{\mathrm{i}}$, respectively,
\begin{equation}
    \phi(q_{\mathrm{i}}) = C_{q} \quad \quad \text{and} \quad \quad \chi(\log_{10}P_{\mathrm{i}}) = C_{p} \quad ,
\end{equation}
with $C_{q}$ and $C_{p}$ being normalisation constants.
The birth probability $p_{\mathrm{birth}}$ of each model is then computed by integrating the product of the initial distribution functions over the parameter size in the model grid,
\begin{align}\label{eq:pbirth}
    p_{\mathrm{birth}} =& \int_{\log_{10}P_{\mathrm{l}}}^{\log_{10}P_{\mathrm{u}}}\int_{q_{\mathrm{l}}}^{q_{\mathrm{u}}}\int_{M_{\mathrm{l}}}^{M_{\mathrm{u}}} \psi(M_{1,\,\mathrm{i}}) \nonumber \\ 
    & \times \phi(q_{\mathrm{i}})\, \chi(\log_{10}P_{\mathrm{i}})\,\mathrm{d}M_{1,\,\mathrm{i}}\,\mathrm{d}q_{\mathrm{i}}\,\mathrm{d}\log_{10}P_{\mathrm{i}} \quad .
\end{align} 
Here, $\log_{10}P_{\mathrm{u,\,l}}$, $q_{\mathrm{u,\,l}}$ and $M_{\mathrm{u,\,l}}$ are the upper- and lower-boundaries of the parameter size of a model in the grid for $\log_{10}P_{\mathrm{i}}$, $q_{\mathrm{i}}$ and $M_{1,\,\mathrm{i}}$, respectively. They are chosen as the midpoints between the initial parameter values of the model and its neighbouring models.

\subsection{Software for analysis and plotting}
For the analysis in this work, we used the following open-source software: \texttt{PyMesaReader} (Wolf \& Schwab, 2017\footnote{Code: \url{https://github.com/wmwolf/py_mesa_reader}. Documentation: \url{https://billwolf.space/py_mesa_reader/}.}), \texttt{NumPy} \citep{harris2020}, \texttt{SciPy} \citep{SciPy2020}, \texttt{MPI for Python} \citep{Dalcin2005,Dalcin2021} and \texttt{Astropy} \citep{astropy2013,astropy2018,astropy2022}. For constructing the figures, we used \texttt{Matplotlib} \citep{Hunter2007}. 

\section{Physical mechanisms leading to contact}\label{sec:mechanisms}

\begin{figure*}
\centering
  \includegraphics[width=16cm]{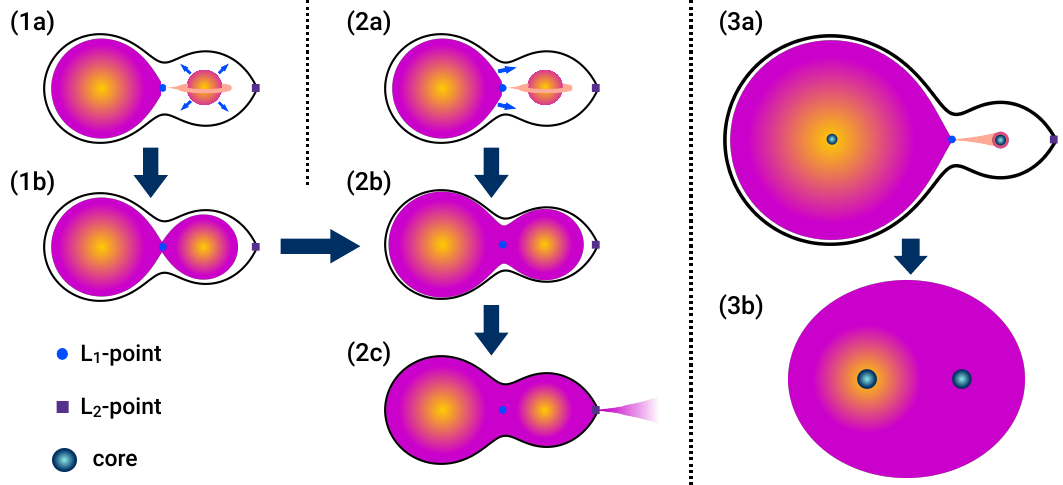}
     \caption{Schematic representation of the physical mechanisms leading to contact (not to scale). The filled blue circles and purple squares indicate the position of the $\mathrm{L}_{1}$- and $\mathrm{L}_{2}$- points, respectively. The filled grey-blue circles represent the stellar cores in Panel (3a) and (3b). Panel (1a) shows the expansion of the accretor leading to a contact binary (1b). This corresponds to the ``Accretor expansion'' mechanism described in Sect.\,\ref{subsec:accr_exp}. Subsequent overfilling of the components' Roche lobes leads to the formation of an overcontact binary (2b). The primary increasingly overfills its Roche lobe (2a) and can eventually fill the secondary's Roche lobe (2b). This can eventually lead to $\mathrm{L}_{2}$-overflow (2c), which likely results in a stellar merger. The scenarios (1b--2b--2c) and (2a--2b--2c) correspond to the ``$\mathrm{L}_{2}$-overflow'' mechanism described in Sect.\,\ref{sec:l2over}. In Panel (3a), runaway mass transfer from a (super-)giant (left) to an MS star (right) leads to the onset of a classical common-envelope phase (3b), where the cores of both stars revolve in the (super-)giant's envelope. This corresponds to the ``Runaway MT'' mechanism described in Sect.\,\ref{sec:unstable_MT}.}
     \label{fig:contact_schematic}
\end{figure*}

We use the following nomenclature of contact phases traced in our grid, based on the physical picture in \citet{Roepke2023} and illustrated in Fig.\,\ref{fig:contact_schematic}. Systems with MS and/or Hertzsprung-gap (HG) stars (i.e. Case~A \& Be) that both (over-)fill their Roche lobes or undergo (unstable) runaway mass transfer are referred to as contact binaries (Fig.\,\ref{fig:contact_schematic}.1b and \ref{fig:contact_schematic}.2b).  Systems with a (super-)giant primary (i.e. Case~Bl \& C) and an MS secondary undergoing (unstable) runaway mass transfer enter a \emph{classical} common-envelope phase (Fig.\,\ref{fig:contact_schematic}.3b). What sets the onset of a classical CE phase apart from the formation of a contact binary is that now the primary has a clear core-envelope boundary and the radius of the secondary is at least an order of magnitude smaller than that of the primary.

In this section, we discuss the different physical mechanisms that can lead to contact phases. As stated before, being in contact does not necessarily mean that the binary will merge. Hence, for each mechanism leading to contact, we discuss the likelihood of a merger or rather a longer-lived contact phase. Additionally, we explain how we trace each mechanism in the binary star models.

\subsection{Expansion of accretor}\label{subsec:accr_exp}
Arguably the most intuitive way to form contact binaries is when the accreting secondary star expands during a mass-transfer phase and (over-)fills its Roche lobe. A schematic representation of this mechanism is shown in Fig.\,\ref{fig:contact_schematic}.1a--b. The timescale on which the secondary expands, generally defined as $\tau_{R/\dot{R}} \equiv R/\dot{R}$ with $R$ the stellar radius and $\dot{R}$ its time derivative, has implications for the evolution of the subsequent contact phase.

When the mass-transfer timescale $\tau_{\mathrm{trans}} \equiv |M_{1}/\dot{M}_{\mathrm{trans}}|$ is shorter than the secondary's thermal or Kelvin-Helmholtz timescale\footnote{Technically one should compare with the \emph{local} thermal timescale in the surface layers of the accreting star, defined in \citet{Kippenhahn2012} (see also \citealt{Temmink2023}) as $\tau_{\mathrm{KH}}^{\mathrm{local}}(m) = \int^{M}_{m}c_{P}(m')T(m')\mathrm{d}m'$, with $c_{P}(m)$ the heat capacity at constant pressure $P$ and $T(m)$ the temperature at mass coordinate $m$. In practice, timescale comparisons should therefore be made at an order-of-magnitude level.} defined as \citep{Kippenhahn2012}

\begin{equation}\label{eq:thermal_timescale}
    \tau_{\mathrm{KH}} = \frac{GM^{2}}{2RL} \approx 1.5\times10^{7}\left(\frac{M}{\mathrm{M}_{\odot}}\right)^{2}\frac{\mathrm{R}_{\odot}}{R}\frac{\mathrm{L}_{\odot}}{L}\mathrm{yr} \quad ,
\end{equation}
the star will be out of thermal equilibrium. In an effort to regain thermal equilibrium, the secondary expands on its thermal timescale \citep{Ulrich1976, Kippenhahn1977, Webbink1976, Neo1977, Pols1994}. A contact binary is formed if the increase in radius is sufficient for the secondary to fill its Roche lobe. Such a contact binary can rapidly merge on a thermal timescale, or evolve back to a semi-detached binary when the accretor regains thermal equilibrium and shrinks.

When the secondary is in thermal equilibrium during mass transfer, it will expand on a nuclear timescale, defined as \citep{Kippenhahn2012}
\begin{equation}\label{eq:nuclear_timescale}
    \tau_{\mathrm{nuc}} = \frac{E_{\mathrm{nuc}}}{L} \approx 10^{10} \frac{M}{\mathrm{M}_{\odot}}\frac{\mathrm{L}_{\odot}}{L}\mathrm{yr} \quad .
\end{equation}
Here, $E_{\mathrm{nuc}}$ is the available nuclear energy. Contact phases driven by the nuclear expansion of the accretor are longer lived since the accretor will not shrink inside its Roche lobe again until the end of the nuclear burning phase. As a result, such contact phases are expected to persist on a nuclear timescale, or until the stars merge.\\

The onset of contact phases through the expansion of the accretor is traced by checking whether at any point the accretor overfills its Roche lobe. These models are labelled as ``Accretor expansion'' in the following sections. This is a stopping condition for models using the \texttt{Kolb} mass-transfer scheme. Models using the \texttt{contact} mass-transfer scheme can evolve through these contact phases. Hence, multiple contact phases might occur throughout the evolution of the system. If more than one contact phase occurs in a model using the \texttt{contact} scheme, only the onset of the first contact phase is considered for a better comparison with models using the \texttt{Kolb} scheme.\\

\begin{figure*}
\centering
  \includegraphics[width=18cm]{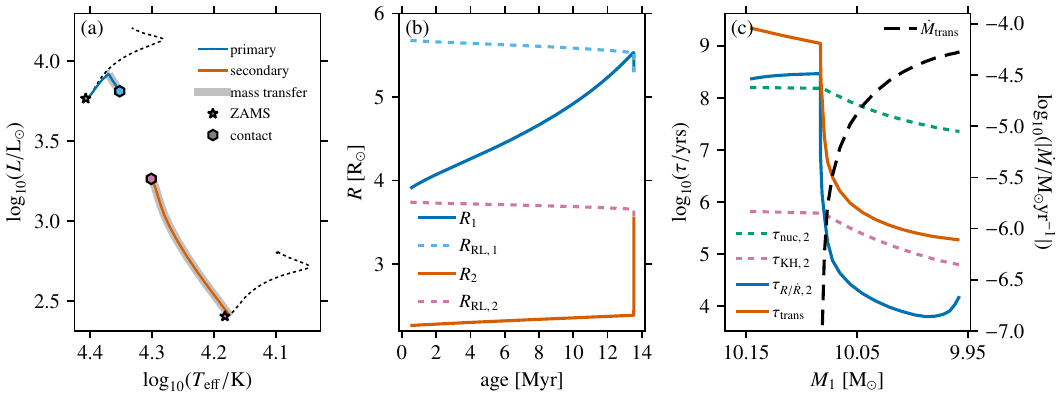}
     \caption{Example of a $M_{1,\,\mathrm{i}}=10.2\,\mathrm{M}_{\odot}$, $q_{\mathrm{i}} = 0.4$ and $a_{\mathrm{i}} = 12.4\,\mathrm{R}_{\odot}$ binary system forming a contact binary during the MS of the primary (Case~A) because of the thermal expansion of the secondary (accretor) star. Panel (a) shows an HRD with the evolutionary tracks of the primary and secondary stars (solid lines). The dashed black lines show the evolutionary tracks of single stars with the same initial masses as the binary components and initial rotation rates of $\Omega/\Omega_{\mathrm{c}} = 0.25$. Panel (b) shows the evolution of the radius $R$ (solid lines) and Roche lobe radius $R_{\mathrm{RL}}$ (dashed lines) of both components. The timescales governing the evolution of the binary and the mass-transfer rate (black dashed line) are shown in Panel (c) as a function of the decreasing primary star mass. The secondary's nuclear ($\tau_{\mathrm{nuc,\,2}}$) and thermal ($\tau_{\mathrm{KH,\,2}}$) timescales are shown with a dashed green and pink line, respectively. The expansion ($\tau_{R/\dot{R},\,2}$) and mass-transfer ($\tau_{\mathrm{trans}}$) timescales are shown with a solid blue and orange line, respectively.}
     \label{fig:example_thermalExp}
\end{figure*}

An example of a system forming a contact binary because of the thermal expansion of the accreting secondary star (shown up to the point of contact) is shown in Fig.\,\ref{fig:example_thermalExp}. From the Hertzsprung-Russell diagram (HRD, Fig.\,\ref{fig:example_thermalExp}a), it can be seen that, as a result of Case-A mass transfer, the primary and secondary become under- and over-luminous compared to their single-star counterparts, respectively. The single-star tracks are shown up to their terminal-age main sequence (TAMS). As a result of mass accretion, the secondary star expands rapidly and fills its Roche lobe (Fig.\,\ref{fig:example_thermalExp}b). The expansion timescale $\tau_{R/\dot{R},\,2}$ becomes about an order of magnitude shorter than the secondary's global thermal timescale $\tau_{\mathrm{KH,\,}2}$ due to the mass-transfer timescale $\tau_{\mathrm{trans}}$ becoming comparable to $\tau_{\mathrm{KH,\,}2}$ (Fig.\,\ref{fig:example_thermalExp}c). An example of a system forming a contact binary because of the nuclear expansion of the accretor is provided in Appendix \ref{app:example}. The accretor expansion timescales of all systems forming contact binaries through accretor expansion are shown in Fig.\,\ref{fig:expansion_timescales}--\ref{fig:expansion_timescales2} in Appendix \ref{app:exp_ts}.

\subsection{Non-conservative mass transfer}\label{subsec:non_cons_mt}
As demonstrated in \citet{Packet1981} and, for example, more recently in \citet{Ghodla2023}, a star only has to accrete between ${\sim}2$ to ${\sim}10$ percent of its own mass to be spun up to the Roche critical rotation rate $\Omega_{\mathrm{c}}$. When a star is rotating at (near) this critical rotation rate, it may not be able to accrete any more material. It is then often assumed that the excess mass is lost through an enhanced stellar wind, or even instantaneously through a so-called ``fast'' wind from the accretor, resulting in non-conservative mass transfer. In our models, instantaneous wind mass loss is invoked to expel the excess matter for accretors rotating near the critical rotation rate and for models in which the accretion timescale $\tau_{\mathrm{acc}}$ is restricted to $10\%$ of the Kelvin-Helmholtz timescale $\tau_{\mathrm{KH}}$.
It is assumed that there is an energy source that can expel this excess mass. Let us consider the energy per unit mass $\varepsilon$ required to drive away the mass to infinity, $\varepsilon = GM_{2}/2R_{2}$. This is under the simplifying assumption that the primary's gravitational potential can be ignored and that the mass is lost from the surface of the accretor. Equating $\varepsilon$ to  $\left(L_{1} + L_{2}\right)/\dot{M}_{\mathrm{max}}$, we eventually find a maximum mass-loss rate $\dot{M}_{\mathrm{max}}$ the accretor can have under the condition that all this mass is driven away to infinity by the combined luminosity $\left(L_{1} + L_{2}\right)$ of the binary \citep{Marchant2017PhD},
\begin{equation}\label{eq:maxmdot}
    \frac{\dot{M}_{\mathrm{max}}}{\mathrm{M}_{\odot} \mathrm{yr}^{-1}} = 10^{-7.19}f_{\mathrm{eff}}\frac{R_{2}}{\mathrm{R}_{\odot}}\frac{\mathrm{M}_{\odot}}{M_{2}}\frac{\left(L_{1} + L_{2}\right)}{\mathrm{L}_{\odot}}\quad .
\end{equation}
The factor $f_{\mathrm{eff}}$ is a free parameter added to take into account the uncertainty on the exact radius at which the non-accreted mass is expelled, the fact that only the gravitational potential of the accretor is accounted for, the unknown fraction of the total luminosity that can be used to expel the non-accreted mass, and the unknown energy of the ejected mass at infinity.

Since the non-accreted mass in systems with $\dot{M}_{2} > \dot{M}_{\mathrm{max}}$ cannot be expelled to infinity, it must remain in or around the binary system. Although it is unclear what exactly happens, the non-accreted mass can potentially lead to a contact phase, for example, if it fills the secondary's Roche lobe or introduces a drag on the binary components. Alternatively, the non-accreted matter can form an accretion disk. In this work, we merely flag such models and discuss the potential consequences for their evolution in Sec.\,\ref{sec:non_ejected_fate}.\\

In the models, we trace the failure to eject non-accreted matter in post-processing by using Eq.\,(\ref{eq:maxmdot}) and assuming $f_{\mathrm{eff}} = 1.0$. Models for which the mass-loss rate of the accretor, defined as $\dot{M}_{\mathrm{ej}}=\left(1-\beta\right)\dot{M}_{\mathrm{trans}}$, at one point exceeds $\dot{M}_{\mathrm{max}}$ are labelled in the following sections as ``Non-conservative MT + cannot eject''.\\

\begin{figure*}
\centering
  \includegraphics[width=18cm]{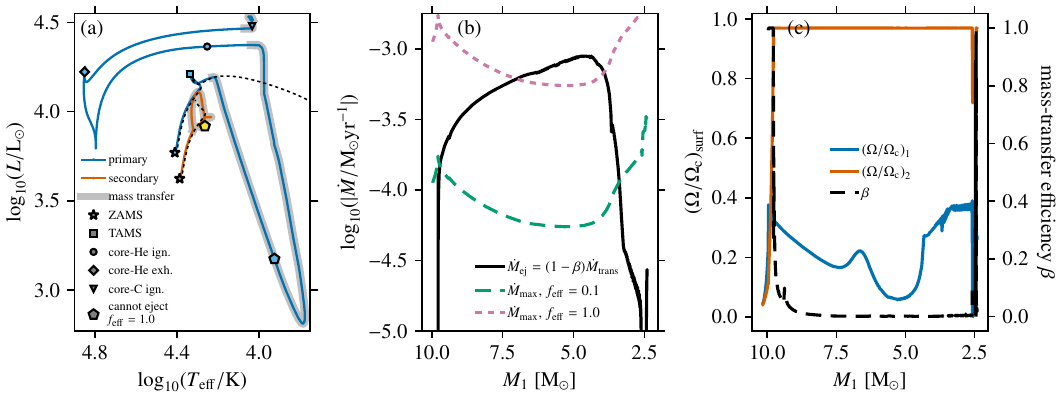}
     \caption{Example of a $M_{1,\,\mathrm{i}}=10.2\,\mathrm{M}_{\odot}$, $q_{\mathrm{i}} = 0.9$ and $a_{\mathrm{i}} = 39.2\,\mathrm{R}_{\odot}$ binary star model in which non-conservative mass transfer occurs and for which the non-accreted matter cannot be driven to infinity. Panel (a) is the same as Fig.\,\ref{fig:example_thermalExp}a (``ign.'' = ``ignition'' and ``exh.'' = ``exhaustion''). In Panel (b), the mass-loss rate of the secondary (solid black line) is compared to the maximum mass-loss rate $\dot{M}_{\mathrm{max}}$ (dashed lines; pink for $f_{\mathrm{eff}} = 1.0$, green for $f_{\mathrm{eff}} = 0.1$) set by Eq.\,(\ref{eq:maxmdot}). The surface rotation rates (solid blue and orange lines) and mass transfer efficiency (dashed black line) are shown in Panel (c) as a function of the primary mass.}
     \label{fig:example_nonConservative}
\end{figure*}

An example of a system in which the accretor cannot eject the non-accreted matter during non-conservative mass transfer is shown in Fig.\,\ref{fig:example_nonConservative}. The system undergoes Case-Be and -C mass transfer. The primary is partially stripped in the former mass-transfer phase, and the remaining H+He envelope amounts to  ${\approx}\,20\%$ of the mass of the post-mass-transfer primary star. After core-He exhaustion, the star expands again, initiating a Case-C mass-transfer phase. During Case-Be mass transfer, the secondary star becomes over-luminous because it is out of thermal equilibrium. However, its luminosity decreases again when its rotation rate $\Omega$ reaches the critical rotation rate $\Omega_{\mathrm{c}}$ (Fig.\,\ref{fig:example_nonConservative}c). At this point ($M_{1} \approx 9.5\,\mathrm{M}_{\odot}$), the mass-transfer efficiency $\beta$ decreases to almost zero. During the non-conservative Case-Be mass transfer, the combined luminosity is insufficient to expel the non-accreted matter to infinity. This holds both for $f_{\mathrm{eff}} = 0.1$ and $f_{\mathrm{eff}} = 1.0$ (Fig.\,\ref{fig:example_nonConservative}b).

\subsection{Outflow from second Lagrange point}\label{sec:l2over}
In the Roche potential, there are three equilibrium points located on the line connecting the centres of the binary components. From lowest to highest potential, these are $\mathrm{L}_{1}$ (through which mass transfer occurs), $\mathrm{L}_{2}$ and $\mathrm{L}_{3}$. The $\mathrm{L}_{2}$-point is always located on the side of the less massive star in the binary, and $\mathrm{L}_{3}$ on the side of the more massive star. In Case-A and -Be binaries, mass loss from the $\mathrm{L}_{2}$-point (Fig\,\ref{fig:contact_schematic}.2c) takes away a significant amount of specific angular momentum from the system, which can lead to rapid orbital shrinkage and a subsequent merger. The rate of orbital shrinkage and hence the time until the merging event depends on the mass-loss rate and the outflow velocity through $\mathrm{L}_{2}$ \citep{Marchant2021}. In Case~Bl and -C binaries, $\mathrm{L}_2$ indicates the onset of a classical CE.\\

$\mathrm{L}_{2}$-overflow is traced in our models when either one (semi-detached) or both (contact) components overfill(s) their Roche lobe. In the latter case, the radii of the components are compared to $\mathrm{L}_{2}$ volume-equivalent radii from \citet{Marchant2016}. This is done during the computation of the model and the moment of $\mathrm{L}_{2}$-overflow is a stopping condition (Sect.\,\ref{subsec:stopping_cond}). In semi-detached systems, we compare the radius of the primary star to the volume-equivalent radius $R_{\mathrm{L}_{2}}$ and the distance to the $\mathrm{L}_{2}$-point $D_{\mathrm{L}_{2}}$ from \citet{Misra2020} (similar fitting formulae for $R_{\mathrm{L}_{2}}$ are derived in \citealt{Ge2020a}). This is done in post-processing, meaning that the outflow from $\mathrm{L}_{2}$ is not modelled and does not affect the evolution of the model. We refer to, for example, \citet{Marchant2021} to see the effect of these outflows. In both cases, if $\mathrm{L}_{2}$-overflow occurs, the models are labelled ``$\mathrm{L}_{2}$-overflow''.\\

\begin{figure*}
\centering
  \includegraphics[width=18cm]{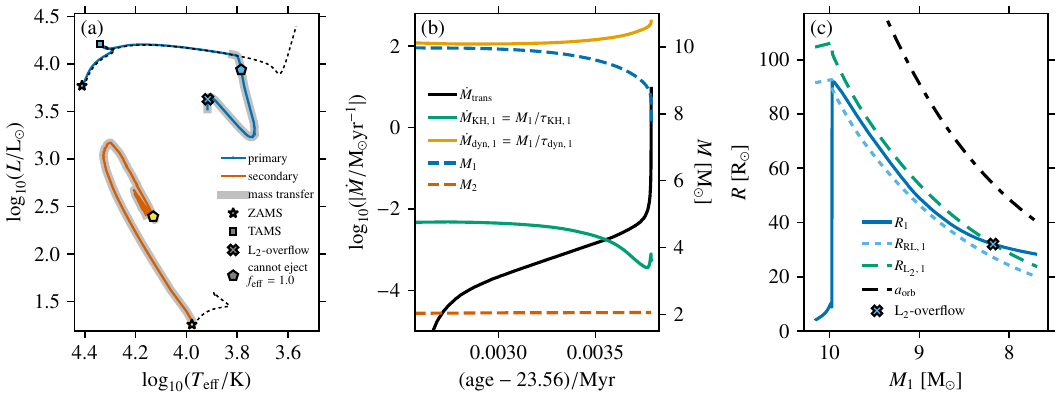}
     \caption{Example of a binary model with $M_{1,\,\mathrm{i}}=10.2\,\mathrm{M}_{\odot}$, $q_{\mathrm{i}} = 0.2$ and $a_{\mathrm{i}} = 175.7\,\mathrm{R}_{\odot}$ going through a phase of (delayed) runaway mass transfer. Panel (a) is the same as Fig.\,\ref{fig:example_thermalExp}a (``ign.'' = ``ignition''). Panel (b) shows the mass-transfer rate $\dot{M}_{\mathrm{trans}}$ (solid black line), thermal mass-transfer rate $\dot{M}_{\mathrm{KH}}$ (solid green line) and dynamical mass-transfer rate $\dot{M}_{\mathrm{dyn}}$ (solid gold line) for the primary star on the left axis. The right axis shows the component mass evolution as a function of age (dashed blue and orange lines). In Panel (c), the radius $R$ (solid blue line), Roche lobe radius $R_{\mathrm{RL}}$ (dashed light-blue line), $\mathrm{L}_{2}$-volume-equivalent radius $R_{\mathrm{L}_{2}}$ \citep[dahsed green line,][Eq.\,15]{Misra2020} and orbital separation $a_{\mathrm{orb}}$ (dashed black line) evolution for the primary are shown as a function of the decreasing primary mass. The blue cross indicates the moment of $\mathrm{L}_{2}$-overflow.}
     \label{fig:example_runawayMT}
\end{figure*}

The evolution of a binary system experiencing $\mathrm{L}_{2}$-overflow is shown in Fig.\,\ref{fig:example_runawayMT}. The primary and secondary become under- and over-luminous, respectively, during Case-Be mass transfer (Fig.\,\ref{fig:example_runawayMT}a). As in the previous example, mass transfer is non-conservative and early on during the mass transfer the non-accreted matter cannot be driven to infinity (Fig.\,\ref{fig:example_runawayMT}b). As the binary orbit shrinks during mass transfer, also the Roche lobe radius $R_{\mathrm{RL,\,1}}$ and the volume-equivalent radius $R_{\mathrm{L}_{2},\,1}$ decrease (Fig.\,\ref{fig:example_runawayMT}c ). After about $1.5\,\mathrm{M}_{\odot}$ is lost from the primary star, its radius decreases slower than $R_{\mathrm{RL,\,1}}$ and $R_{\mathrm{L}_{2},\,1}$. As a consequence, the star increasingly overfills its Roche lobe until it reaches the point of $\mathrm{L}_{2}$-overflow. The subsequent loss of mass and angular momentum from the $\mathrm{L}_{2}$-point will result in an accelerated orbital shrinkage, which is not accounted for in this model. The onset of $\mathrm{L}_{2}$-overflow is shown schematically in Fig.\,\ref{fig:contact_schematic}. In the evolutionary scenario depicted in Fig.\,\ref{fig:contact_schematic}.1a--1b--2b--2c, a contact binary is formed because of the expansion of the accretor. The components continuously overfill their Roche lobes (Fig.\,\ref{fig:contact_schematic}.2b) until the overcontact binary fills the $\mathrm{L}_{2}$-lobe (Fig.\,\ref{fig:contact_schematic}.2c). In the second scenario, from Fig.\,\ref{fig:contact_schematic}.2a--2c, the primary overfills its own Roche lobe (Fig.\,\ref{fig:contact_schematic}.2a) and later fills the Roche lobe of the secondary, forming an overcontact binary (Fig.\,\ref{fig:contact_schematic}.2b). Eventually, this overcontact binary fills the $\mathrm{L}_{2}$-lobe and $\mathrm{L}_{2}$-overflow occurs (Fig.\,\ref{fig:contact_schematic}.2c).

\subsection{Tidally-driven contact}

\begin{figure}
    \centering
    \resizebox{0.95\hsize}{!}{\includegraphics{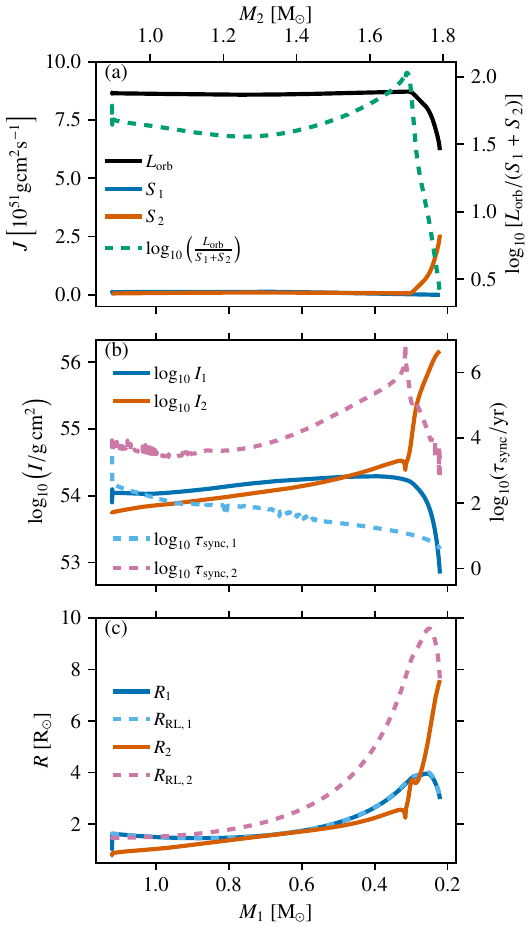}}
    \caption{Example of a $M_{1,\,\mathrm{i}}=1.1\,\mathrm{M}_{\odot}$, $q_{\mathrm{i}} = 0.8$ and $a_{\mathrm{i}} = 4.1\,\mathrm{R}_{\odot}$ binary system in which tides lead to contact. In Panel (a) we show the exchange between the orbital angular momentum $L_{\mathrm{orb}}$ (solid black line) and the secondary's spin angular momentum $S_{2}$ (solid orange line), which coincides with the decrease in $L_{\mathrm{orb}}/(S_{1} + S_{2})$ (dashed green line). The primary's spin angular momentum $S_{1}$ is shown with a solid blue line.  Panel (b) shows the evolution of the moment of inertia (solid lines) and tidal synchronisation timescale (dashed lines) for both components. While the moment of inertia of the primary decreases around $M_{1}=0.3\,\mathrm{M}_{\odot}$, it sharply increases for the secondary. At the same time, the secondary star's tidal synchronisation timescale decreases by approximately two orders of magnitude. Panel (c) shows the evolution of the primary's and secondary's radii (solid lines) and Roche lobe radii (dashed lines). While the former fills its Roche lobe, tides cause orbital shrinkage, which results in the secondary also filling its Roche lobe.}
    \label{fig:tidally_driven_contact}
\end{figure}

When a single star ascends the (super-)giant branch, its moment of inertia $I$ increases, such that its surface rotation velocity decreases. In a binary and if the tidal synchronisation timescale $\tau_{\mathrm{sync}}$ \citep[][Eq.\,27]{Hurley2002} is shorter than the expansion timescale $\tau_{R/\dot{R}}$, the star is tidally locked to the orbit and does not spin down. The transfer of angular momentum from the orbit to the spin of the star shrinks the orbit. Under certain circumstances, tidally driven orbital decay can lead to contact phases, which demonstrates the importance of considering the effect of tides in binaries.

In the most extreme case, the binary system becomes Darwin unstable \citep{Darwin1879}. This instability arises if 
\begin{align}\label{eq:classical_Darwin}
    \frac{L_{\mathrm{orb}}}{(I_{1} + I_{2})\Omega} < 3\quad ,
\end{align}
where $L_{\mathrm{orb}}$ is the orbital angular momentum, $\Omega$ is the orbital angular rotation velocity, and $I_{1}$ and $I_{2}$ are the moment of inertia of the primary and secondary star, respectively. This condition is derived under the assumption of a circular ($e=0$, with $e$ the eccentricity), coplanar (spins of components are aligned) and synchronised ($\omega_{1} = \omega_{2} = \Omega$) system. Here, $\omega_{1}$ and $\omega_{2}$ are the angular rotation rates of the primary and secondary, respectively. Moreover, solid-body rotation is assumed for the individual components, allowing the spin angular momenta $S_{1,\,2}$ to be written as $S_{1,\,2}=\omega_{1,\,2} I_{1,\,2}$. 
Especially for close or contact binaries, where $\tau_{\mathrm{sync}}$ is typically of the order of a few years, the Darwin instability can lead to a dynamical inspiral of the components, resulting in a merger.\\

Models labelled ``Tidally-driven contact'' experience orbital decay caused by tides before the onset of contact, while orbital widening is expected from mass transfer ($q > 1$). We trace this condition in post-processing. This label is not used for systems which are at one point during their evolution Darwin unstable. Although these systems will experience orbital decay, this can happen on timescales longer than the evolutionary timescales of the system. However, most of the systems that are driven into contact by tides do eventually become Darwin unstable.\\

An example of a system in which tides lead to the onset of contact is shown in Fig.\,\ref{fig:tidally_driven_contact}. In this low-mass binary, the primary overfills its Roche lobe near the end of the MS. During mass transfer, the primary star reaches the TAMS, but does not become a red giant because of the continuous stripping of its envelope. During the mass-transfer phase, the secondary star overtakes in evolution, leaves the MS and turns to the giant branch. At this point, the mass ratio of the system has already reversed, and mass transfer is almost conservative, widening the orbit. Two important changes occur when the secondary becomes a giant. First, there is the development of a deep convective envelope and an increase in radius, which lead to increased tidal coupling of the star: $\tau_{\mathrm{sync},\,2}$ drops by several orders of magnitude (Fig.\,\ref{fig:tidally_driven_contact}b). Because of this, the orbital and rotation period of the secondary synchronise. Secondly, the increase in radius and density redistribution (the envelope now has a deep convective zone) both increase the moment of inertia $I_{2}$ of the star (Fig.\,\ref{fig:tidally_driven_contact}b). Following the conservation of angular momentum, the secondary spins down, yet it is immediately spun up again by tides. Hence, the spin angular momentum of the secondary $S_{2}$ increases and the orbital angular momentum $L_{\mathrm{orb}}$ decreases (Fig.\,\ref{fig:tidally_driven_contact}a). Eventually, the system becomes Darwin unstable, as it fulfils the condition\footnote{Since the stars in our models are in general not solid body rotators, we cannot assume that $S_{1,\,2}=\omega_{1,\,2} I_{1,\,2}$. Therefore, we use a modified version of the classical criterion for the Darwin instability: $L_{\mathrm{orb}}/(S_{1}+S_{2}) \lesssim 3$.} $L_{\mathrm{orb}}/(S_{1}+S_{2}) \lesssim 3$. Because of the decrease in $L_{\mathrm{orb}}$, the orbit, which was widening from mass transfer, starts shrinking again. As a consequence of the shrinking Roche lobe, the primary star is stripped of its outer envelope increasingly rapidly, leading to a decrease in radius and an increase in mass-transfer rate (not shown).  The shrinking of the Roche lobes eventually leads to the formation of a contact binary once the secondary also fills its Roche lobe (Fig.\,\ref{fig:tidally_driven_contact}c). At the point when the contact binary is formed, the helium core mass $M_{\mathrm{He}}$ has increased beyond that of the primary star. This means that the secondary is closer to core helium ignition, that is, it has overtaken the primary in evolution.\\

\subsection{Unstable mass transfer and CE phases}\label{sec:unstable_MT}
During mass transfer, the donor star's radius and Roche lobe can both either shrink or expand. Mass transfer is generally stable when the donor star shrinks faster or expands slower than its Roche lobe. However, when the donor star overfills its Roche lobe by increasing amounts as a reaction to mass loss, the binary enters an unstable runaway situation \citep[e.g.][]{Soberman1997}. This happens, for example, when the Roche lobe shrinks while the donor expands (e.g. during mass transfer with $q < 1$ and/or because of orbital angular momentum loss). The responses of the donor's radius and Roche lobe radius are quantified in the mass-radius exponents $\xi_{R_{1}} \equiv \mathrm{d}\ln R_{1}/\mathrm{d} \ln M_{1}$ and $\xi_{\mathrm{RL}} \equiv \mathrm{d}\ln R_{\mathrm{RL},\,1}/\mathrm{d} \ln M_{1}$, respectively \citep{Webbink1984}. If $\xi_{R_{1}} < \xi_{\mathrm{RL}}$, mass transfer is unstable and the mass-transfer rate keeps increasing, potentially reaching values larger than $1\,\mathrm{M}_{\odot}\,\mathrm{yr}^{-1}$.  During runaway (unstable) mass transfer, the primary star increasingly overfills its Roche lobe and the secondary's expansion timescale becomes orders of magnitude lower than its thermal timescale. When the primary star is an MS or HG star (Case-A or -Be), a contact binary will form and merge on a timescale shorter than the primary's thermal timescale because of the runaway expansion of both components. In Case-Bl and -C binaries, runaway mass transfer leads to the engulfment of the secondary in the primary's envelope, that is, a classical CE phase. A classical CE phase can lead to a merger or leave a close binary behind \citep[see, e.g.][]{Roepke2023}. The onset of a classical CE through runaway mass transfer is shown schematically in Fig.\,\ref{fig:contact_schematic}.3a--b.\\

We look for runaway mass transfer in post-processing and label such models ``Runaway MT'' given the following three criteria. Firstly, the mass-transfer rate $\dot{M}_{\mathrm{trans}}$ needs to exceed the thermal-timescale mass-transfer rate, set by $\dot{M}_{\mathrm{KH}} = M_{1}/\tau_{\mathrm{KH,\,1}}$. Secondly, the second time derivative of $\log_{10}\dot{M}_{\mathrm{trans}}$ has to be positive. When this is the case, the rate of change in $\dot{M}_{\mathrm{trans}}$ is increasing, which is indicative of a runaway situation. Moreover, this condition also ensures that $\dot{M}_{\mathrm{trans}}$ does not decrease again, as is observed for stable Case-C mass transfer (see Sect.\,\ref{sec:stable_caseC}). Thirdly, the condition for unstable mass transfer described above needs to be fulfilled: $\xi_{R_{1}} < \xi_{\mathrm{RL}}$. For semi-detached models using \texttt{MESA}'s \texttt{contact} mass-transfer scheme, $\xi_{R_{1}} = \xi_{\mathrm{RL}}$ by definition. Hence, in this case, only the first two conditions are evaluated. Note that $\mathrm{L}_{2}$-overflow is not a necessary condition for the onset of runaway mass transfer. However, it does often occur in systems with runaway mass transfer.\\

An example of a binary system experiencing delayed runaway mass transfer is shown in Fig.\,\ref{fig:example_runawayMT} and was previously discussed to demonstrate $\mathrm{L}_{2}$-overflow in Sect.\,\ref{sec:l2over}. The evolution of the mass-transfer rate $\dot{M}_{\mathrm{trans}}$ as a function of the binary system's age is shown in Fig.\,\ref{fig:example_runawayMT}b, where it is also compared to the thermal and dynamical mass-transfer rates. Shortly after $\dot{M}_{\mathrm{trans}}$ exceeds the former, the runaway nature of the evolution is observed, during which $\dot{M}_{\mathrm{trans}}$ starts nearing the dynamical mass-transfer rate. Simultaneously, we see in Fig.\,\ref{fig:example_runawayMT}c that the slope of $R_{1}$ as a function of the decreasing $M_{1}$ becomes larger than that of $R_{\mathrm{RL},\,1}$, which means that the primary increasingly overfills its Roche lobe. This continues up to the point where $\mathrm{L}_{2}$-overflow occurs, after which an accelerated orbital shrinkage is expected (see Sect.\,\ref{sec:l2over}). In this example, the binary does not enter a runaway phase immediately at the onset of mass transfer. Therefore, it is an example of a \emph{delayed} runaway mass-transfer phase \citep[e.g.][]{Hjellming1987, Han2006, Pavlovskii2015, Ge2015, Ge2020}. Since in this particular example the primary star is not yet a supergiant at the onset of mass transfer, this system is not expected to enter a classical common-envelope phase, but rather become a contact binary that will most likely eventually merge.

\section{Occurrence of contact phases}\label{sec:results1}
In this section, we present our contact tracing results for three initial primary masses (Sect.\,\ref{subsec:res_contact_phases}). Then, we focus on a particular region of the initial binary parameter space in which Case-Bl~and~-C mass transfer has been found to be stable (Sect.\,\ref{sec:stable_caseC}). Next, we look at the incidence of the different physical mechanisms leading to contact in a population of binary systems (Sect.\,\ref{sec:population}), and put lower limits on the stellar merger and classical CE  fractions of mass-transferring binaries (Sect.\,\ref{sec:merger_CE_frac}). Lastly, we compare the properties of Case-A contact systems found in our grid with those of observed systems (Sect.\,\ref{sec:comparison}).

\begin{figure*}
\sidecaption
  \includegraphics[width=12cm]{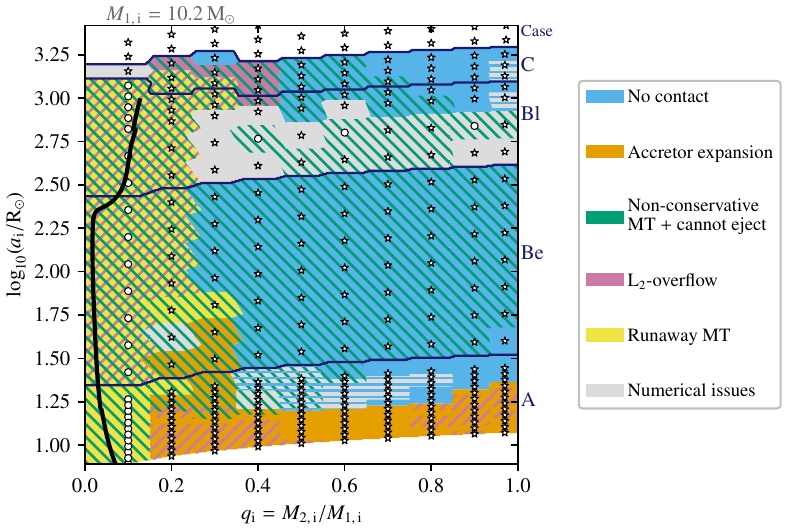}
     \caption{Occurrence of contact phases for models with initial primary masses $M_{1,\,\mathrm{i}}=10.2\,\mathrm{M}_{\odot}$ on the initial mass ratio--separation plane. Models marked with a dot are those for which accretion was limited to $0.1$ times the secondary's global thermal timescale $\tau_{\mathrm{KH}}$ (see Sect.\,\ref{methods:mass_transfer}). The other ones are marked with a star symbol. The dark-blue quasi-horizontal lines indicate the initial mass transfer cases, which can be read from the right side. Systems on the left of the solid black line are Darwin unstable at the onset of mass transfer according to Eq.\,(\ref{eq:classical_Darwin}) and assuming $R_{1} = R_{\mathrm{RL,\,1}}$.}
     \label{fig:contact_Mp1016}
\end{figure*}

\subsection{Contact phases for different initial primary masses}\label{subsec:res_contact_phases}

\subsubsection{Initial primary mass of $10.2\,\mathrm{M}_{\odot}$}\label{sec:contact_1016}
In Fig.\,\ref{fig:contact_Mp1016}, we show our binary models in the initial mass ratio--separation ($q_{\mathrm{i}}$-$\log_{10}a_{\mathrm{i}}$) plane for a fixed initial primary mass $M_{1,\,\mathrm{i}}$ of $10.2\,\mathrm{M}_{\odot}$. Systems with  $M_{1,\,\mathrm{i}} = 20.0$ and $1.6\,\mathrm{M}_{\odot}$ are shown in Fig.\,\ref{fig:contact_Mp20.0-1.58}, and described in Sect.\,\ref{sec:contact_200} and \ref{sec:contact_158}, respectively. The results for the other $M_{1,\,\mathrm{i}}$ are shown in Appendix \ref{app:contact}. A table with the evolutionary outcomes of all computed \texttt{MESA} models as well as the model input and output files are available online\footnote{\url{https://zenodo.org/doi/10.5281/zenodo.10148634}}. An extract of this table can be found in Appendix \ref{app:table}.

The coloured regions in Fig.\,\ref{fig:contact_Mp1016} indicate the different physical mechanisms leading to contact (``Accretor expansion'', ``Non-conservative MT + cannot eject'', ``$\mathrm{L}_{2}$-overflow'', ``Tidally-driven contact'' and ``Runaway MT'', see Sect.\,\ref{sec:mechanisms}). They have been constructed using nearest neighbour interpolation. Models labelled ``No contact'' undergo at least one phase of mass transfer but manage to avoid any form of contact until the end of our computations (see Sect.\,\ref{subsec:stopping_cond}). In some models, reverse mass transfer occurs from the initially less massive secondary to the primary, but we do not consider this here for contact tracing. These models are also labelled ``No contact''. The potential fates of reverse mass transfer systems are briefly described later in this section. Models labelled ``Numerical issues'' were numerically unable to reach the desired stopping conditions of our computations (see Sect.\,\ref{subsec:stopping_cond}). Lastly, models labelled ``MT after TP'' are those for which Case-C mass transfer occurs after the TP-AGB phase. Their final fate is not further interpreted (see Sect.\,\ref{sec:wind_mass_loss}).

In some regions of the initial binary parameter space, more than one of the above conditions are met, and we indicate this by the corresponding hatching (ancillary outcome). For the background colour (principal outcome), priority is given in the following order (highest to lowest priority): ``Tidally driven contact'', ``Runaway MT'', ``Accretor expansion'', ``$\mathrm{L}_{2}$-overflow''  ``No contact'', and ``Numerical issues''. Regions in which ``No contact'' is indicated by hatching contain models that make it to the end of core-He burning and are therefore strong candidates for avoiding contact but encountered numerical difficulties before they could reach the end of core-C burning (see Sect.\,\ref{subsec:stopping_cond}). We use ``Non-conservative MT + cannot eject'' exclusively as an ancillary outcome (hatching). Although this is a viable mechanism leading to contact (Sect.\,\ref{subsec:non_cons_mt}), it is uncertain to what evolutionary outcome it leads. For example, a system with ``No contact'' as its principal outcome and ``Non-conservative MT + cannot eject'' as its ancillary outcome can either avoid contact when, for example, an accretion or circumbinary disk is formed, or form a contact binary when the non-accreted matter fills the secondary's Roche lobe.  \\

For systems with  $M_{1,\,\mathrm{i}}=10.2\,\mathrm{M}_{\odot}$, contact binaries form because of the expansion of the accreting secondary star for $q_{\mathrm{i}} = 0.15\text{--}1.00$ in the initially closest Case-A binary systems (``Accretor expansion''; Fig.\,\ref{fig:contact_Mp1016}). For $q_{\mathrm{i}} = 0.15\text{--}0.45$ with $\log_{10}\left(a_{\mathrm{i}}/\mathrm{R}_{\odot}\right) \lesssim 1.23$ and $q_{\mathrm{i}} = 0.75\text{--}1.00$, these contact systems additionally experience $\mathrm{L}_{2}$-overflow, which leads to orbital angular momentum loss and subsequent stellar mergers. For $q_{\mathrm{i}} = 0.45\text{--}0.75$, the contact binaries do not expand beyond the $\mathrm{L}_{2}$-lobe. This behaviour is found consistently throughout the range of $M_{1,\,\mathrm{i}}=2.6\text{--}20.0\,\mathrm{M}_{\odot}$ for $q_{\mathrm{i}}$ between 0.45 and $0.65\text{--}0.75$, where the upper boundary decreases with decreasing $M_{1,\,\mathrm{i}}$. By comparing with the expansion timescales of the accretors (Figs.\,\ref{fig:expansion_timescales}--\ref{fig:expansion_timescales2}) in systems avoiding $\mathrm{L}_{2}$-overflow, we find that systems with $q_{\mathrm{i}} \gtrsim 0.5$ likely produce longer-lived ($\tau_{\mathrm{nuc,\,1}}$) contact binaries, which may be observable. The contact binaries with $q_{\mathrm{i}} \lesssim 0.5$ form through the thermal-timescale expansion of the accretor and merge or detach again on the accretor's thermal timescale.

The initially wider Case-A ($\log_{10}\left(a_{\mathrm{i}}/\mathrm{R}_{\odot}\right) \gtrsim 1.23$) systems with $q_{\mathrm{i}} = 0.15\text{--}0.35$ that avoid $\mathrm{L}_{2}$-overflow reach mass-transfer rates $\dot{M}_{\mathrm{trans}}$ orders of magnitude larger than their thermal mass-transfer rate $\dot{M}_{\mathrm{KH}}$. $\dot{M}_{\mathrm{trans}}$ eventually reaches the stopping condition value of $10\,\mathrm{M}_{\odot}\,\mathrm{yr}^{-1}$ (see Sect.\,\ref{subsec:stopping_cond}). In the Case-Be models labelled ``Accretor expansion'', the primary is an HG star and the secondary is expanding on a timescale orders of magnitude shorter than its thermal timescale. The resulting contact binary will likely experience $\mathrm{L}_{2}$-overflow and merge.

The Case-A binary systems with $q_{\mathrm{i}} \leq 0.15$ and $\log_{10}\left(a_{\mathrm{i}}/\mathrm{R}_{\odot}\right) \lesssim 1.37$ experience a phase of runaway mass transfer (``Runaway MT''). In addition, the mass-accretion rate of the secondaries is limited to $10\dot{M}_{\mathrm{KH},\,2}$, which brings the mass-transfer efficiency $\beta$ almost down to zero right after the onset of mass transfer. The lack of further accretion quenches initially rapid expansion ($\tau_{R/\dot{R},\,2} < \tau_{\mathrm{KH,\,2}}$) of the secondary. Assuming that in reality this rapid expansion continues (the accretion rate in these models is limited for numerical reasons), the two MS stars are likely to form contact binaries.

In the region of $q_{\mathrm{i}}=0.35\text{--}1.00$ and $\log_{10}\left(a_{\mathrm{i}}/\mathrm{R}_{\odot}\right) \approx 1.2\text{--}1.5$, we find Case-A systems that avoid contact (``No contact''). In these systems, the primary stars are stripped in Case-A, Case-AB, and in some models even Case-ABC mass-transfer phases. Eventually, they reach or are expected to reach (horizontal hatching) the end of core carbon burning, or enter a phase of reverse mass transfer. The latter occurs for some of the binary systems with $q_{\mathrm{i}} = 0.65\text{--}1.00$ close to the border between forming contact binaries and avoiding contact. Here, the secondary (over-)fills its Roche lobe after the primary has detached, leading to a phase of accretion onto a stripped primary star. Other models that experience reverse mass-transfer phases are found at $q_{\mathrm{i}} = 0.97$ and $\log_{10}\left(a_{\mathrm{i}}/\mathrm{R}_{\odot}\right) \approx 1.52\text{--}2.30$. Such phases are not considered further in our contact tracing.\\

For Case-Be systems, contact is likely avoided for $q_{\mathrm{i}}\gtrsim 0.25\text{--}0.35$. The primary stars in these systems all follow the same evolutionary pathway as the example described in Fig.\,\ref{fig:example_nonConservative}: the HG primaries are stripped and reach core-C exhaustion. The secondary stars are spun up by accretion and mass transfer becomes non-conservative. Tidal synchronisation timescales are longer than the spin-up timescales, hence tides are not able to prevent spin-up to the critical rotation rate as is the case in the closer-orbit Case-A systems. Virtually all systems fail to eject the non-accreted matter at one point during this phase of non-conservative mass transfer (``Non-conservative MT + cannot eject'').

At $q_{\mathrm{i}} \leq 0.25-0.35$, Case-Be binaries experience runaway mass transfer. All systems experience non-conservative mass transfer and fail to eject the non-accreted matter. Except for systems at $\log_{10}\left(a_{\mathrm{i}}/\mathrm{R}_{\odot}\right) \lesssim 1.78$ and $q_{\mathrm{i}} = 0.15\text{--}0.35$, all of them also experience $\mathrm{L}_{2}$-overflow. The secondary stars are still on the MS. The primaries are HG stars without a clear core-envelope boundary (see Fig.\,\ref{fig:contact_configs}--\ref{fig:contact_configs3} in Appendix \ref{app:contact_configs} for the evolutionary state of all models at contact/termination). Hence, these systems are expected to evolve into contact binaries and not result in classical CEs. \\

We find runaway mass transfer in Case-Bl systems with $q_{\mathrm{i}} \leq 0.25$ and $\log_{10}\left(a_{\mathrm{i}}/\mathrm{R}_{\odot}\right) \lesssim 2.92$, and $q_{\mathrm{i}} \leq 0.35$ and $\log_{10}\left(a_{\mathrm{i}}/\mathrm{R}_{\odot}\right) \gtrsim 2.92$ as well as a small set of Case-C systems with $q_{\mathrm{i}} = 0.15\text{--}0.35$. The secondary stars in these systems are MS stars. The difference compared to the Case-Be systems experiencing runaway mass transfer is that all primary stars have now evolved into supergiants with a clear core-envelope boundary by the time mass transfer starts (this can be observed from, for example, the steeper drop in density and binding energy around the core-envelope boundary). As a consequence, all these systems are expected to enter a classical CE phase. 

Following \citet{Rasio1995a}, we have computed the mass ratios below which binary systems are Darwin unstable at the onset of mass transfer (black solid line in Fig.\,\ref{fig:contact_Mp1016}). Systems with supergiant donors with deep convective envelopes (Case~Bl and C) are Darwin unstable at the onset of mass transfer for $q_{\mathrm{i}} \lesssim 0.1$. In Case-A and -Be binaries, this is the case for $q_{\mathrm{i}} \lesssim 0.05$.

At $q_{\mathrm{i}} = 0.15\text{--}0.45$, we find Case-C systems (and one Case-Bl system) with runaway mass-transfer where the primary stars overfill their $\mathrm{L}_{2}$-lobes by a factor of more than one. The primaries have (almost) engulfed their companion and we witness the beginning of a classical CE phase.

For $q_{\mathrm{i}} \geq 0.45$, Case-Bl and -C binary systems evolve through a stable phase of mass transfer. The Case-Bl systems are actually found to go through two stable mass-transfer phases, Case-Bl and Case-BC. Thanks to the stability of these mass-transfer phases, contact is avoided. The stability of Case~Bl and -C mass transfer is described further in Sect.\,\ref{sec:stable_caseC}.\\

The region around the transition between early and late Case-B systems contains models for which \texttt{MESA} does not converge numerically because of the rapid spin-up of the accretor. These issues with the spin-up of the accretor occur in systems with initial primary masses down to $8.6\,\mathrm{M}_{\odot}$. For systems with initial primary masses of $\geq 13.1\,\mathrm{M}_{\odot}$, models with Case~Be mass transfer are also affected. This can be seen in Figs.\,\ref{fig:contact_Mp20.0-1.58} and \ref{fig:appendix_contact_tracing1}. 

\subsubsection{Initial primary mass of $20.0\,\mathrm{M}_{\odot}$}\label{sec:contact_200}

\begin{figure*}
\centering
  \includegraphics[width=18cm]{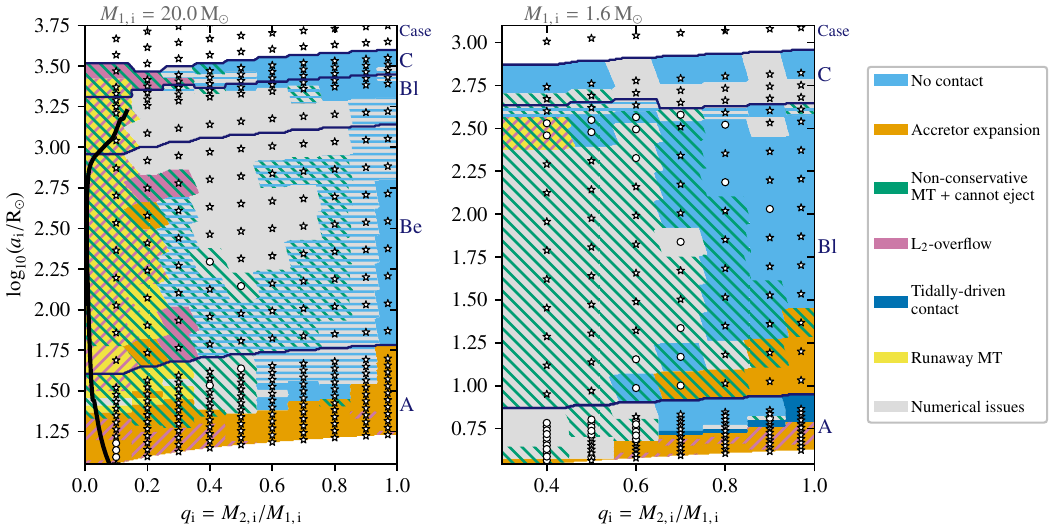}
     \caption{Same as Fig.\,\ref{fig:contact_Mp1016}, but for $M_{1,\,\mathrm{i}} = 20.0\,\mathrm{M}_{\odot}$ (left) and $M_{1,\,\mathrm{i}} = 1.6\,\mathrm{M}_{\odot}$ (right).}
     \label{fig:contact_Mp20.0-1.58}
\end{figure*}

For $M_{1,\,\mathrm{i}}=20.0\,\mathrm{M}_{\odot}$ (left panel in Fig.\,\ref{fig:contact_Mp20.0-1.58}), the general picture is similar to that of models with 
$M_{1,\,\mathrm{i}}=10.2\,\mathrm{M}_{\odot}$ presented above. However, there are some notable differences.\\

Similarly to the models with $M_{1,\,\mathrm{i}}=10.2\,\mathrm{M}_{\odot}$, the initially closest Case-A systems form contact binaries through the expansion of the secondary star during mass transfer. For the twin systems ($q_{\mathrm{i}} \approx 0.97$), this even happens for all the Case-A systems. Binary systems with $q_{\mathrm{i}} \leq 0.15$ and $\log_{10}\left(a_{\mathrm{i}}/\mathrm{R}_{\odot}\right) \lesssim 1.37$ enter contact through accretor expansion, whereas binary systems in the equivalent region for $M_{1,\,\mathrm{i}}=10.2\,\mathrm{M}_{\odot}$ do so through a phase of runaway mass transfer. In general, it is found that the former holds for models with $M_{1,\,\mathrm{i}}=12.6\text{--}20.0\,\mathrm{M}_{\odot}$ and the latter for $M_{1,\,\mathrm{i}}=5.2\text{--}10.2\,\mathrm{M}_{\odot}$. For $M_{1,\,\mathrm{i}} < 5.0\,\mathrm{M}_{\odot}$, models with $q_{\mathrm{i}} = 0.1$ are not computed, so information about this region is not available.

Just as for the systems with $M_{1,\,\mathrm{i}}=10.2\,\mathrm{M}_{\odot}$, contact binaries with $q_{\mathrm{i}} \leq 0.45$ formed through accretor expansion experience $\mathrm{L}_{2}$-overflow (Fig.\,\ref{fig:contact_Mp20.0-1.58}, left). However, at $q_{\mathrm{i}} \geq 0.75$, $\mathrm{L}_{2}$-overflow is found to be largely avoided during the first phase of contact, contrary to what is found for systems with $M_{1,\,\mathrm{i}}=10.2\,\mathrm{M}_{\odot}$. This, however, does not imply that these contact phases are long-lived ($\tau_{\mathrm{contact}} \sim \tau_{\mathrm{nuc,\,1}}$) since the secondary has been found to shrink again after regaining thermal equilibrium ($\tau_{\mathrm{contact}} \sim \tau_{\mathrm{KH,\,2}})$. In almost all of these systems, a second contact phase follows later in the evolution because of the nuclear timescale expansion of the secondary star. The primary stars are at his point either MS or post-MS stars. In the former case, the models have again been computed through the contact phase, which almost exclusively results in $\mathrm{L}_{2}$-overflow. In the latter case, the evolution is terminated when the accretor fills its Roche lobe (see Sect.\,\ref{subsec:stopping_cond}).

Initially wider Case-A systems, with $\log_{10}\left(a_{\mathrm{i}}/\mathrm{R}_{\odot}\right) = 1.37\text{--}1.68$ and $q_{\mathrm{i}} \leq 0.35$, go through a phase of runaway mass transfer and thus likely form contact binaries. Compared to the systems with $M_{1,\,\mathrm{i}}=10.2\,\mathrm{M}_{\odot}$, this region extends to higher values of $q_{\mathrm{i}}$. At similar initial separations and $q_{\mathrm{i}}=0.35\text{--}0.55$, \texttt{MESA} does not converge numerically. As in the equivalent region for $M_{1,\,\mathrm{i}}=10.2\,\mathrm{M}_{\odot}$ models, the solver fails to find a suitable solution for the accretor such that the rotation rate remains below the critical rotation rate.

Primary stars in Case-A systems with $q_{\mathrm{i}}=0.550\text{--}0.935$ and $\log_{10}\left(a_{\mathrm{i}}/\mathrm{R}_{\odot}\right) = 1.38\text{--}1.77$ are stripped in Case-A, Case-AB and Case-ABC mass-transfer phases and avoid/are expected to avoid contact. Contrarily to similar systems with $M_{1,\,\mathrm{i}}=10.2\,\mathrm{M}_{\odot}$, none experience a phase of reverse mass transfer. There is, however, a set of models of twin systems with $q_{\mathrm{i}} = 0.97$ and $\log_{10}\left(a_{\mathrm{i}}/\mathrm{R}_{\odot}\right) \approx 1.78\text{--}2.80$ where reverse mass transfer does occur, as in the binaries with $M_{1,\,\mathrm{i}}=10.2\,\mathrm{M}_{\odot}$.\\

In Case-Be binaries, the differences with respect to the equivalent $M_{1,\,\mathrm{i}}=10.2\,\mathrm{M}_{\odot}$ systems are more pronounced. Firstly, a few systems \emph{only} experience $\mathrm{L}_{2}$-overflow and thus avoid runaway mass transfer. Closer inspection shows that in these systems, the radius of the primary star exceeds $R_{\mathrm{L}_{2}}$ by ${\lesssim}\,20\%$ for ${\lesssim}\,10^{3}\,\mathrm{yrs}$. It is uncertain whether these contact binaries (see Fig.\,\ref{fig:contact_schematic}.2c) are stable, given that the phase of $\mathrm{L}_{2}$-overflow is short and might not lead to sufficient angular momentum loss to cause significant orbital decay and hence a merger. Moreover, the primary star shrinks again rapidly afterwards, causing the binary to become semi-detached again.
Secondly, Case-Be models have a harder time converging numerically. In these models, the primary stars are stripped in one or more mass-transfer phase(s), and continue core-He burning and core-C burning as stripped stars with a thin hydrogen layer ($\lesssim 1\,\mathrm{M}_{\odot}$). At these masses, \texttt{MESA} runs into numerical difficulties and the timesteps of the simulations drop well below one year. This is a known issue for this kind of stripped stars (Y. G\"otberg, 2022, priv. comm.) and has prevented us from computing the evolution to the end of core-C exhaustion for all models. A select number of models are computed with small timesteps until core-C exhaustion. In these models, the hydrogen surface layers expand during core-C burning and drive a stable and short-lived (${\sim}10^{3}\,\mathrm{yrs}$) Case-C mass-transfer phase. Mass transfer is stable because the donor stars shrink again when nearing core-C exhaustion and the core-mass fraction is ${>}\,0.5$, which typically signals stability \citep{Temmink2023}. At lower initial primary masses, such as for models with $M_{1,\,\mathrm{i}} = 18.4\,\mathrm{M}_{\odot}$ (see Fig.\,\ref{fig:appendix_contact_tracing1} in Appendix \ref{app:contact}), the aforementioned numerical difficulties are less severe, allowing most stripped primaries to reach core-C exhaustion. Most of them avoid Case-C mass transfer, such that we do not expect contact phases. \\

From the left panel of Fig.\,\ref{fig:contact_Mp20.0-1.58}, we see that for Case-Be, -Bl and -C systems with $q_{\mathrm{i}} \leq 0.25$ and $\log_{10}\left(a_{\mathrm{i}}/\mathrm{R}_{\odot}\right) \lesssim 2.5$, and $q_{\mathrm{i}} \leq 0.15$ and $\log_{10}\left(a_{\mathrm{i}}/\mathrm{R}_{\odot}\right) =  2.50\text{--}3.42$, contact is reached again through runaway mass transfer. Among these, Case-Bl and Case-C systems are likely to enter classical CE phases. We also see two models at $q_{\mathrm{i}} = 0.2$ with $\log_{10}\left(a_{\mathrm{i}}/\mathrm{R}_{\odot}\right) \approx  1.72$ and $\log_{10}\left(a_{\mathrm{i}}/\mathrm{R}_{\odot}\right) \approx  2.57$, respectively, that enter contact through accretor expansion. However, the accretor expansion is driven by numerical difficulties with finding accretors that spin below critical. Contact appears likely in these two models, albeit for another reason (unstable mass transfer or $\mathrm{L}_{2}$-overflow).\\

Just as for the $M_{1,\,\mathrm{i}} = 10.2\,\mathrm{M}_{\odot}$ systems, there is a small region in the initial binary parameter space at $q_{\mathrm{i}} \lesssim 0.35$ and $\log_{10}\left(a_{\mathrm{i}}/\mathrm{R}_{\odot}\right) \approx  3.4\text{--}3.5$ where runaway mass transfer does not occur, but a classical CE phase is expected from $\mathrm{L}_{2}$-overflow. For $q_{\mathrm{i}} \gtrsim 0.35$, a similar region of stable Case-Bl and Case-C mass transfer as for the $M_{1,\,\mathrm{i}} = 10.2\,\mathrm{M}_{\odot}$ systems is found (see Sect.\,\ref{sec:stable_caseC}).\\

In the same way as the $10.2\,\mathrm{M}_{\odot}$ initial primary mass models, the spin-up of the accretor leads to numerical convergence problems in both the Case~Bl and -Be regions. At these higher masses, the problems persist down to initial separations of a few $100\,\mathrm{R}_{\odot}$.

\subsubsection{Initial primary mass of $1.6\,\mathrm{M}_{\odot}$}\label{sec:contact_158}
At lower initial primary masses, more specifically $M_{1,\,\mathrm{i}} = 1.6\,\mathrm{M}_{\odot}$, the situation is different than at higher masses (right panel in Fig.\,\ref{fig:contact_Mp20.0-1.58})\footnote{Note that the lower limit for initial secondary star masses is $0.5\,\mathrm{M}_{\odot}$. Hence, there are no models at $q_{\mathrm{i}} < 0.4$.}. Although the primary stars have radiative envelopes during the MS, none of our systems undergo Case-Be mass transfer, because of the resolution in $a_{\mathrm{i}}$\footnote{$1.6\,\mathrm{M}_{\odot}$ stars only expand about $20\%$ in radius ($\lesssim 1\,\mathrm{R}_{\odot}$) before the envelope becomes predominantly convective and Case-Bl mass transfer ensues when the star fills its Roche lobe. Given the average resolution of ${\sim}5\,\mathrm{R}_{\odot}$ in the Case-B region of the grid for these initial primary masses, a narrow Case-Be region is not resolved.}.\\

Case-A models at $q_{\mathrm{i}} = 0.35\text{--}0.65$ encounter numerical issues when the accreting secondary star reaches critical rotation. Also in Case-Bl systems with mass ratios $q < 0.65\text{--}0.75$, the models do not converge numerically. As can be seen from Figs.\,\ref{fig:appendix_contact_tracing2}--\ref{fig:appendix_contact_tracing3}, this is an issue that occurs for $M_{1,\,\mathrm{i}} \leq 2.2\,\mathrm{M}_{\odot}$. Although this prevents us from including binary systems with $M_{1,\,\mathrm{i}} \leq 2.2\,\mathrm{M}_{\odot}$ in further population analysis, specific regions of the initial binary parameter space can still be described and hold valuable information.\\

The expansion of the accretor star is leading to the formation of contact binaries in Case-A systems with $\log_{10}\left(a_{\mathrm{i}}/\mathrm{R}_{\odot}\right) \lesssim 0.57$ and $q_{\mathrm{i}} = 0.35\text{--}0.55$, and $\log_{10}\left(a_{\mathrm{i}}/\mathrm{R}_{\odot}\right) =  0.57\text{--}0.79$ and $q_{\mathrm{i}} = 0.55\text{--}1.00$ (Fig.\,\ref{fig:contact_Mp20.0-1.58}). Similar to contact systems with $M_{1,\,\mathrm{i}}=10.2\,\mathrm{M}_{\odot}$, systems with $q_{\mathrm{i}} \gtrsim 0.80$ experience $\mathrm{L}_{2}$-overflow.

Case-A systems with $q_{\mathrm{i}} = 0.65\text{--}0.94$ and $\log_{10}\left(a_{\mathrm{i}}/\mathrm{R}_{\odot}\right) =  0.73\text{--}0.94$ evolve through similar pathways as those in equivalent regions of the initial parameters space with $M_{1,\,\mathrm{i}} = 10.2\,\mathrm{M}_{\odot}$ and $M_{1,\,\mathrm{i}} = 20.0\,\mathrm{M}_{\odot}$. The primaries are stripped in Case-A and Case-B mass-transfer phases, after which reverse mass transfer occurs.

In between the Case-A systems at $q_{\mathrm{i}} = 0.65\text{--}1.00$ that form contact binaries through accretor expansion and those that avoid contact, contact binaries form by tidal interaction. In our grid, such tidally-driven contact in Case-A systems can be found for initial primary masses between $0.8$ and $1.8\,\mathrm{M}_{\odot}$. For initial primary masses between $0.8$ and $1.1\,\mathrm{M}_{\odot}$ tidally driven contact also occurs in the initially closest Case-B systems with $q_{\mathrm{i}} = 0.97$ (Fig.\,\ref{fig:appendix_contact_tracing3}).\\

Case-Bl binary systems at $q_{\mathrm{i}} = 0.65\text{--}1.00$ and $\log_{10}\left(a_{\mathrm{i}}/\mathrm{R}_{\odot}\right) =  0.91\text{--}1.45$ evolve into contact binaries because of the expansion of the accretor. The donor stars in these binaries have deep convective envelopes, resulting in mass-transfer rates ${\gtrsim}\,10^{-5}\,\mathrm{M}_{\odot}/\mathrm{yr}$. Because of the relatively high mass transfer rates, the accretors are out of thermal equilibrium and expand with $\tau_{\mathrm{dyn,\,2}} < \tau_{R/\dot{R},\,2} < \tau_{\mathrm{KH,\,2}}$ and fill their Roche lobes. In comparison, the Case-A binaries at $\log_{10}\left(a_{\mathrm{i}}/\mathrm{R}_{\odot}\right) <  0.91$, which avoid contact, have mass transfer rates ${\lesssim}\,10^{-7}\,\mathrm{M}_{\odot}/\mathrm{yr}$, resulting in a nuclear-timescale expansion of the accretor star. The three models at $q_{\mathrm{i}} = 0.85\text{--}1.00$ not labelled ``Non-conservative mass transfer + cannot eject'' have conservative mass transfer because they manage to reach contact before the secondary star rotates at its critical rotation rate. 

At larger initial separations ($\log_{10}\left(a_{\mathrm{i}}/\mathrm{R}_{\odot}\right) =  1.07\text{--}2.64$), Case-Bl systems with $q_{\mathrm{i}} > 0.65\text{--}0.75$ avoid contact. The primary stars have their envelopes removed during the red giant (RG) phase and end up as (partially) stripped stars. Mass transfer is stable because of the relatively high initial mass ratios (the mass ratio becomes greater than one early on during mass transfer). The mass transfer stops when the entire hydrogen envelope is removed. A pure helium WD remains if the hydrogen-burning shell is stripped before the helium core grows to a mass above roughly $0.45\,\mathrm{M}_{\odot}$. If this is not the case, central helium ignition occurs and the primary ends up as a carbon-oxygen white dwarf. Systems with $\log_{10}\left(a_{\mathrm{i}}/\mathrm{R}_{\odot}\right) = 2.42,\,2.27,\,2.28,\,2.12$ do not ignite helium in the centre, whereas the initially wider systems do. We find reverse mass transfer in all cases once the secondary star becomes an RG and fills its Roche lobe. Given that the primary stars are WDs, these systems might be observable as symbiotic binaries. 

Case-Bl systems with $q_{\mathrm{i}} = 0.35\text{--}0.45$ and $\log_{10}\left(a_{\mathrm{i}}/\mathrm{R}_{\odot}\right) =  2.38\text{--}2.56$ go through a phase of runaway mass transfer and experience $\mathrm{L}_{2}$-overflow. Since the primary star is an RG with a deep convective envelope and the secondary star an MS star at the onset of mass transfer, we expect a classical CE phase. Initially slightly wider binary systems ($q_{\mathrm{i}} = 0.35\text{--}0.65$ and $\log_{10}\left(a_{\mathrm{i}}/\mathrm{R}_{\odot}\right) \approx 2.60$) avoid runaway Case-Bl mass transfer. After core-He exhaustion, Case-BC mass transfer starts and the model quickly fails to converge numerically. While there is no clear indication for future runaway mass transfer, we cannot rule it out. Binary systems with $q_{\mathrm{i}} = 0.45\text{--}0.65$ and $\log_{10}\left(a_{\mathrm{i}}/\mathrm{R}_{\odot}\right) =  2.39\text{--}2.60$ only go through Case-Bl mass transfer. The models stop when the primary stars evolve into a pure helium WD, and the secondaries climb the giant branch.\\

Finally, we find that in the initially widest Case-C systems, contact phases are avoided. Here, mass transfer is stable because the primary star is in its TP-AGB phase, during which it experiences enhanced mass loss in our models (Sect.\,\ref{sec:wind_mass_loss}). Moreover, the star's radius periodically decreases again, preventing a runaway situation. Models that failed to converge numerically experience Case-C mass transfer before or early on during the TP-AGB phase of the primary. Their outcomes are uncertain.

\subsection{Stable Case-Bl and -C mass transfer}\label{sec:stable_caseC}

\begin{figure*}
\centering
  \includegraphics[width=18cm]{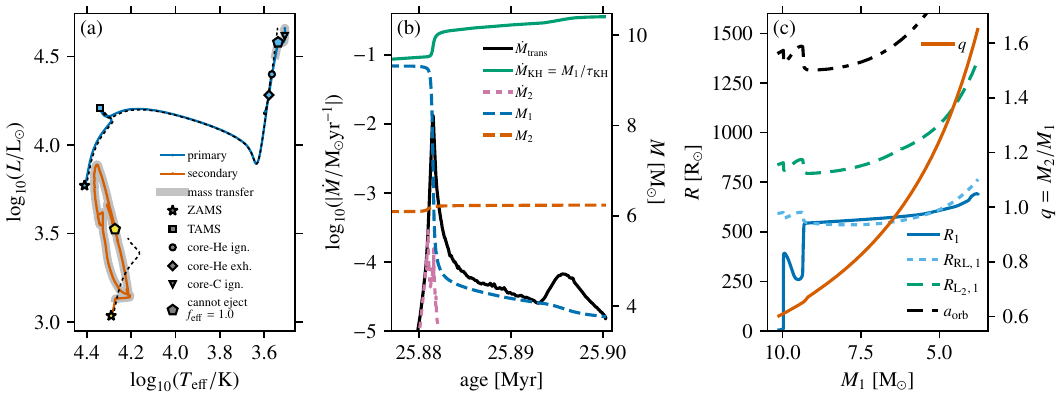}
     \caption{Example of a $M_{1,\,\mathrm{i}}=10.16\,\mathrm{M}_{\odot}$, $q_{\mathrm{i}} = 0.6$ and $a_{\mathrm{i}} = 1398.2\,\mathrm{R}_{\odot}$ binary model with stable Case-C mass transfer. Panel (a) is the same as Fig.\,\ref{fig:example_thermalExp}a (``ign.'' = ``ignition''). Panel (b) is the same as Fig.\,\ref{fig:example_runawayMT}b. Panel (c) is the same as Fig.\,\ref{fig:example_runawayMT}c, with the addition of a solid orange line indicating the evolution of the mass ratio.}
     \label{fig:example_stableCaseC}
\end{figure*}

For all initial primary masses considered in this work, Case-C mass transfer has been found to be stable over a wide range of initial mass ratios, even down to $q_{\mathrm{i}} \approx 0.2$ in some cases (e.g. for $M_{1,\,\mathrm{i}}=15.5\,\mathrm{M}_{\odot}$, see Fig.\,\ref{fig:appendix_contact_tracing1}). This also applies to the initially widest Case-Bl systems, but the exact extent of this region in the initial binary parameter space is unknown because of the aforementioned numerical difficulties.\\

The response of the donor star's radius to mass loss has been studied using models with polytropic equations of state in \citet{Hjellming1987} and detailed adiabatic mass loss computations such as in \citet{Ge2010,Ge2015,Ge2020}. It is found that donors with radiative or convective envelopes with core-mass fractions\footnote{The core-mass fraction is defined as the mass of the core over the total mass of the star.} greater than 0.5 shrink and hence have stable mass transfer \citep{Temmink2023}\footnote{As discussed in \citet{Temmink2023}, a major caveat of this simplified picture is the assumption that the response of the donor star is fully adiabatic, as assumed in, for example, \citet{Ge2010,Ge2015,Ge2020}. They show that even in giants with convective envelopes, the subsurface layers, right below the layers stripped by mass transfer, thermally readjust on timescales shorter than the dynamical timescale. Considering the \emph{local} thermal timescale of these subsurface layers, they derive a maximum mass-loss rate for the donor for which it can thermally readjust and avoid the (unstable) adiabatic response.}. Donors with convective envelopes and core-mass fractions below 0.5 typically expand in response to mass loss and cause unstable mass transfer.\\

In our models, we identify two stabilising effects during Case-Bl and -C mass transfer. First, the primary star's envelope is partially lost due to winds prior to mass transfer. This increases the core-mass fraction. At core helium ignition, values of the core-mass fraction in our stable Case-C models are 0.22, 0.15, 0.19, 0.26 and 0.31 for $M_{1,\,\mathrm{i}} = 1.9,\,4.4,\,8.6,\,14.3$ and $20.0\,\mathrm{M}_{\odot}$, respectively. At core helium exhaustion, the values for the core-mass fraction are 0.26, 0.22, 0.28, 0.36 and 0.44, respectively. Only for the most massive primary stars with $M_{\mathrm{1,\,i}}=20.0\,\mathrm{M}_{\odot}$ at core helium exhaustion, the core-mass fraction approaches the stabilising value of 0.5. Hence, the increased core-mass fraction alone is insufficient to explain the mass transfer stability. However, stellar wind mass loss increases the mass ratio $q$ towards unity before the onset of mass transfer. With $q$'s approaching or exceeding one, orbits will (soon) widen, stabilising mass transfer.

Second, orbital widening can stabilise mass transfer. Because the mass transfer in these systems is non-conservative, it results in less orbital shrinkage $\dot{a}/a$ than in the conservative case. Furthermore, this also causes orbital widening already at mass ratios $q < 1$, whereas this only occurs for $q \geq 1$ in conservative mass transfer \citep{Tauris2006}. How orbital widening stabilises mass transfer can be seen in the example of a $M_{1,\,\mathrm{i}}=10.2\,\mathrm{M}_{\odot}$, $q_{\mathrm{i}} = 0.6$ and $a_{\mathrm{i}} = 1398.2\,\mathrm{R}_{\odot}$ system undergoing stable Case-C mass transfer in Fig.\,\ref{fig:example_stableCaseC}. The orbital separation drops sharply before the onset of mass transfer because of an enhanced spin-orbit coupling when the primary becomes a supergiant. After the primary loses about $0.1\,\mathrm{M}_{\odot}$, the mass-transfer efficiency $\beta$ becomes zero when the secondary star reaches critical rotation. The orbit starts to widen when the mass ratio reaches a value of $0.67$ and the primary star's Roche lobe radius stays nearly constant, preventing a situation of runaway mass transfer as described in Sect.\,\ref{sec:unstable_MT}. Further on, the primary star's Roche lobe radius increases faster than the star's radius, such that $R_{1} < R_{\mathrm{RL,\,1}}$ at $q=1.21$. At $M_{1} = 7.2\,\mathrm{M}_{\odot}$ ($q=0.87$, age $=25.8815\,$Myr) the mass-transfer rate reaches its peak, after which it decreases sharply. This model ends its evolution when the primary reaches core-C exhaustion, which happens during mass transfer as in most of our stable Case-C models.

\subsection{Population properties}\label{sec:population}

\begin{figure*}
\centering
  \includegraphics[width=18cm]{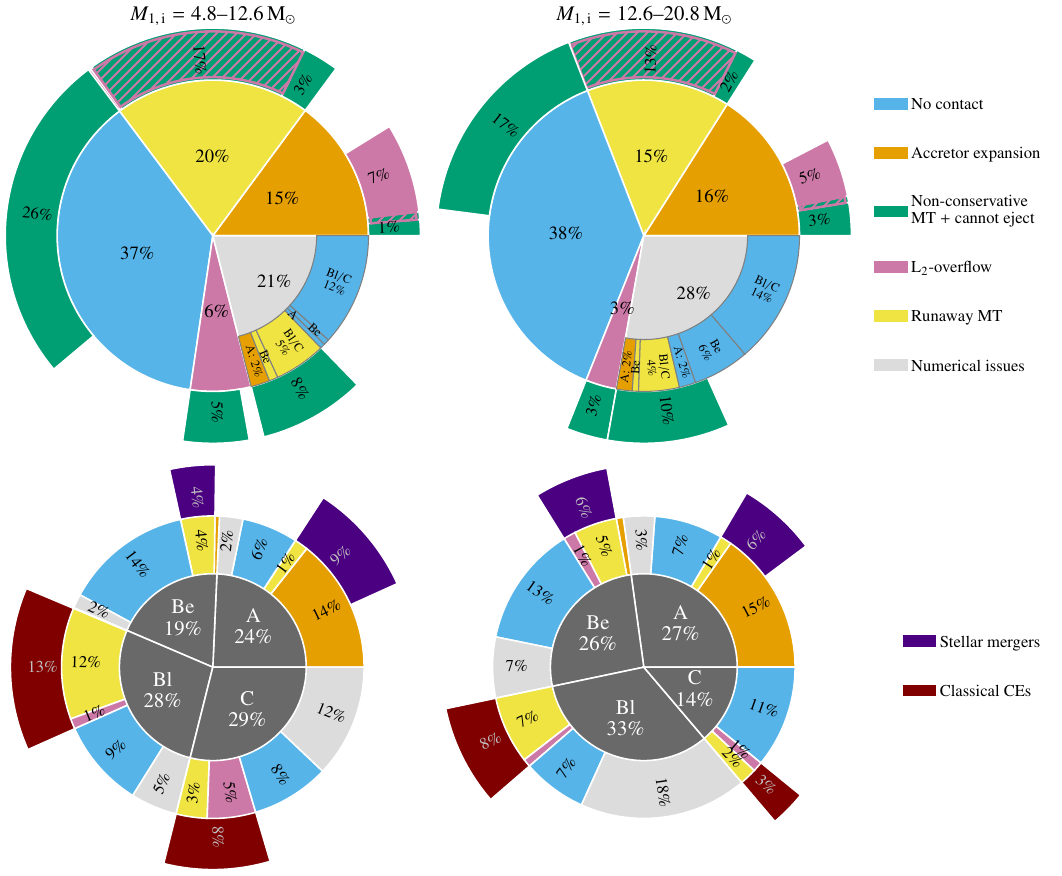}
     \caption{Sunburst charts displaying the fractions of evolutionary outcomes for mass-transferring binary systems in the grid over initial primary mass ranges of $\left[4.8;\,12.6\right)\,\mathrm{M}_{\odot}$ (left) and $\left[12.6;\,20.8\right]\,\mathrm{M}_{\odot}$ (right). Wedges with a percentage ${<}\,1\%$ are not labelled. \emph{(Top row)} The inner level shows the principal outcome of the evolution (``Accretor expansion'', ``Runaway MT'', ``$\mathrm{L}_{2}$-overflow'' and ``No contact''), while the outer level shows the ancillary outcome (``$\mathrm{L}_{2}$-overflow'', ``Non-conservative MT + cannot eject''). Ancillary ``No contact'' outcomes are incorporated in the inner ``No contact'' category. Models in the ``Numerical issues'' category are assigned a likely evolutionary outcome based on their initial mass ratio $q_{\mathrm{i}}$ and mass-transfer case, displayed on the sunburst chart's intermediate level. The labels ``A'', ``Be'' and ``Bl/C'' refer to Case-A, Case-Be and Case-Bl or -C mass transfer, respectively. \emph{(Bottom row)} The inner level shows the percentage of Case-A, -Be, -Bl and -C systems. We show the principal outcome per case on the middle level. The outer level shows the lower limits of the total fraction of stellar mergers and classical CEs per mass-transfer case.}
     \label{fig:pies}
\end{figure*}

\begin{figure*}
\centering
  \includegraphics[width=18cm]{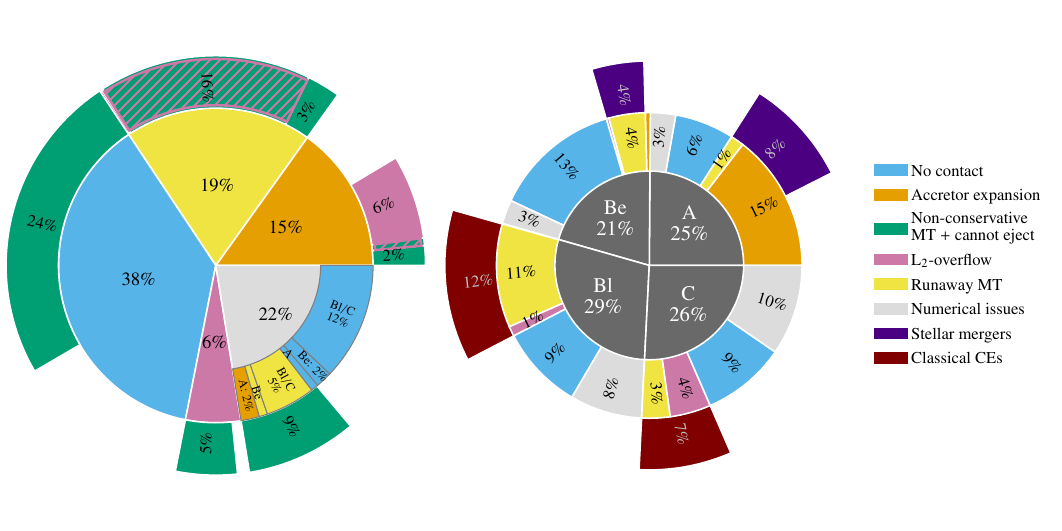}
     \caption{Sunburst charts displaying the fractions of evolutionary outcomes for mass-transferring binary systems in the grid over initial primary mass ranges of $\left[4.8;\,20.8\right]\,\mathrm{M}_{\odot}$. The left and right charts are equivalent to those in the top and bottom row of Fig.\,\ref{fig:pies}, respectively.}
     \label{fig:pies_allmasses}
\end{figure*}

To understand what fraction of systems avoid or end up in contact and through which mechanism, we count the systems in each category (``No contact'', ``Accretor expansion'' etc.) and weigh them with their birth probability $p_{\mathrm{birth}}$ (see Sect.\,\ref{sec:method_probs}). We show the results for initial primary mass ranges of $\left[4.8;\,12.6\right)\,\mathrm{M}_{\odot}$ and $\left[12.6;\,20.8\right]\,\mathrm{M}_{\odot}$ in so-called sunburst charts in Fig.\,\ref{fig:pies}. Fig.\,\ref{fig:pies_allmasses} shows the results for the whole mass range of $M_{1,\,\mathrm{i}}\,\in \,\left[4.8; 20.8\right]\,\mathrm{M}_{\odot}$\footnote{To compute $p_{\mathrm{birth}}$, the initial mass function is integrated from $M_{\mathrm{l}}$ to $M_{\mathrm{u}}$ (see Sect.\,\ref{sec:method_probs}). For the models with $M_{\mathrm{1,\,i}} = 20.0\,\mathrm{M}_{\odot}$, $M_{\mathrm{u}} = 20.8\,\mathrm{M}_{\odot}$. This explains why the initial primary mass range extends to $20.8\,\mathrm{M}_{\odot}$.}. We only consider binaries with $M_{1,\,\mathrm{i}} \geq 4.8\,\mathrm{M}_{\odot}$ because models with lower masses have not been computed for the entire $q_{\mathrm{i}}=0.1\text{--}0.97$ range (Sect.\,\ref{subsec:binary_grid}). The sunburst charts in the top row show the incidences of the physical mechanisms leading to contact for mass-transferring binaries. The inner level of these charts shows the fraction of all mass-transferring binary systems in the specified mass range that end their evolution with the indicated outcome. In other words, they represent the principal outcomes, ``Accretor expansion'', ``Runaway MT'', ``$\mathrm{L}_{2}$-overflow'' and ``No contact'' (see Sect.\,\ref{sec:contact_1016}), which correspond to the outcomes indicated by the background colours in Figs.\,\ref{fig:contact_Mp1016}--\ref{fig:contact_Mp20.0-1.58}. Models expected to avoid contact but which encountered numerical issues before the primaries reached core-C exhaustion (``No contact'' hatching in Figs.\,\ref{fig:contact_Mp1016}--\ref{fig:contact_Mp20.0-1.58}) are included in the principal ``No contact'' outcomes. The outer level of the sunburst charts shows the incidence of ancillary outcomes, that is, the hatching in Figs.\,\ref{fig:contact_Mp1016}--\ref{fig:contact_Mp20.0-1.58} (``$\mathrm{L}_{2}$-overflow'' and ``Non-conservative MT + cannot eject''). We assign systems that experienced numerical issues a likely evolutionary outcome based on their initial mass ratio and first mass-transfer case. For more information, we refer to Appendix \ref{app:assignments}. These outcomes are shown at the intermediate level of the charts in the ``Numerical issues'' slice. On the bottom row of Fig.\,\ref{fig:pies}, we show the incidences of the principal outcomes per initial mass transfer case. The outer level of these charts shows the lower limits of the incidences for stellar mergers and classical CEs (see Sect.\,\ref{sec:merger_CE_frac}).\\

We find that the fractions for $M_{1,\,\mathrm{i}}\,\in \,\left[4.8; 12.6\right)\,\mathrm{M}_{\odot}$ and $M_{1,\,\mathrm{i}}\,\in \,\left[12.6; 20.8\right]\,\mathrm{M}_{\odot}$ are relatively similar (Fig.\,\ref{fig:pies}). The most noticeable differences are found for the systems going through a phase of runaway mass transfer (``Runaway MT'') and systems failing to eject non-accreted matter (``Non-conservative MT + cannot eject''). The lower fraction of systems that simultaneously avoid contact and fail to eject non-accreted matter for higher primary masses can be traced back to the mass-luminosity relation. In general, $L\sim M^{\alpha}$, where the exponent $\alpha > 1$\, \citep{Kippenhahn2012}. So, even though the gravitational potential increases linearly with increasing mass, the luminosity increase is steeper since $\alpha > 1$. Hence, it is easier for higher mass systems to expel non-accreted matter.

The fraction of systems with runaway mass transfer in the lower mass range is $5\%$ higher than in the higher mass range. At the same time, the fraction of systems with numerical issues is $7\%$ higher in the higher mass range. However, the total fraction of systems with numerical issues that have runaway mass transfer as their most likely outcome is $4\text{--}6\%$ in both mass ranges. The higher fraction of systems with runaway mass transfer in the lower mass range is thus not caused by increased numerical difficulties at higher masses but is physical.\\

Excluding the systems with numerical issues, we find that ${\geq}\,41\%$ of binaries with $M_{1,\,\mathrm{i}}\,\in \,\left[4.8; 12.6\right)\,\mathrm{M}_{\odot}$ and ${\geq}\,34\%$ of binaries with $M_{1,\,\mathrm{i}}\,\in \,\left[12.6; 20.8\right]\,\mathrm{M}_{\odot}$ enter a contact phase (Fig.\,\ref{fig:pies}). Over the whole initial primary mass range of $M_{1,\,\mathrm{i}}\,\in \,\left[4.8; 20.8\right]\,\mathrm{M}_{\odot}$, the percentage of binaries entering a contact phase is ${\geq}\,40\%$ (Fig.\,\ref{fig:pies_allmasses}). These are lower limits because, in addition to excluding the systems with numerical issues, we do not take into account the binaries that avoid contact in our models but fail to eject non-accreted matter.

\subsection{Stellar merger and classical CE incidence}\label{sec:merger_CE_frac}

We assume that all binaries experiencing runaway mass transfer and/or $\mathrm{L}_{2}$-overflow merge or enter a classical CE phase. Following the physical picture described in Sect.\,\ref{sec:mechanisms}, we make the distinction based on the structure of the primary star. This means that Case-A and -Be binaries with runaway mass transfer and/or $\mathrm{L}_{2}$-overflow lead to stellar mergers, and Case-Bl and -C binaries to classical CEs. Based on this, we compute lower limits on the incidences of stellar mergers and classical CEs, shown in the bottom row of Fig.\,\ref{fig:pies} for initial primary mass ranges of $\left[4.8;\,12.6\right)\,\mathrm{M}_{\odot}$ and $\left[12.6;\,20.8\right]\,\mathrm{M}_{\odot}$, and for each initial primary mass in Table \ref{tab:fractions}. This table also lists the critical mass ratios $q_{\mathrm{crit}}$ for which binaries with $q_{\mathrm{i}} < q_{\mathrm{crit}}$ merge or enter classical CE phases. The stellar merger and classical CE incidences are ${\geq}\,13\%$ and ${\geq}\,21\%$, respectively, for the primary mass range of $\left[4.8;\,12.6\right)\,\mathrm{M}_{\odot}$, and ${\geq}\,12\%$ and ${\geq}\,11\%$, respectively, for the primary mass range of $\left[12.6;\,20.8\right]\,\mathrm{M}_{\odot}$. Figure \ref{fig:pies_allmasses} shows similar charts for the total initial primary mass range of $\left[4.8;\,20.8\right]\,\mathrm{M}_{\odot}$, and we find that ${\geq}\,12\%$ of mass-transferring binaries merge and ${\geq}\,19\%$ evolve towards a classical CE phase. The stellar merger incidence for Case-A binaries of $8\%$ (Fig.\,\ref{fig:pies_allmasses}) is similar to the incidence of massive stars with strong, large-scale magnetic fields of ${\sim}\,10\%$ \citep{Donati2009, Fossati2015, Grunhut2017}, which are likely formed by (pre-)MS stellar mergers \citep{Schneider2019}. 

The incidences reported in Figs.\,\ref{fig:pies}--\ref{fig:pies_allmasses} and Table \ref{tab:fractions} are lower limits since the criteria given here do not take into account binary systems that merge as a result of a classical CE phase, binary systems that experience $\mathrm{L}_{2}$-overflow in later contact phases (so far if more than one contact phase occurred, we only considered the first one, see Sect.\,\ref{subsec:accr_exp}), and potential mergers among the models that had numerical issues. Furthermore, a certain fraction of models avoiding contact but failing to eject non-accreted matter (``Non-conservative MT + cannot eject'') might in reality also merge (see discussion in Sect.\,\ref{sec:non_ejected_fate}). If we now assume that \emph{all} contact binaries formed through the expansion of the accretor eventually merge (except those which survive contact and reach core-C exhaustion in either component), we find stellar merger incidences of ${\geq}\,18\%$, ${\geq}\,20\%$ and ${\geq}\,19\%$ for initial primary masses of $\left[4.8; 12.6\right)\,\mathrm{M}_{\odot}$, $\left[12.6; 20.8\right]\,\mathrm{M}_{\odot}$ and $\left[4.8; 20.8\right]\,\mathrm{M}_{\odot}$, respectively. For comparison, the merger incidence for O-type stars ($M_{1} \gtrsim 15\,\mathrm{M}_{\odot}$) found by interpreting observations in the context of binary evolution and reported in \citet{Sana2012} is $20\text{--}30\%$.

We find that the incidences of stellar mergers of ${\sim}\,12\%$ are similar for all initial primary masses (bottom row of Fig.\,\ref{fig:pies} and Table \ref{tab:fractions}). The classical CE incidence varies more significantly, from $12\%$ at $M_{1,\,\mathrm{i}} = 5.2\,\mathrm{M}_{\odot}$, to $33\%$ at $M_{1,\,\mathrm{i}} = 10.2\,\mathrm{M}_{\odot}$ and $9\%$ at $M_{1,\,\mathrm{i}} = 20.0\,\mathrm{M}_{\odot}$. The decrease in classical CE incidence from $10.2\,\mathrm{M}_{\odot}$ to $20.0\,\mathrm{M}_{\odot}$ is linked to the decrease in Case-C systems (bottom row of Fig.\,\ref{fig:pies}). Our values of $q_{\mathrm{crit}}=0.15\text{--}0.35$, correspond reasonably well with those for HG star donors from \citet{Temmink2023}.

\begin{table}[]
\caption{Stellar merger and classical CE incidences, and critical mass ratios $q_{\mathrm{i,\,crit}}$ for mass-transferring binaries with $M_{1,\,\mathrm{i}} \in [4.8;\,20.8]\,\mathrm{M}_{\odot}$. For each initial primary mass, the incidence fractions are given per mass-transfer case.}
\label{tab:fractions}
\resizebox{0.65\columnwidth}{!}{\begin{tabular}{c|c|c|c}
\hline \hline
$\boldsymbol{M_{1,\,\mathrm{i}}}$  & \textbf{Mergers}  & \textbf{Class. CEs}  & $\boldsymbol{q_{\mathrm{i,\,crit}}}$\tablefootmark{a} \\ 
$\mathbf{[\mathrm{M}_{\odot}]}$ & [$\%$] & [$\%$] & \\ \hline
\textbf{5.2} & \textbf{12.8} & \textbf{11.9} &  \\
A & 10.2 & - & 0.35 \\
Be & 2.6 & - & 0.35 \\
Bl & - & 4.3 & 0.35--0.55 \\
C & - & 7.5 & 0.55 \\ \hline
\textbf{6.1} & \textbf{14.8} & \textbf{16.1} &  \\
A & 12.0 & - & 0.35 \\
Be & 2.9 & - & 0.35 \\
Bl & - & 9.1 & 0.25--0.45 \\
C & - & 6.9 & 0.55 \\ \hline
\textbf{7.2} & \textbf{13.0} & \textbf{18.4} &  \\
A & 8.2 & - & 0.35 \\
Be & 4.8 & - & 0.15--0.25 \\
Bl & 0 & 12.3 & 0.35--0.45 \\
C & 0 & 6.1 & 0.25 \\ \hline
\textbf{8.6} & \textbf{11.3} & \textbf{31.2} &  \\
A & 7.3 & - & 0.35 \\
Be & 4.0 & - & 0.25--0.35 \\
Bl & - & 19.7 & 0.25--0.35 \\
C & - & 11.8 & 0.45 \\ \hline
\textbf{10.2} & \textbf{12.2} & \textbf{33.0} &  \\
A & 8.0 & - & 0.25--0.35 \\
Be & 4.2 & - & 0.25--0.35 \\
Bl & - & 19.9 & 0.25--0.35 \\
C & - & 13.0 & 0.45 \\ \hline
\textbf{12.0} & \textbf{11.3} & \textbf{23.1} &  \\
A & 7.0 & - & 0.35 \\
Be & 4.3 & - & 0.25--0.35 \\
Bl & - & 19.1 & 0.15--0.35 \\
C & - & 4.0 & 0.15--0.45 \\ \hline
\textbf{13.1} & \textbf{8.1} & \textbf{12.4} &  \\
A & 5.7 & - & 0.35 \\
Be & 2.4 & - & 0.25--0.35 \\
Bl & - & 8.7 & 0.35--0.45 \\
C & - & 3.7 & 0.15--0.35 \\ \hline
\textbf{14.2} & \textbf{11.4} & \textbf{16.4} &  \\
A & 5.6 & - & 0.35 \\
Be & 5.8 & - & 0.25--0.35 \\
Bl & - & 12.4 & 0.15--0.45 \\
C & - & 4.0 & 0.15--0.35 \\ \hline
\textbf{15.6} & \textbf{13.9} & \textbf{9.2} &  \\
A & 6.9 & - & 0.35 \\
Be & 7.0 & - & 0.25--0.35 \\
Bl & - & 8.2 & 0.15--0.35 \\
C & - & 0.9 & 0.15 \\ \hline
\textbf{16.9} & \textbf{14.3} & \textbf{8.7} &  \\
A & 6.8 & - & 0.35 \\
Be & 7.6 & - & 0.15--0.35 \\
Bl & - & 7.7 & 0.15--0.35 \\
C & - & 1.0 & 0.15 \\ \hline
\textbf{18.4} & \textbf{14.0} & \textbf{8.5} & \\
A & 6.5 & - & 0.35 \\
Be & 7.5 & - & 0.15--0.35 \\
Bl & - & 5.8 & 0.15--0.35 \\
C & - & 2.7 & 0.15--0.25 \\ \hline
\textbf{20.0} & \textbf{13.6} & \textbf{8.6} &  \\
A & 6.8 & - & 0.35 \\
Be & 6.8 & - & 0.15--0.35 \\
Bl & - & 4.9 & 0.15 \\
C & - & 8.6 & 0.25--0.35\\ \hline
\end{tabular}}
\\ \tablefoottext{a}{Case~A and -Be systems with $q_{\mathrm{i}} < q_{\mathrm{i,\,crit}}$ form contact binaries through runaway mass transfer and/or $\mathrm{L}_{2}$-overflow, and merge. Case-Bl and -C systems form classical CEs through runaway mass transfer and/or $\mathrm{L}_{2}$-overflow.}
\end{table}

\subsection{Comparison with observed contact binaries}\label{sec:comparison}

\begin{figure*}
\sidecaption
  \includegraphics[width=12cm]{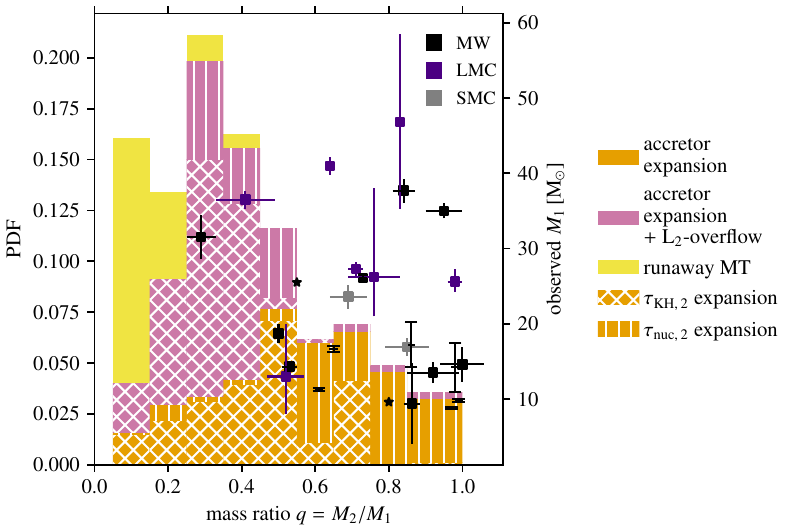}
     \caption{Probability density function (PDF) of contact systems formed in Case-A binaries with $M_{1,\,\mathrm{i}}=7.9\text{--}20.8\,\mathrm{M}_{\odot}$ as a function of the mass ratio $q$ at the onset of contact. Data points (filled squares) show the observed mass ratio and primary mass (right axis) for all observed MW, LMC and SMC contact and near-contact systems compiled in \citet{Menon2021}, including the uncertainties on their values. Systems without reported uncertainties on $q$ are indicated with a dash symbol, and those without reported uncertainties on $q$ and $M_{1}$ are indicated with a star symbol.}
     \label{fig:observations}
\end{figure*}

In Fig.\,\ref{fig:observations}, our simulated population of Case-A contact binaries with initial primary masses of $4.8\text{--}20.8\,\mathrm{M}_{\odot}$ is compared to observed near-contact\footnote{Near-contact systems are systems in which for both components $R/R_{\mathrm{RL}} \geq 0.9$ \citep{Menon2021}.} and contact systems in the Milky Way (MW), Large Magellanic Cloud (LMC) and Small Magellanic Cloud (SMC) from the compilation in \citet{Menon2021}. The sample consists of MS O+O, B+O and B+B massive contact systems, which is why we have chosen to compare only to Case-A binaries from our grid with initial primary masses ${\geq}\,4.8\,\mathrm{M}_{\odot}$ (this is as low as we can go in terms of the initial primary mass while still having models at all $q_{\mathrm{i}}$). The mass ratios of the observed systems are compared to the probability distribution function (PDF) of the models as a function of the mass ratio $q$ at contact. For models that enter a contact phase through the expansion of the accretor, contact is defined as the moment when both the primary and the secondary overfill their respective Roche lobes. When runaway mass transfer is responsible for the onset of a contact phase, the moment at which $\dot{M}_{\mathrm{trans}} > \dot{M}_{\mathrm{KH,\,1}}$ is taken as the moment of contact. To have a meaningful comparison with observed contact systems, the mass ratios at contact of the observed systems are defined here as the mass of the \emph{currently} less massive component over the mass of the currently more massive one (whereas before $q$ has always been defined as the mass of the \emph{initially} less massive component over the mass of the initially more massive one). The probabilities used to construct the PDF are the birth probabilities $p_{\mathrm{birth}}$ computed via Eq.\,(\ref{eq:pbirth}) in Sect.\,\ref{sec:method_probs}. We highlight the following contributions to the PDF: ``accretor expansion'' systems are those that enter contact because of accretor expansion but do not experience $\mathrm{L}_{2}$-overflow, while the ``accretor expansion + $\mathrm{L}_{2}$-overflow'' do. The third contribution comes from runaway mass transfer systems (``runaway MT''). The ``accretor expansion'' and ``accretor expansion + $\mathrm{L}_{2}$-overflow'' systems are further divided into systems where the accretor expands on a nuclear and a thermal timescale prior to filling its Roche lobe. This division is made by comparing the mean of the expansion timescale $\tau_{R/\dot{R}}$ from the onset of mass transfer to the formation of a contact system with the mean of the thermal and nuclear timescales respectively (see Appendix \ref{app:exp_ts}).\\

There is a striking difference between the PDF from the model systems and the observations. From the models, we expect two to three times more contact systems at $q < 0.5$ than at $q \geq 0.5$. Contrarily, observations show a dearth of contact binaries at $q < 0.5$. There are only a few models at $q \geq 0.5$ experiencing $\mathrm{L}_2$-overflow, while such systems are dominant at $q < 0.5$. For $q < 0.45$, there is an additional contribution of contact systems formed through runaway mass transfer. Among the contact binaries that do not experience $\mathrm{L}_{2}$-overflow or runaway mass transfer (``accretor expansion''), it can be seen that at $q < 0.5$ contact is formed mostly on a thermal timescale, while for $q \geq 0.5$ contact is formed on the longer, nuclear timescale. Overall, $29\%$ of the Case-A contact binaries in the mass range considered here form through the nuclear timescale expansion of the accretor. $59\%$ of those ($17\%$ of the total) do not yet experience $\mathrm{L}_{2}$-overflow and are expected to be longer-lived.\\

We find a number of ways in which this (apparent) discrepancy can be resolved. Firstly, it should be noted that the sample of observed systems is small, especially if one disregards the near-contact systems, which have been considered in this comparison. The absence of observed systems at mass ratios $q < 0.5$ might be because they have simply not been observed yet. As argued in \citet{AbdulMasih2022a}, a larger sample size, which requires a dedicated search for these massive contact binary systems, should shed light on whether this is the case.

A second reason for the discrepancy could be that contact systems rapidly evolve to equal-mass systems, that is, to $q=1$, and are therefore not observed as unequal-mass systems. However, following the same argument as in \citet{AbdulMasih2022a}, the observed systems are uniformly distributed for $q \geq 0.5$. Should contact systems all equalise in mass, the bulk of them would be expected to be observed at $q \approx 1$. Furthermore, they find that based on the observed stability of the orbit, the systems in their sample will continue to evolve as unequal-mass contact binaries on a nuclear timescale.

Based on the different contributions to the PDF in Fig.\,\ref{fig:observations}, a third reason for the lack of contact systems at $q < 0.5$ can be found. The largest contribution to the PDF at these mass ratios comes from contact binaries that experience $\mathrm{L}_{2}$-overflow. As described in Sect.\,\ref{sec:l2over}, this can lead to considerable orbital angular momentum loss and thereby likely stellar mergers. Case-A contact binaries formed through runaway mass transfer, which we find at $q < 0.45$, also lead to mergers (see Sect.\,\ref{sec:unstable_MT}). In other words, contact systems are almost exclusively observed at $q \geq 0.5$ because those at $q < 0.5$ merge shortly after they get into contact. 

Considering that binaries with $\mathrm{L}_{2}$-overflow or runaway mass transfer merge quickly, we are left with a PDF that at $q < 0.5$ is dominated by contact systems formed by the thermal-timescale expansion of the accretor. As explained in Sect.\,\ref{subsec:accr_exp}, such contact binaries either detach or merge quickly. In the former case, these systems would be observed as semi-detached binaries. In the latter case, we argue that if the thermal expansion of the accretor continues during contact, it eventually leads to $\mathrm{L}_{2}$-overflow and a merger. \citet{Ge2023} find that the critical mass ratio $q_{\mathrm{crit}}$ below which binaries experience runaway mass transfer increases with decreasing metallicities. Therefore, at lower metallicities, we expect an even more significant fraction of binaries to experience runaway mass transfer and a subsequent quick merger.  

In conclusion, we find that contact systems are almost exclusively observed at $q \geq 0.5$ because those at $q < 0.5$ merge or detach shortly after they get into contact.

\section{Discussion}\label{sec:discussion}
In this section, we put into perspective the influence of mass-transfer efficiency on our contact tracing results (Sect.\,\ref{sec:discussion_efficiency}), the occurrence of stable mass transfer from (super-)giant donors (Sect.\,\ref{sec:discussion_stable}), and the potential fates of binaries with non-ejected matter (Sect.\,\ref{sec:non_ejected_fate}).

\subsection{The role of mass-transfer efficiency}\label{sec:discussion_efficiency}
The mean mass-transfer efficiencies $\bar{\beta}$ of binaries in our grid (Fig.\,\ref{fig:mt_efficiency} in Appendix \ref{app:mt_eff}) are relatively low for Case-B and -C mass transfer. In these systems, the secondary usually reaches critical rotation after accreting $M_{\mathrm{accreted}}/M_{2,\,\mathrm{i}} <\,3\%$ of its initial mass (with $M_{\mathrm{accreted}}$ the accreted mass), which is lower than the ${\sim}5\text{--}10\%$ from \citet{Packet1981} and consistent with the ${\sim}2\%$ found by \citet{Ghodla2023}. It should be noted that only a few percent of the secondary's outer envelope in terms of its mass is rotating near the critical rotation rate. With more efficient angular momentum transport, the relative mass accretion fraction would be higher than $3\%$ because of a delay of critical surface rotation. 

Because the secondary stars in our models accrete less matter than those in more conservative models, they also expand less. Compared to models with higher mass-transfer efficiencies, such as those from \citet{Pols1994}, \citet{Wellstein2001}, \citet{deMink2007}, \citet{Claeys2011} and \citet{Menon2021}, we find fewer contact systems from accretor expansion. The difference is particularly noticeable for Case-B binaries, where virtually none of such systems are found in our models. Our results generally agree better with the model grids from \citet{Fragos2023}. This is expected since similar assumptions with regard to mass-transfer efficiency are made in their models.\\

From observations of Be X-ray binaries in the SMC, \citet{Vinciguerra2020} infer a mass transfer efficiency of ${\sim}30\%$ for the progenitor systems (i.e. the systems before the primary has formed a compact object). Potential progenitors of Be X-ray binaries in our grid, those with post-CHeB (core helium burning) and MS components (see Figs.\,\ref{fig:contact_configs}--\ref{fig:contact_configs3} in Appendix \ref{app:contact_configs}), have a wide range of mass-transfer efficiencies (see Fig.\,\ref{fig:mt_efficiency}). For Case-A binaries with $M_{1,\,\mathrm{i}} = 6.1\text{--}8.6\,\mathrm{M}_{\odot}$ and Case~C binaries with $M_{1,\,\mathrm{i}} = 5.2\text{--}14.2\,\mathrm{M}_{\odot}$ the mass-mass transfer efficiency is consistent with the observationally constrained value of ${\sim}30\%$.

Using observations of the classical Algol system $\delta$~Librae, \citet{Dervisoglu2018} found a mass-transfer efficiency of ${\sim}100\%$. Based on the component masses derived for $\delta$~Librae, the initial primary mass would have been between ${\approx}\,2.6\,\mathrm{M}_{\odot}$ ($q_{i} = 1.0$) and ${\approx}\,4.1\,\mathrm{M}_{\odot}$ ($q_{\mathrm{i}}=0.25$, stable mass transfer). Our Case-A mass-transfer efficiency for $M_{1,\,\mathrm{i}} \approx 2.6\,\mathrm{M}_{\odot}$ is ${\sim}\,95\%$, which is consistent with the observationally constrained value. For $M_{1,\,\mathrm{i}} \approx 4.1\,\mathrm{M}_{\odot}$, we find a Case-A mass-transfer efficiency of ${\sim}\,25\%$, which is inconsistent with the value inferred for $\delta$~Librae.\\

\citet{Sen2022} computed a grid of Case-A systems with $M_{1,\,\mathrm{i}} = 10\text{--}40\,\mathrm{M}_{\odot}$ at LMC metallicity ($Z = 0.0047$) and similar assumptions as ours with regards to rotation, tides and accretion of angular momentum. For thermal-timescale mass-transfer phases in systems avoiding contact on the MS, they find relatively low (time-averaged) mass-transfer efficiencies (peak around ${\sim}\,0.06$). During the slower, nuclear-timescale mass transfer phases, the mass-transfer efficiency peaks around ${\sim}\,0.94$. This is consistent with our models, in which the rapid spin-up of the accretor during thermal-timescale mass transfer results in non-conservative mass transfer. For $M_{1,\,\mathrm{i}} > 10\,\mathrm{M}_{\odot}$, the time-averaged mass-transfer efficiencies in our models range from ${\sim}\,0.4$ to ${\sim}\,0.65$ (Fig.\,\ref{fig:mt_efficiency}). The thermal timescale is several orders of magnitude shorter than the nuclear timescale (Eq.\,\ref{eq:thermal_timescale} and \ref{eq:nuclear_timescale}). From estimating the time-averaged mass-transfer efficiencies over the whole Case-A phase (thermal- and nuclear-timescale mass-transfer phases) in the models of \citet{Sen2022}, we see that their values of the mass-transfer efficiency are closer to unity. The slightly different definitions of the mass-transfer efficiency and the different metallicities (and, therefore, wind mass-loss rates) between our grids likely explain this difference.\\

Our models' current assumption that stars cannot accrete anymore once they are at breakup velocities might be up for debate. Interactions with an accretion disk have been proposed to spin down the surface of the accretor and allow for additional accretion of material \citep{Paczynski1991, Popham1991}. This contributes to the fact that the incidences of contact systems reported in Sect.\,\ref{sec:merger_CE_frac} are lower limits because more accretion allows the accretor to expand more and potentially fill its Roche lobe. This could in turn increase the stellar merger incidences.

\subsection{The stability of Case-B and -C mass transfer}\label{sec:discussion_stable}
In Sect.\,\ref{sec:stable_caseC} we have described how non-conservative mass transfer and stellar wind mass loss from the donor prior to mass transfer can lead to stable mass transfer from (super-)giant donors (Case~Bl and C) with initial mass ratios down to 0.1. Following the discussion in Sect.\,\ref{sec:discussion_efficiency}, these outcomes might also change with different assumptions regarding the mass-transfer efficiency. Just as for the formation of contact binaries, an increase in the mass-transfer efficiency leads to a higher incidence of classical CEs.

The presence of stable Case-Bl and -C mass transfer in our models agrees with what was found by \citet{Chen2008} for stars with $M_{1,\,\mathrm{i}} \lesssim 8\,\mathrm{M}_{\odot}$, who also note that the effect of wind mass loss prior to the onset of mass transfer only has a minor effect on the stability.

We find that our critical mass ratios $q_{\mathrm{i,\,crit}}$ for Case-Bl and -C mass transfer (Table \ref{tab:fractions}) are higher than those from the adiabatic mass loss computations of \citet{Ge2010} for $M_{1,\,\mathrm{i}} > 10\,\mathrm{M}_{\odot}$, while being in relatively good agreement for stars with lower initial primary masses. These differences are likely caused by the fact that contrary to \citet{Ge2010}, we also take the response of the accretor and orbit into account. 
Another difference is that \citet{Ge2010,Ge2015,Ge2020} derive the critical mass ratios under the assumption of fully conservative mass transfer. Our critical mass ratio ranges agree relatively well with similar simulations with fully non-conservative mass transfer (H.~Ge, 2023, priv. comm.).

\citet{Picco2023} use the adiabatic mass-radius exponents $\xi_{R,\,\mathrm{ad}}$ (see Sect.\,\ref{sec:unstable_MT}) from \citet{Ge2020} to determine the stability of mass transfer in their detailed binary evolution models. Our critical mass ratios agree relatively well with the ones from their simulations of fully non-conservative mass transfer in binaries with $M_{\mathrm{1,\,i}}=8.0\,\mathrm{M}_{\odot}$.

The adiabatic mass-radius exponents $\xi_{R,\,\mathrm{ad}}$ are also used by \citet{Li2023} in their binary population synthesis computations. For binaries with $M_{\mathrm{1,\,i}} = 8.0\,\mathrm{M}_{\odot}$ with non-conservative ($\beta = 0\text{--}0.5$) Case-Bl and -C mass transfer, they find values for $q_{\mathrm{i,\,crit}} \approx 0.37\text{--}0.66$, which is in broad agreement with our values.\\ 

Similar stable Case-Bl and -C mass transfer has been found by \citet{Ercolino2023} for $M_{1,\,\mathrm{i}} = 12.6\,\mathrm{M}_{\odot}$. The reported critical mass ratio $q_{\mathrm{crit}} = 0.525\text{--}0.625$ is slightly higher than what is found in our models for stars with similar masses (Table \ref{tab:fractions}). This difference can be attributed to the different stability criteria for mass transfer used in their work.

\subsection{The fate of binaries with non-ejected matter}\label{sec:non_ejected_fate}
In the majority of our models with $\beta < 1$, the non-accreted mass cannot be ejected to infinity. This raises the question of where this excess material resides. Should the excess matter remain in the accretor's Roche lobe or the binary's $\mathrm{L}_{2}$-lobe, it can fill it up and lead to a situation similar to a contact binary. In this case, our grid would contain significantly more contact systems. For the orbital evolution, this scenario would essentially correspond to that of systems with higher mass-transfer efficiencies. Should the excess matter overfill the $\mathrm{L}_{2}$-lobe and hydrodynamic drag becomes significant, a classical CE could form. Alternatively, the matter could settle in a circumbinary disk or shell, which can significantly influence the further evolution of the binary \citep{Wei2023}.

Alternatively, a circumstellar disk may form, through which the secondary can potentially continue to accrete mass \citep{Paczynski1991, Popham1991}. In this scenario, systems might show signs of circumstellar disks such as H$\alpha$ excess and emission features in Be/Oe stars. This could explain the fast-rotating H$\alpha$ emitters in the extended MS turnoff observed in young clusters \citep{Milone2018}. Moreover, \citet{Bodensteiner2021} has found a significant close-binary fraction of $34^{+8}_{-7}\%$ for one of the clusters analysed in \citet{Milone2018}, NGC~330. 

\subsection{The onset of contact through runaway mass transfer and $\mathrm{L}_{2}$-overflow}
In Sect.\,\ref{subsec:res_contact_phases}, we find that runaway mass transfer and/or $\mathrm{L}_{2}$-overflow are responsible for the onset of contact for a significant fraction of the Case-A, -Be, -Bl, and -C binaries. In Case-A binaries, $\mathrm{L}_{2}$-overflow is of lesser importance for the onset of contact but leads to stellar mergers in contact binaries formed through the expansion of the accretor. Runaway mass transfer in Case-A systems leads to the formation of unstable contact binaries and subsequent stellar mergers. The same happens in Case-Be binaries, but now in unison with $\mathrm{L}_{2}$-overflow. The added effect of the orbital angular momentum loss associated with $\mathrm{L}_{2}$-overflow contributes to the instability of the contact binary formed through runaway mass transfer. In Case-Bl and -C binaries, the onset of runaway mass transfer and $\mathrm{L}_{2}$-overflow often occur quasi simultaneously. In these systems, the $\mathrm{L}_{2}$-overflow serves as an additional indication that the secondary is being engulfed by the rapidly expanding envelope of the (super-)giant primary star during runaway mass transfer. Lastly, some Case-Bl and C systems do not experience runaway mass transfer but do have primary stars that extend far beyond the $\mathrm{L}_{2}$-lobe. In both cases, we expect the onset of a classical CE phase.

\section{Summary and conclusions}\label{sec:conclusions}
Using a grid of ${\sim}\,6000$ detailed binary evolution models including rotation, tidal interactions, the evolution of both components, and with component masses between $0.5$ and $20.0\,\mathrm{M}_{\odot}$, we examine in which regions of the initial binary parameter space we expect contact phases, such as contact binaries and classical common-envelope (CE) phases, to occur. We identify five mechanisms that lead to contact: the expansion of the accretor, runaway mass transfer, $\mathrm{L}_{2}$-overflow, orbital decay because of tides, and non-conservative mass transfer.

We find that accretor expansion, $\mathrm{L}_{2}$-overflow, and runaway mass transfer lead to the formation of contact binaries in Case-A and -Be systems, and $\mathrm{L}_{2}$-overflow and runaway mass transfer to the onset of classical CEs in Case-Bl and -C systems. This distinction stems from the fact that primary stars in Case-Bl and Case-C systems have extended envelopes with a clear core-envelope boundary, which engulfs the more compact MS companion upon contact. Case-A binaries with initial primary masses blow $2\,\mathrm{M}_{\odot}$ also form contact binaries because of the orbital decay caused by tides. Overall, the incidences of mass-transferring binaries forming contact binaries or entering a classical CE phase are ${\geq}\,41\%$ and ${\geq}\,34\%$ for $M_{1,\,\mathrm{i}}\,\in \,\left[4.8; 12.6\right)\,\mathrm{M}_{\odot}$ and $M_{1,\,\mathrm{i}}\,\in \,\left[12.6; 20.8\right]\,\mathrm{M}_{\odot}$, respectively. Over the entire mass range of $M_{1,\,\mathrm{i}}\,\in \,\left[4.8; 20.8\right]\,\mathrm{M}_{\odot}$, the incidence is ${\geq}\,40\%$. These numbers are lower limits because they do not take into account the binaries that might enter a contact phase because of the interaction with non-accreted matter. Such systems are fairly common in our grid, and could alternatively result in systems with circumstellar disks and H$\alpha$ emission features. Potential observational counterparts could be found in the extended MS turnoff in young stellar clusters.

We find that mass transfer is non-conservative in a large part of the initial binary parameter space, which is caused by the spin-up of the accretors to critical rotation after accreting ${<}\,3\%$ of their own mass. The mass transfer efficiencies are $15\text{--}65\%$, $5\text{--}25\%$ and $25\text{--}50\%$ in Case-A,-B and -C mass transfer, respectively, for primary-star masses above $3\,\mathrm{M}_{\odot}$. The incidence of systems entering a contact phase is sensitive to the mass-transfer efficiency. Given the relatively low mass-transfer efficiencies in our models, we might be underestimating this incidence. Another consequence of non-conservative mass transfer is that Case-Bl and -C mass transfer is stable for mass ratios ${\geq}\,0.15\text{--}0.35$.

By assuming that systems Case-A and -Be systems with $M_{1,\,\mathrm{i}}\,\in \,\left[4.8; 20.8\right]\,\mathrm{M}_{\odot}$ that experience $\mathrm{L}_{2}$-overflow and/or runaway mass transfer merge, and Case-Bl and -C systems experiencing this enter a classical CE, we find stellar merger and classical CE incidences among mass-transferring binaries of ${\geq}\,12\%$ and ${\geq}\,19\%$, respectively. If we relax this assumption and assume that all contact binaries, except those for which either component reaches core-C exhaustion, eventually merge, the stellar merger incidence increases to ${\geq}\,19\%$. Just as for the incidence of mass-transferring binaries reaching a contact phase, the stellar merger and classical CE incidences are lower limits, which are sensitive to the mass-transfer efficiency.

Lastly, we compare our population of massive ($M_{1,\,\mathrm{i}} \gtrsim 5\,\mathrm{M}_{\odot}$) Case-A contact systems with observed (near-)contact systems. We find from our models that approximately two to three times as many contact binaries form with mass ratios $q < 0.5$ upon contact than with $q > 0.5$. Moreover, the majority of the systems with $q < 0.5$ form contact binaries because of runaway mass transfer or the thermal-timescale expansion of the accretor, with subsequent $\mathrm{L}_{2}$-overflow in more than half of the cases. Therefore, most of these systems quickly merge or detach. This is in agreement with the observations since almost no (near-)contact systems are observed with mass ratios below 0.5. Out of all systems forming contact binaries, $17\%$ do so on the accretor's nuclear timescale and avoid $\mathrm{L}_{2}$-overflow, and are expected to be stable and long-lived. The majority of these systems have $q > 0.5$ upon contact, which is again in excellent agreement with the observations.

Mergers and classical CEs are some of the most complex and fascinating outcomes of binary evolution. While we have identified the physical mechanisms that can lead to contact, the outcome of mergers and classical CEs still need to be explored in more detail and leave many open questions.

\begin{acknowledgements}
We wish to thank the referee, Hongwei Ge, for his valuable comments and suggestions, which have helped us further improve this work. We thank S.~Hekker, Ph.~Podsiadlowski, D.~Wei, V.~Bronner, P.~Marchant, and C.~Wang for the meaningful discussions, suggestions, and comments.
The authors acknowledge support from the Klaus Tschira Foundation.
This work has received funding from the European Research Council (ERC) under the European Union’s Horizon 2020 research and innovation programme (Grant agreement No.\ 945806). This work is supported by the Deutsche Forschungsgemeinschaft (DFG, German Research Foundation) under Germany’s Excellence Strategy EXC 2181/1-390900948 (the Heidelberg STRUCTURES Excellence Cluster).
\end{acknowledgements}

\bibliographystyle{aa}
\bibliography{phd_library}

\clearpage

\begin{appendix}

\section{Nuclear timescale expansion of the accretor}\label{app:example}
\begin{figure*}
\centering
  \includegraphics[width=18cm]{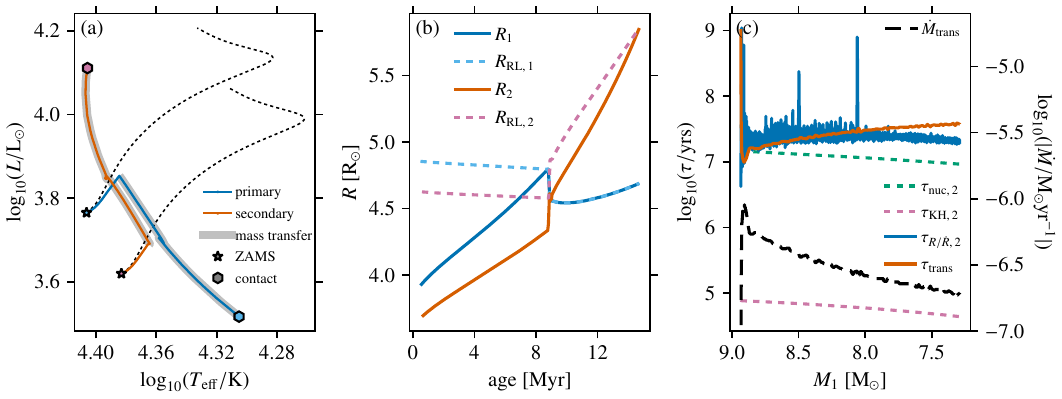}
     \caption{Same as Fig.\,\ref{fig:example_thermalExp} but for a $M_{1,\,\mathrm{i}}=10.2\,\mathrm{M}_{\odot}$, $q_{\mathrm{i}} = 0.9$ and $a_{\mathrm{i}} = 12.5\,\mathrm{R}_{\odot}$ binary.}
     \label{fig:example_nuclearExp}
\end{figure*}

Here, we present an example of a model forming a contact binary from the nuclear timescale expansion of the secondary (accretor) star. The evolution of the Case-A system with $M_{1,\,\mathrm{i}}=10.2\,\mathrm{M}_{\odot}$, $q_{\mathrm{i}} = 0.9$ and $a_{\mathrm{i}} = 12.5\,\mathrm{R}_{\odot}$ is shown in Fig.\,\ref{fig:example_nuclearExp}. After regaining thermal equilibrium, the primary star detaches and later fills its Roche lobe again because of nuclear expansion. The detachment phase is short compared to the evolutionary timescale and only visible by a slight wiggle in the HRD (Fig.\,\ref{fig:example_nuclearExp}a). During the first, thermal-timescale mass-transfer phase, the secondary star expands on its thermal timescale (Fig.\,\ref{fig:example_nuclearExp}b). However, during the second, nuclear-timescale mass-transfer phase, the secondary remains in thermal equilibrium and its radius evolves on a nuclear timescale (Fig.\,\ref{fig:example_nuclearExp}c) and increases with mass accretion. Because $R_{2}$ grows faster in time than $R_{\mathrm{RL},\,2}$, the secondary eventually fills its Roche lobe simultaneously with the primary, and a contact binary is formed. 

\section{Contact tracing results for other $M_{1,\,\mathrm{i}}$}\label{app:contact}
\begin{figure*}
\centering
  \includegraphics[width=18cm]{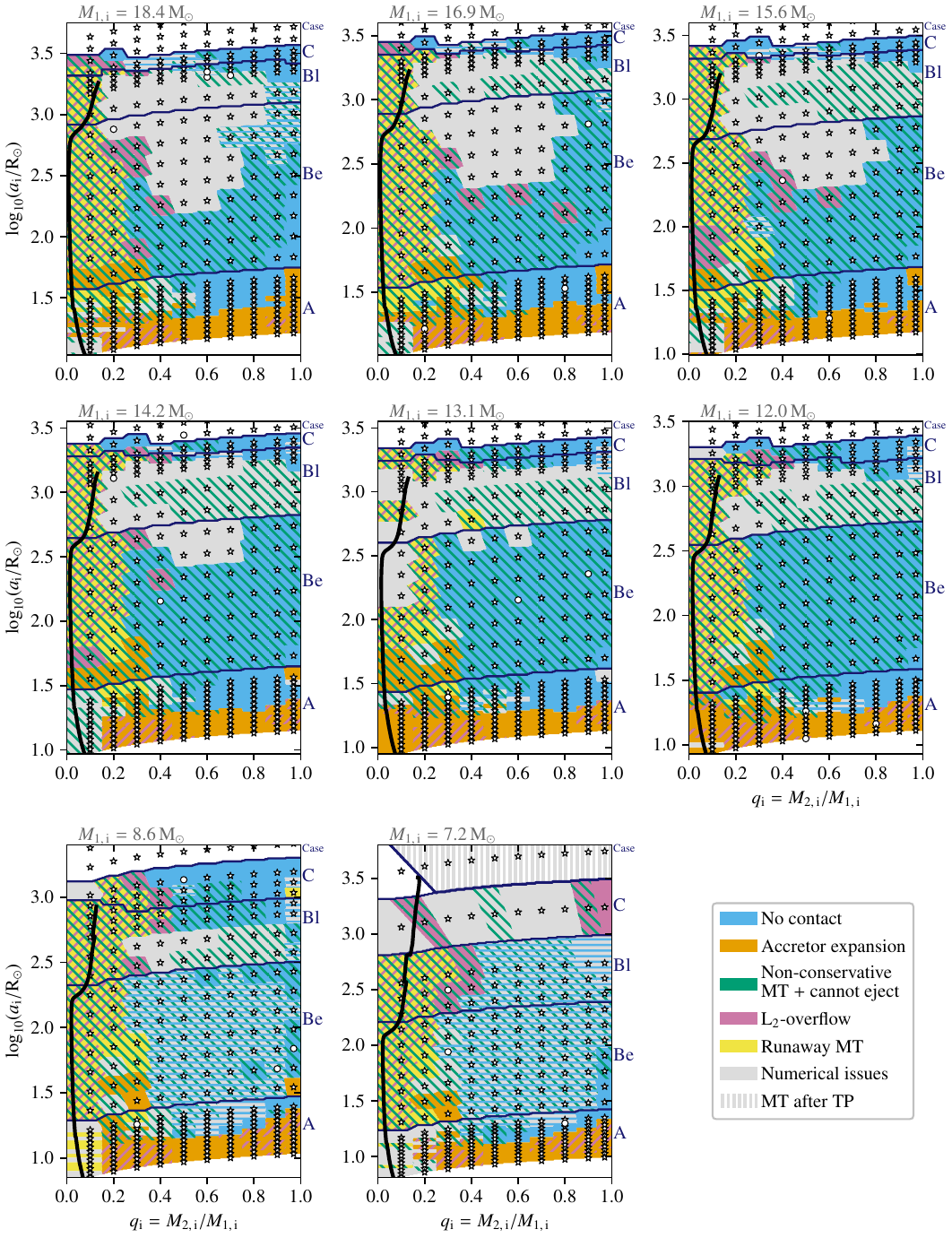}
    \caption{Contact tracing results for $M_{1,\,\mathrm{i}}=7.2\text{--}18.4\,\mathrm{M}_{\odot}$. Models with  $M_{1,\,\mathrm{i}}=7.2\,\mathrm{M}_{\odot}$ and initial separations larger than the Case-C systems experience numerical issues after the TP-AGB phase (see Sect.\,\ref{sec:wind_mass_loss}). Only the model with $q_{\mathrm{i}} = 0.1$ avoids these issues and does not initiate mass transfer.}
     \label{fig:appendix_contact_tracing1}
\end{figure*}

\begin{figure*}
\centering
  \includegraphics[width=18cm]{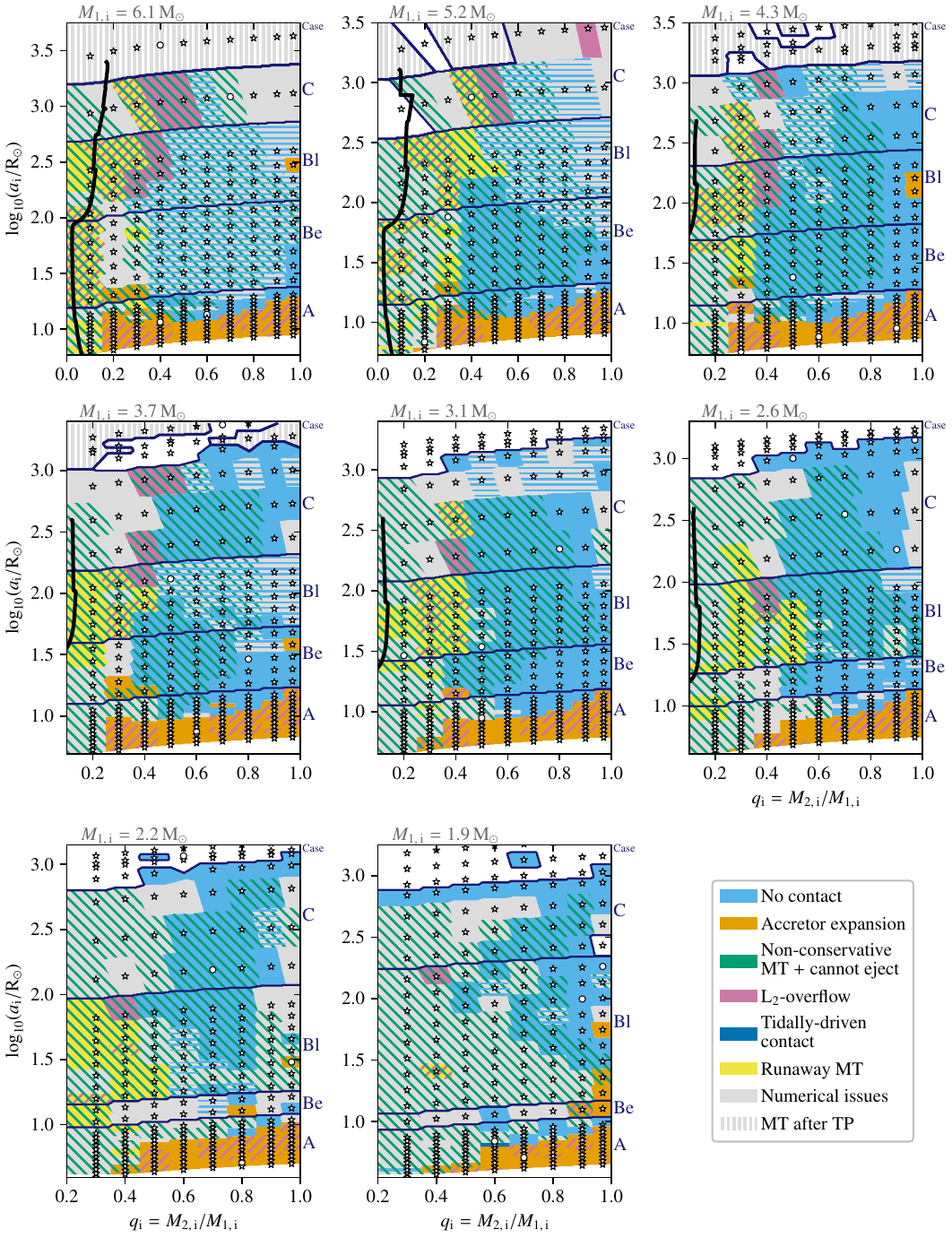}
    \caption{Contact tracing results for $M_{1,\,\mathrm{i}}=1.9\text{--}6.1\,\mathrm{M}_{\odot}$. Models with $M_{1,\,\mathrm{i}}=3.7\text{--}6.1\,\mathrm{M}_{\odot}$ experience numerical issues after the TP-AGB phase (see Sect.\,\ref{sec:wind_mass_loss}). This explains the unexpected onset of mass transfer at initial separations larger than those of systems avoiding mass transfer. For $M_{1,\,\mathrm{i}}=1.9\text{--}2.2\,\mathrm{M}_{\odot}$ we find certain models where mass transfer starts when the primary is on the WD cooling track and experiences sudden radial expansion.}
     \label{fig:appendix_contact_tracing2}
\end{figure*}

\begin{figure*}
\centering
  \includegraphics[width=18cm]{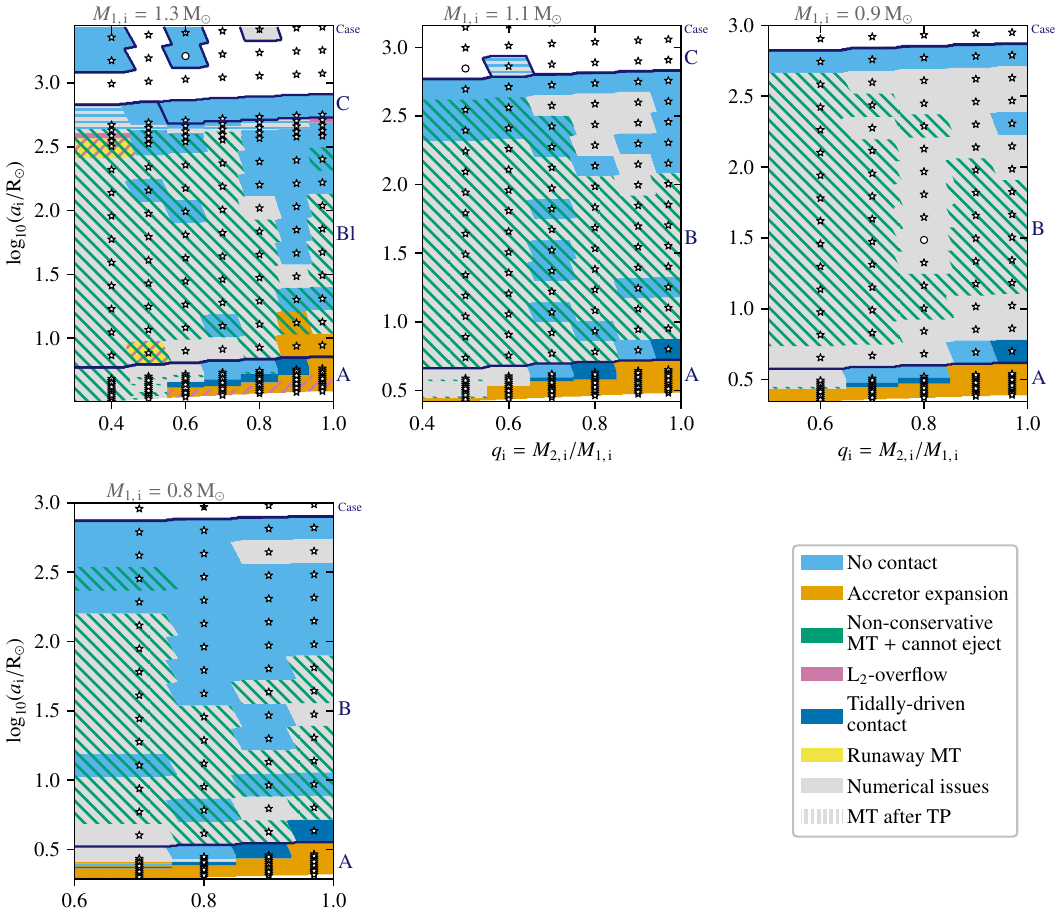}
    \caption{Contact tracing results for $M_{1,\,\mathrm{i}}=0.8\text{--}1.3\,\mathrm{M}_{\odot}$. For $M_{1,\,\mathrm{i}}=1.1\text{--}1.3\,\mathrm{M}_{\odot}$ we find certain models where mass transfer starts when the primary is on the WD cooling track and experiences sudden radial expansion.}
     \label{fig:appendix_contact_tracing3}
\end{figure*}

Figs.\,\ref{fig:appendix_contact_tracing1}--\ref{fig:appendix_contact_tracing3} contain the contact tracing results for the initial primary masses $M_{1,\,\mathrm{i}} = 0.8$, $0.9$, $1.1$, $1.3$, $1.9$, $2.2$, $2.6$, $3.1$, $3.7$, $4.3$, $5.2$, $6.1$, $7.2$, $8.6$, $12.0$, $13.1$, $14.2$, $15.6$, $16.9$ and $18.4\,\mathrm{M}_{\odot}$.

\section{Evolutionary states at contact or termination of models}\label{app:contact_configs}
\begin{figure*}
\centering
  \includegraphics[width=18cm]{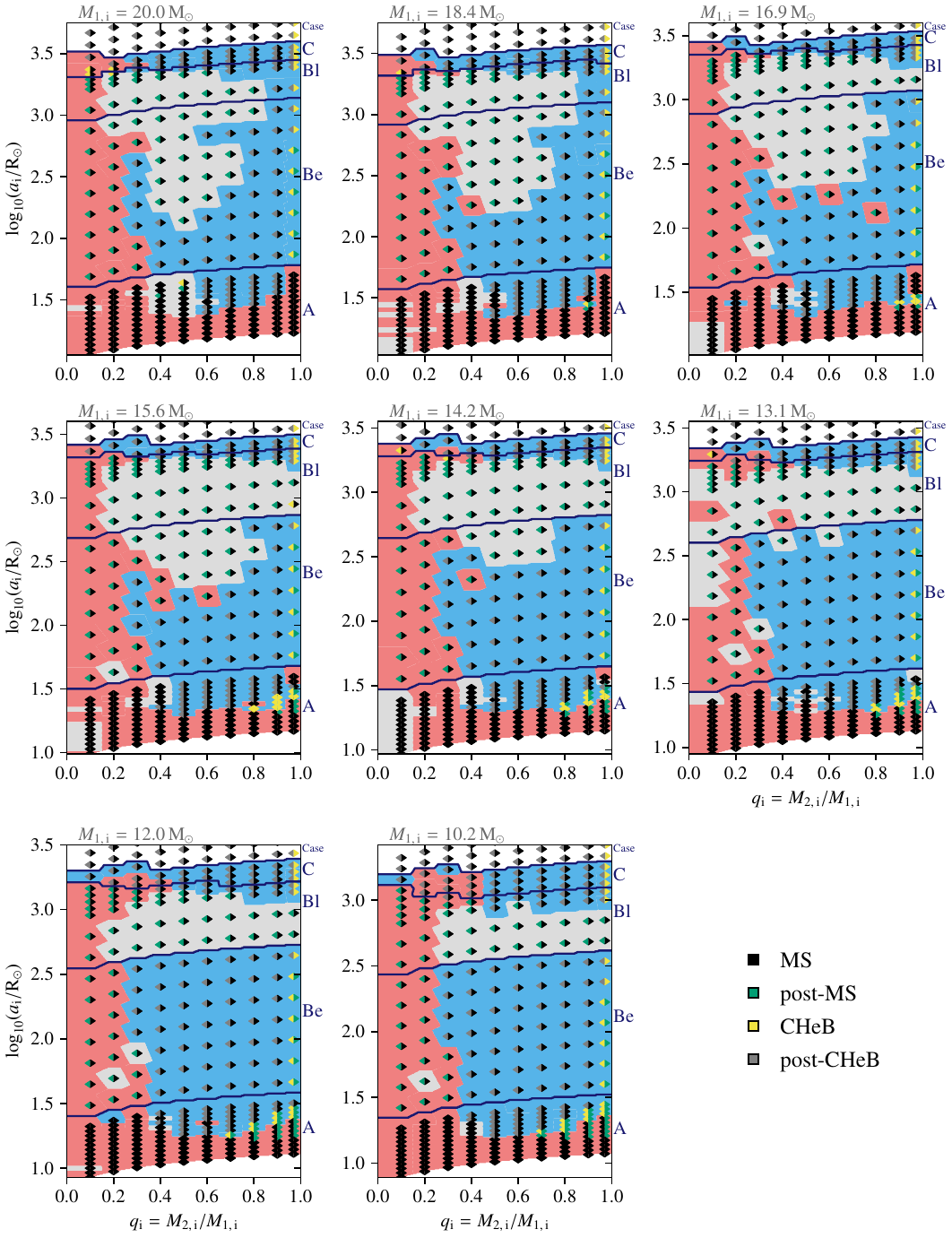}
    \caption{Evolutionary state of the primary (left triangle) and secondary (right triangle) at contact or termination for $M_{1,\,\mathrm{i}}=8.6\text{--}20.0\,\mathrm{M}_{\odot}$. The pink, blue and grey background colours indicate systems that get into contact, avoid contact and fail to converge numerically, respectively. Post-MS stars are those that have exhausted hydrogen but have not yet ignited helium in their core. ``CHeB'' stands for core helium burning.}
     \label{fig:contact_configs}
\end{figure*}

\begin{figure*}
\centering
  \includegraphics[width=18cm]{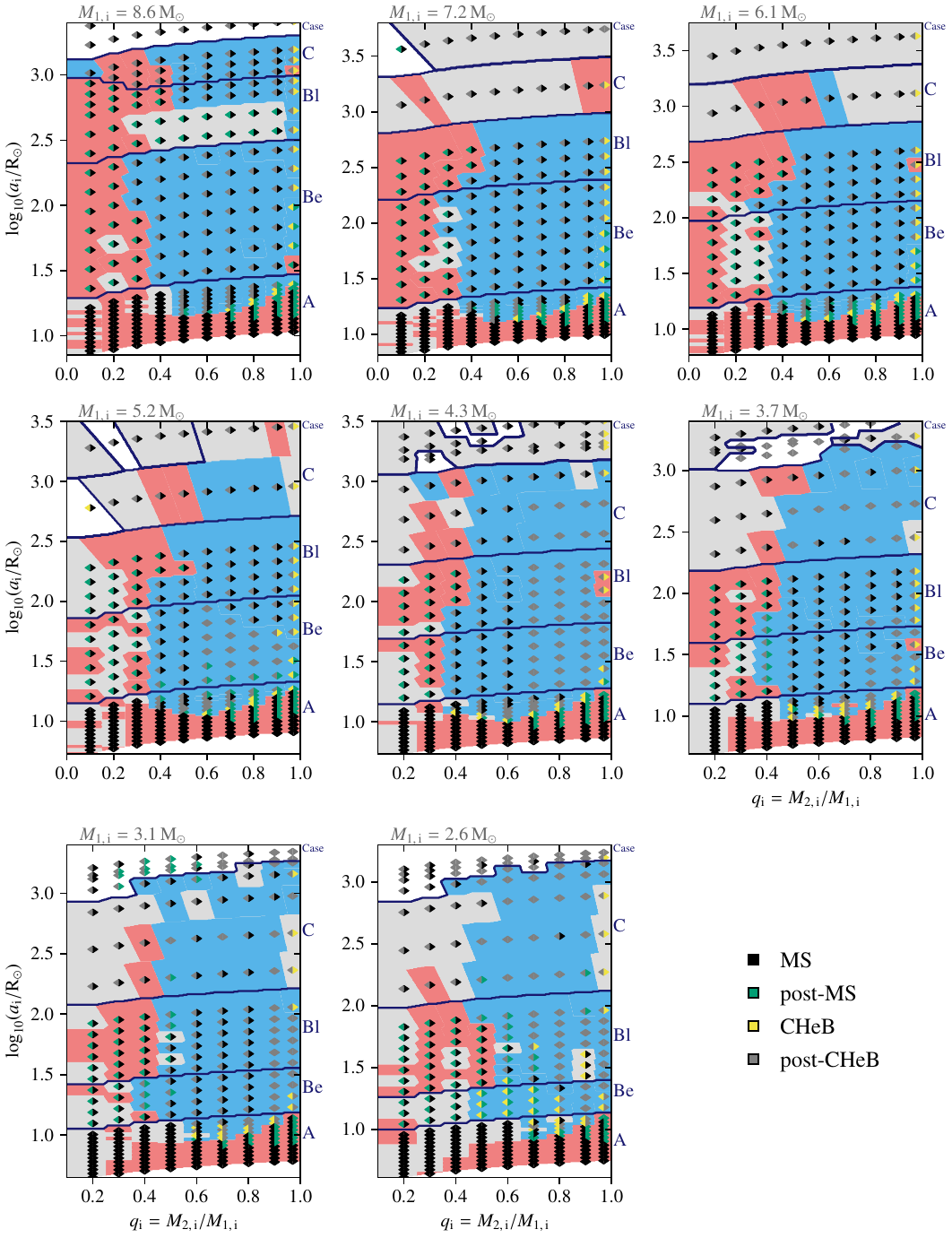}
    \caption{Same as for Fig.\,\ref{fig:contact_configs} but for $M_{1,\,\mathrm{i}}=1.9\text{--}7.2\,\mathrm{M}_{\odot}$.}
     \label{fig:contact_configs2}
\end{figure*}

\begin{figure*}
\centering
  \includegraphics[width=18cm]{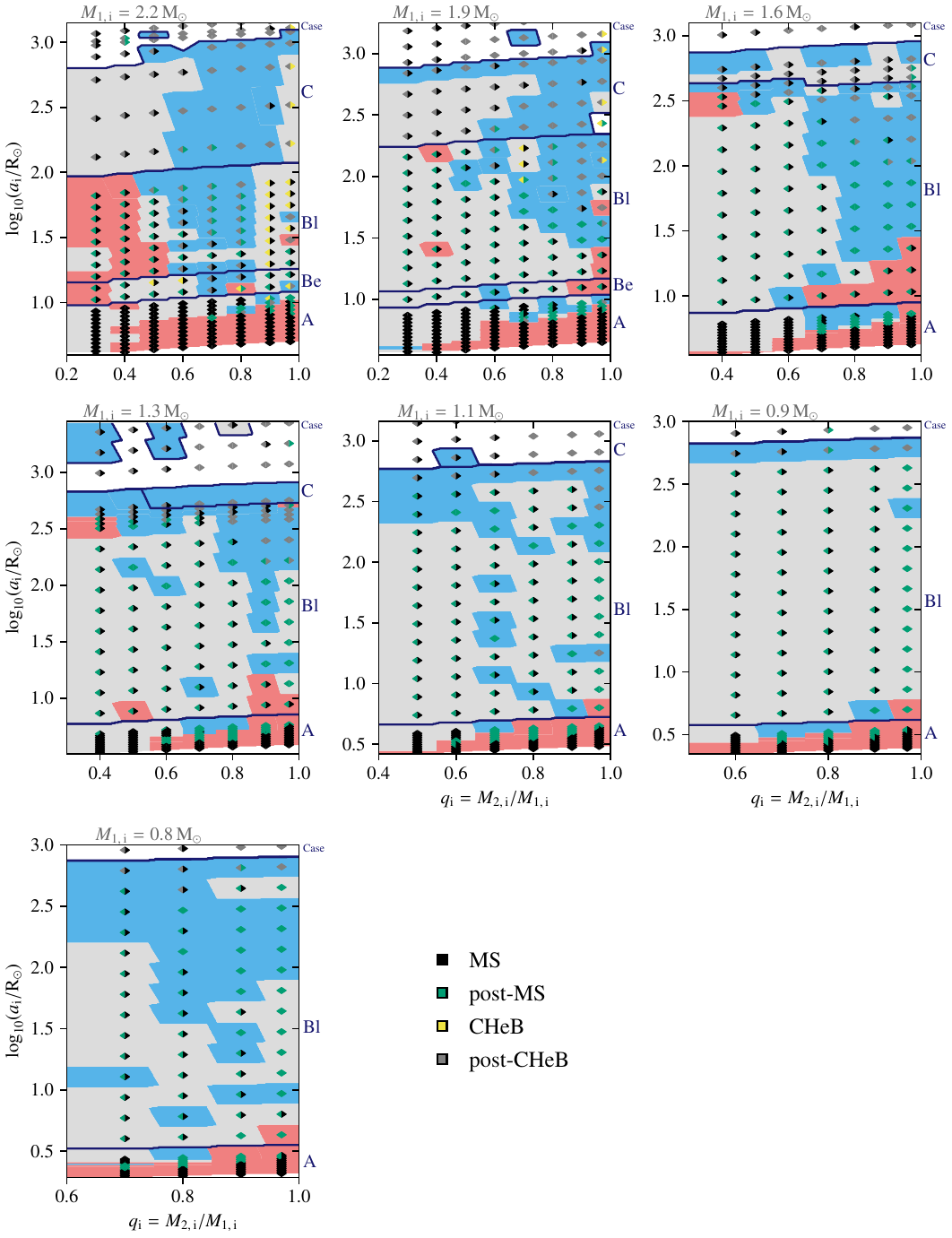}
    \caption{Same as for Fig.\,\ref{fig:contact_configs} but for $M_{1,\,\mathrm{i}}=0.8\text{--}1.6\,\mathrm{M}_{\odot}$.}
     \label{fig:contact_configs3}
\end{figure*}

In Figs.\,\ref{fig:contact_configs}--\ref{fig:contact_configs3} we show the evolutionary state of the binary components for all systems in our grid at contact, or at termination for systems that avoid contact.

\section{Expansion timescales for Case-A contact systems}\label{app:exp_ts}
\begin{figure*}
\centering
  \includegraphics[width=18cm]{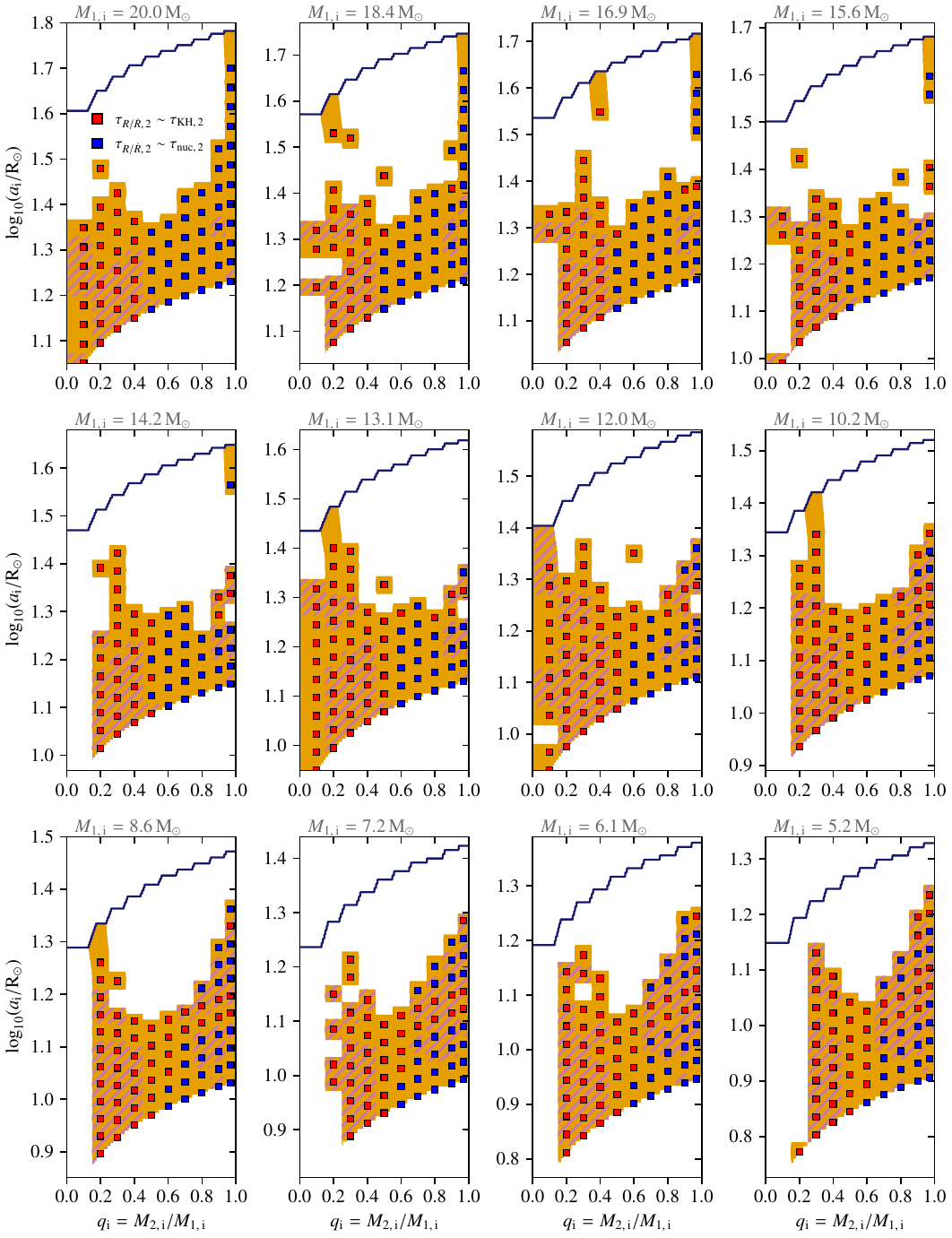}
    \caption{Expansion timescales for Case-A contact systems formed through the expansion of the accretor star with $M_{1,\,\mathrm{i}}=5.2\text{--}20.0\,\mathrm{M}_{\odot}$. In models marked with filled red (blue) squares, the accretor expands on its thermal (nuclear) timescale prior to the onset of contact. The colour scheme is the same as in Fig.\,\ref{fig:contact_Mp1016}. The dark blue solid line indicates the division between Case-A and -Be systems.}
     \label{fig:expansion_timescales}
\end{figure*}

\begin{figure*}
\centering
  \includegraphics[width=18cm]{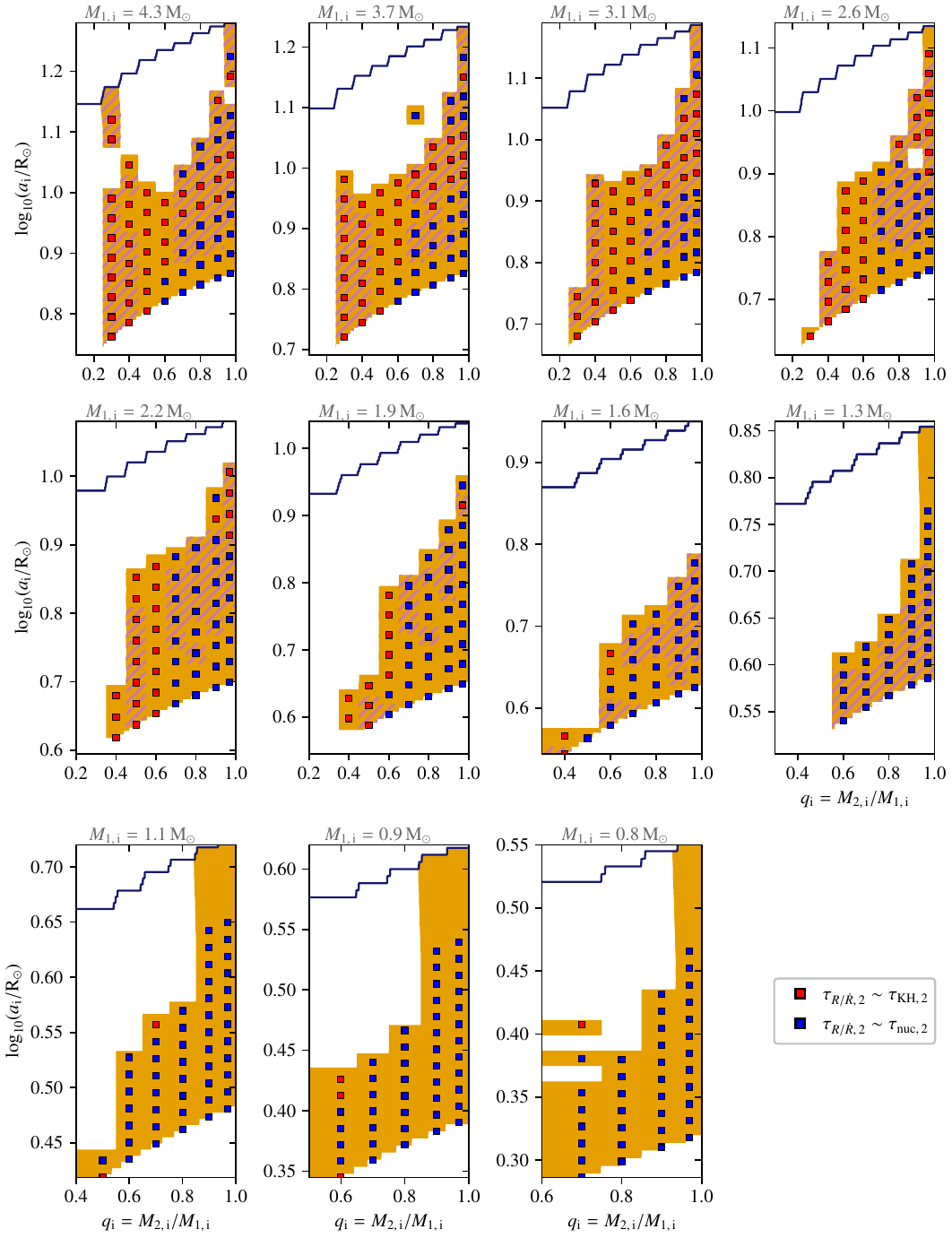}
    \caption{Same as Fig.\,\ref{fig:expansion_timescales} but for $M_{1,\,\mathrm{i}}=0.8\text{--}4.3\,\mathrm{M}_{\odot}$.}
     \label{fig:expansion_timescales2}
\end{figure*}

In Figs.\,\ref{fig:expansion_timescales}--\ref{fig:expansion_timescales2}, the Case-A region of the initial binary parameter space is shown for all initial primary masses of the grid. The contact systems formed through the expansion of the accretor and those which experience $\mathrm{L}_{2}$-overflow are indicated with the same colour scheme as in Fig.\,\ref{fig:contact_Mp1016}. Each model is marked with a red or blue square, based on whether the mean of the accretor's expansion timescale is of the order of its mean thermal or nuclear timescale, respectively. 

\section{Assignment criteria for evolutionary outcomes of models with numerical issues}\label{app:assignments}
Table \ref{tab:crashed} lists the criteria used to determine the evolutionary outcome of models that experience numerical issues. The outcome assignment criteria for these models are based on the outcomes of neighbouring models and information from equivalent systems at different initial primary masses. 
\begin{table}[h!]
\caption{Evolutionary outcome assignment criteria for models with numerical issues.}
\label{tab:crashed}
\resizebox{\columnwidth}{!}{%
\begin{tabular}{l|ccc}
\hline \hline
              $q_{\mathrm{i}}$-range      & Case~A & Case~Be & Case~Bl/C \\ \hline
$q_{\mathrm{i}} < 0.3$ & Accr. exp.\tablefootmark{a}       & Runaway MT         & Runaway MT           \\
$q_{\mathrm{i}} \geq 0.3$   & No contact        & No contact         & No contact      \\  \hline  
\end{tabular}%
}
\tablefoottext{a}{ ``Accr. exp.'' = ``Accretor expansion''.}
\end{table}

\section{Mass-transfer efficiency}\label{app:mt_eff}

\begin{figure}
     \centering
     \resizebox{\hsize}{!}{\includegraphics{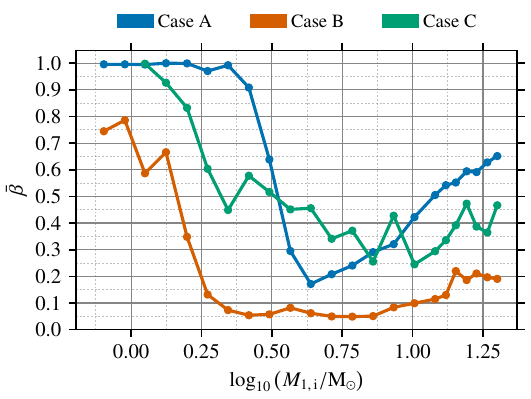}}
     \caption{Mean mass-transfer efficiency $\bar{\beta}$ per mass-transfer case for each initial primary mass $M_{1,\,\mathrm{i}}$.}
     \label{fig:mt_efficiency}
 \end{figure}

We show the mean mass-transfer efficiency $\bar{\beta}$ for Case-A, -B and -C mass transfer as a function of the initial primary mass $M_{1,\,\mathrm{i}}$ in Fig.\,\ref{fig:mt_efficiency}. We use the birth probabilities $p_{\mathrm{birth}}$ (Sect.\,\ref{sec:method_probs}) of each model as weights for the computation of these mean values. The value of the mass-transfer efficiency for each individual model is averaged over time. We only consider mass transfer in systems that avoid contact. \\

First, we see that for all mass-transfer cases the mean mass-transfer efficiency $\bar{\beta}$ is relatively high for $\log_{10}(M_{1,\,\mathrm{i}}) < 0.25\text{--}0.45$. For $\log_{10}(M_{1,\,\mathrm{i}}) > 0.25\text{--}0.45$, $\bar{\beta}$ has moderate to low values. The high values of $\bar{\beta}$ for binaries with lower initial primary masses should be regarded as an approximation since for all mass-transfer cases the models avoiding contact are sparse due to numerical issues (Fig.\,\ref{fig:appendix_contact_tracing2}--\ref{fig:appendix_contact_tracing3}). The models that avoided numerical issues have high mass-transfer efficiencies.

Case-A mass transfer has low to moderate mean mass-transfer efficiencies for $\log_{10}(M_{1,\,\mathrm{i}}) > 0.45$. The initially closest Case-A binaries go through conservative mass transfer before they form contact binaries and are, therefore, not taken into account for the computation of $\bar{\beta}$. Case-A binaries that avoid contact are on initially wider orbits. The accretors in these systems have longer tidal synchronisation times, so tides are unable to prevent them from rotating critically. As discussed in Sect.\,\ref{sec:discussion_efficiency}, Case-A mass transfer consists of a short thermal-timescale phase, which is mostly non-conservative, and a more conservative, longer nuclear-timescale phase. This results in values of $\bar{\beta}$ between ${\sim}\,0.15$ and ${\sim}\,0.65$.

Case-B mass transfer in binaries with  $\log_{10}(M_{1,\,\mathrm{i}}) > 0.25$ have relatively low values of $\bar{\beta}$, with values between ${\sim}\,0.05$ and ${\sim}\,0.25$. In virtually all systems with Case-B mass transfer, the accretors reach critical rotation after accreting a few percent of their own mass, which quenches accretion (Sect.\,\ref{subsec:res_contact_phases} and \ref{sec:discussion_efficiency}).

Case-C mass transfer is generally highly non-conservative (Sect.\,\ref{sec:stable_caseC}). However, the contribution of Case-C systems with relatively short mass-transfer phases before core-C exhaustion in which the accretor is not spun up to critical rotation, increases the value of $\bar{\beta}$ for $\log_{10}(M_{1,\,\mathrm{i}}) > 0.25$.

The contribution of both conservative and non-conservative mass transfer in Case-A and -C mass transfer cases, for the former in terms of mass-transfer phases (thermal and nuclear) and for the latter in terms of systems (highly non-conservative models and conservative models), results in somewhat similar mean mass-transfer efficiencies. 

\section{Table with contact tracing results}\label{app:table}
Table \ref{tab:contact_tracing_results} contains an extract of the table containing the contact tracing results for all our models.
\begin{landscape}
\begin{table}[]
\caption{Extract of the table with the contact tracing results of all 5957 binary \texttt{MESA} models. The full table is available online at \url{https://zenodo.org/doi/10.5281/zenodo.10148634}.}
\label{tab:contact_tracing_results}
\resizebox{\columnwidth}{!}{%
\begin{tabular}{cccccccccccccccccccc}
\hline \hline
\begin{tabular}[c]{@{}c@{}}$M_{1,\,\mathrm{i}}$\tablefootmark{a}\\ $[\mathrm{M}_{\odot}]$\end{tabular} & \begin{tabular}[c]{@{}c@{}}$M_{2,\,\mathrm{i}}$\\ $[\mathrm{M}_{\odot}]$\end{tabular} & $\log_{10}(a_{\mathrm{i}}/\mathrm{R}_{\odot})$ & $\log_{10}(P_{\mathrm{i}}/\mathrm{d})$ & \begin{tabular}[c]{@{}c@{}}$M_{1,\,\mathrm{f}}$\tablefootmark{b}\\ $[\mathrm{M}_{\odot}]$\end{tabular} & \begin{tabular}[c]{@{}c@{}}$M_{2,\,\mathrm{f}}$\\ $[\mathrm{M}_{\odot}]$\end{tabular} & $\log_{10}(a_{\mathrm{f}}/\mathrm{R}_{\odot})$ & $\log_{10}(P_{\mathrm{f}}/\mathrm{d})$ & $\log_{10}(\mathrm{age}_{\mathrm{f}}/\mathrm{yrs})$ & AE\tablefootmark{c} & RMT\tablefootmark{d} & NCCE\tablefootmark{e} & L2O\tablefootmark{f} & TDC\tablefootmark{g} & NC\tablefootmark{h} & MTTP\tablefootmark{i} & NI\tablefootmark{j} &  \begin{tabular}[c]{@{}c@{}}Case\\ {[}A,B,C{]}\end{tabular} & $\mathrm{ES}_{1}$\tablefootmark{k} & $\mathrm{ES}_{2}$\tablefootmark{l} \\ \hline
0.80 & 0.56 & 0.407 & -0.392 & 0.65 & 0.71 & 0.324 & -0.517 & 10.343 & 1 & 0 & 0 & 0 & 0 & 0 & 0 & 0 & [1,0,0] & MS & MS \\
0.95 & 0.57 & 2.745 & 3.090 & 0.53 & 0.57 & 2.865 & 3.342 & 10.481 & 0 & 0 & 0 & 0 & 0 & 1 & 0 & 0 & [0,1,0] & post-CHeB & MS\\
1.12 & 0.90 & 0.616 & -0.165 & 0.22 & 1.79 & 1.130 & 0.607 & 9.953 & 0 & 0 & 0 & 0 & 1 & 0 & 0 & 0 & [1,1,0] & post-MS & post-MS\\
1.33 & 0.53 & 2.629 & 2.872 & 0.78 & 0.59 & 2.641 & 2.958 & 9.631 & 0 & 0 & 1 & 0 & 0 & 1 & 0 & 1 & [0,1,1] & post-CHeB & MS\\
1.58 & 1.11 & 0.681 & -0.130 & 1.13 & 1.55 & 0.668 & -0.148 & 9.387 & 1 & 0 & 0 & 1 & 0 & 0 & 0 & 0 & [1,0,0] & MS & MS\\
1.87 & 1.68 & 2.254 & 2.169 & 0.50 & 1.55 & 2.660 & 2.898 & 9.308 & 0 & 0 & 0 & 0 & 0 & 1 & 0 & 0 & [0,1,1] & post-CHeB & post-CHeB\\
2.21 & 0.66 & 1.554 & 1.166 & 2.21 & 0.67 & 1.438 & 0.992 & 8.926 & 0 & 1 & 1 & 0 & 0 & 0 & 0 & 0 & [0,1,0] & post-MS & MS\\
2.62 & 1.83 & 0.934 & 0.140 & 0.30 & 2.61 & 1.821 & 1.563 & 8.986 & 0 & 0 & 0 & 0 & 0 & 1 & 0 & 0 & [1,1,0] & post-MS & post-MS\\
3.10 & 1.24 & 0.832 & -0.007 & 3.07 & 1.27 & 0.809 & -0.041 & 8.339 & 1 & 0 & 0 & 1 & 0 & 0 & 0 & 0 & [1,0,0] & MS & MS\\
3.68 & 3.57 & 0.988 & 0.116 & 2.12 & 5.12 & 1.141 & 0.346 & 8.300 & 1 & 0 & 0 & 1 & 0 & 0 & 0 & 0 & [1,0,0] & MS & post-MS\\
4.35 & 2.17 & 2.035 & 1.710 & 0.79 & 2.22 & 2.432 & 2.472 & 8.834 & 0 & 0 & 1 & 0 & 0 & 1 & 0 & 0 & [0,1,1] & post-CHeB & MS\\
5.16 & 0.52 & 1.207 & 0.497 & 5.11 & 0.52 & 1.111 & 0.355 & 7.976 & 0 & 1 & 1 & 1 & 0 & 0 & 0 & 0 & [0,1,0] & post-MS & MS\\
6.11 & 3.67 & 4.357 & 5.105 & 1.60 & 3.67 & 4.625 & 5.641 & 7.860 & 0 & 0 & 0 & 0 & 0 & 0 & 1 & 0 & [0,0,1] & post-CHeB & MS\\
7.24 & 5.79 & 1.202 & 0.309 & 1.70 & 9.10 & 1.881 & 1.369 & 7.770 & 1 & 0 & 0 & 0 & 0 & 0 & 0 & 0 & [1,1,0] & post-MS & post-MS\\
8.57 & 5.14 & 2.385 & 2.072 & 1.24 & 5.27 & 2.871 & 2.963 & 7.557 & 0 & 0 & 1 & 0 & 0 & 1 & 0 & 0 & [0,1,0] & post-CHeB & MS\\
10.16 & 6.10 & 1.399 & 0.557 & 1.30 & 6.34 & 2.298 & 2.070 & 7.448 & 0 & 0 & 1 & 0 & 0 & 1 & 0 & 0 & [1,1,1] & post-CHeB & MS\\
12.03 & 10.83 & 1.246 & 0.253 & 5.72 & 16.68 & 1.489 & 0.622 & 7.282 & 1 & 0 & 0 & 1 & 0 & 0 & 0 & 0 & [1,0,0] & MS & post-MS\\
13.14 & 6.57 & 2.637 & 2.371 & 3.80 & 6.72 & 2.784 & 2.729 & 7.227 & 0 & 0 & 1 & 0 & 0 & 1 & 0 & 0 & [0,1,1] & post-CHeB & MS\\
14.25 & 11.40 & 1.243 & 0.224 & 12.72 & 12.88 & 1.222 & 0.192 & 6.910 & 1 & 0 & 0 & 0 & 0 & 0 & 0 & 0 & [1,0,0] & MS & MS\\
15.57 & 15.10 & 1.403 & 0.425 & 7.01 & 22.92 & 1.691 & 0.862 & 7.129 & 1 & 0 & 0 & 0 & 0 & 0 & 0 & 0 & [1,0,0] & MS & post-MS\\
16.88 & 1.69 & 3.262 & 3.323 & 16.19 & 1.69 & 3.066 & 3.036 & 7.037 & 0 & 1 & 1 & 1 & 0 & 0 & 0 & 0 & [0,1,0] & post-MS & MS\\
18.44 & 14.75 & 2.152 & 1.532 & 5.89 & 14.84 & 2.601 & 2.307 & 7.027 & 0 & 0 & 1 & 0 & 0 & 1 & 0 & 0 & [0,1,0] & post-CHeB & MS\\
20.00 & 19.40 & 3.510 & 3.532 & 14.61 & 16.50 & 3.562 & 3.660 & 6.986 & 0 & 0 & 0 & 0 & 0 & 1 & 0 & 0 & [0,0,1] & post-CHeB & CHeB\\
... & ... & ... & ... & ... & ... & ... & ... & ... & ... & ... & ... & ... & ... & ... & ... & ... & ... & ... & ...\\
\hline
\end{tabular}%
}
\tablefoottext{a}{``i'' = ``initial''}
\tablefoottext{b}{``f'' = ``final'' -- at contact or termination}
\tablefoottext{c}{Accretor expansion}
\tablefoottext{d}{Runaway mass transfer}
\tablefoottext{e}{Non conservative mass transfer + cannot eject}
\tablefoottext{f}{$\mathrm{L}_{2}$-overflow}
\tablefoottext{g}{Tidally-driven contact}
\tablefoottext{h}{No contact}
\tablefoottext{i}{Mass transfer after thermal pulses}
\tablefoottext{j}{Numerical issues}
\tablefoottext{k}{Primary's evolutionary state at contact or termination. MS: before core-H exhaustion. Post-MS: after core-H exhaustion and before core-He ignition. CHeB: after core-He ignition and before core-He exhaustion. Post-CHeB: after core-He exhaustion.}
\tablefoottext{l}{Secondary's evolutionary state at contact or termination.}
\end{table}
\end{landscape}

\end{appendix}

\end{document}